\PassOptionsToPackage{table,dvipsnames}{xcolor}
\documentclass[sigconf, nonacm]{acmart}

\usepackage{amsmath,amsfonts,amsthm}
\usepackage{graphicx}
\usepackage{booktabs}
\usepackage{tablefootnote}
\usepackage{hhline}
\usepackage{subcaption}
\usepackage{textcomp}
\usepackage{xcolor}
\usepackage{scalerel}
\usepackage[frozencache,cachedir=minted-cache]{minted}
\usepackage{forest}
\usepackage{url}
\usepackage{mathtools}
\usepackage{todonotes}
\usepackage{multicol}
\usetikzlibrary {arrows.meta}
\usetikzlibrary{tikzmark}
\usepackage[ruled,vlined]{algorithm2e}
\usepackage{enumitem}

\newcommand{\nop}[1]{}
\newcommand{\Yann}{\mathit{Yann}}
\newcommand{\Count}{\mathit{Freq}}
\newtheorem{example}{Example}[section]
\newcommand{\pwg}{piece\-wise-guarded\xspace}
\newcommand{\Pwg}{Piece\-wise-guarded\xspace}
\newcommand{\PWG}{Piece\-wise-Guarded\xspace}
\newcommand{\Attr}{\mathit{Att}}
\newcommand{\Aggj}{\mathit{Agg}_j}
\newcommand{\Agg}{\mathit{Agg}}
\newcommand{\MIN}{\texttt{MIN}\xspace}
\newcommand{\MAX}{\texttt{MAX}\xspace}
\newcommand{\SUM}{\texttt{SUM}\xspace}
\newcommand{\COUNT}{\texttt{COUNT}}
\newcommand{\Aj}{A_j(f_j(\bar B_j))}

\newcommand{\As}{A_s(f_s(\bar B_s))}
\newcommand{\fBj}{f_j(\bar B_j)}
\newcommand{\sd}[1]{{\scriptsize{$\color{darkgray}  \pm #1$}}}

\newcommand\vldbdoi{XX.XX/XXX.XX}

\newcommand\vldbavailabilityurl{URL_TO_YOUR_ARTIFACTS}

\newif\ifArxiv
\Arxivtrue

\begin{document}
\title{Avoiding Materialisation for 
Guarded Aggregate Queries}

\author{Matthias Lanzinger}
\email{matthias.lanzinger@tuwien.ac.at}
\orcid{XXXX-XXXX-XXXX}
\affiliation{%
  \institution{TU Wien}
  \city{Vienna}
  \country{Austria}
}
\author{Reinhard Pichler}
\email{reinhard.pichler@tuwien.ac.at}
\orcid{XXXX-XXXX-XXXX}
\affiliation{%
  \institution{TU Wien}
  \city{Vienna}
  \country{Austria}
}
\author{Alexander Selzer}
\email{alexander.selzer@tuwien.ac.at}
\orcid{XXXX-XXXX-XXXX}
\affiliation{%
  \institution{TU Wien}
  \city{Vienna}
  \country{Austria}
}

\begin{abstract}
Optimising queries with many joins is known to be a hard problem. 
The explosion of intermediate results as opposed to a much smaller final result 
poses a serious challenge to modern database management systems (DBMSs). 
This is particularly glaring in case of analytical queries that 
join many tables but ultimately 
only output comparatively small aggregate information.
Analogous problems are faced by graph database systems when processing 
analytical queries with aggregates on top of complex path queries.

In this work, 
we propose novel optimisation techniques both, on the logical and physical level, 
that allow us to avoid the materialisation of join results for certain types of aggregate queries.
The key to these optimisations is the notion of {\em guardedness},
by which we impose restrictions on the occurrence of attributes in 
\texttt{GROUP}~\texttt{BY} clauses and in aggregate expressions.
The efficacy of our optimisations is validated 
through their implementation  in Spark SQL and extensive empirical evaluation
on various standard benchmarks.
\end{abstract}

\maketitle



\section{Introduction}
\label{sect:Introduction}

As the amounts of data to be processed increase, 
the limitations of established query evaluation methods become apparent.
While modern DBMSs, such as Spark SQL, provide powerful frameworks for processing massive datasets, 
they may still struggle with many different types of complex queries.
A key issue 
is the potential explosion of intermediate results, even if the final output is much smaller. This is particularly glaring in the context of analytical queries that combine data from 
many tables but 
ultimately produce comparatively small aggregate results.
Analogous problems are faced by graph database systems when processing 
analytical queries with aggregates on top of complex path queries.

 

Traditionally, database engines try to avoid expensive intermediate blow-up by optimising the order in which joins are processed. More recently, worst-case optimal join techniques, which
limit the blow-up to the theoretical worst-case, have gained popularity as an alternative approach for reducing intermediate materialisation.
However, while these techniques may help to {\em alleviate}  
the problem of (unnecessarily) big intermediate results in certain cases,
they do not {\em eliminate} the problem ~\cite{DBLP:journals/siamcomp/AtseriasGM13}. 
Furthermore, the problem of big intermediate results holds all the same even if joins are made only along foreign-key 
relationships~\cite{DBLP:conf/sigmod/ManciniKCMA22}.
\nop{**************
Moreover, we note that 
the problem of big intermediate results as opposed to small final results
also applies to 
graph databases when executing analytical queries that involve 
aggregates over complex path queries.
**************}

For queries that exhibit certain favourable structural properties, Yannakakis~\cite{DBLP:conf/vldb/Yannakakis81} showed that it is possible to avoid the materialisation of unnecessary intermediate results.
More specifically, if a query is acyclic -- that is, if it has a join tree  
(formal definitions of acyclicity and join trees will be provided in Section \ref{sect:ACQs})
-- then one can eliminate all dangling tuples 
(i.e., tuples not contributing to the final join result) 
via semi-joins before the actual join computation starts. 
However, even if dangling tuples have been eliminated and joins are evaluated in an
optimal order, the intermediate results thus produced may still become prohibitively big. 
Especially in aggregate queries, where only a restricted amount of information is ultimately output, 
it would be highly desirable to reduce or, ideally, 
avoid altogether the materialisation of intermediate join results.

Actually, it is well known that, in case of Boolean acyclic queries (e.g., if we are only interested whether 
the result of a join query is non-empty), the final answer can be determined by carrying out only semi-joins and skipping the entire join step.
So-called 0MA (= zero-materialisation aggregate) queries have recently been identified as a special class of queries with
\texttt{MIN} or  \texttt{MAX} aggregates over acyclic join queries,  
which can be evaluated without materialising any (intermediate or final) join results \cite{DBLP:conf/amw/GottlobLLOPS23,DBLP:journals/corr/abs-2303-02723}  
(a formal definition of 0MA queries is given in Section \ref{sect:ACQs}).
%
Several works~\cite{DBLP:journals/jcss/PichlerS13,DBLP:journals/mst/0001M15} investigated how variations 
of the same algorithmic idea also apply to join queries with \texttt{COUNT} aggregates. 
Subsequently, these ideas were extended to more general aggregate queries in the FAQ-framework (Functional Aggregate Queries)~\cite{DBLP:conf/pods/KhamisNR16} and, similarly, under the name AJAR (Aggregations and Joins over Annotated Relations)~\cite{DBLP:conf/pods/JoglekarPR16}. We offer a more detailed account of related work in Section~\ref{sect:RelatedWork}. 

However, previous works in this area have left a gap: Most approaches (such as FAQ and AJAR mentioned above)
aim at {\em reducing} (not eliminating) 
the number of joins  and/or the cost of computing them by applying sophisticated join techniques. But the computation and materialisation of joins
remains the dominating cost factor. On the other hand, approaches that avoid the computation or materialisation of intermediate join results
depend on severe restrictions of the class of queries, such as Boolean queries or 0MA queries. Indeed, as we will show in our empirical evaluation in Section \ref{sect:Experiments}, only a small fraction of the queries in 
the standard benchmarks considered here satisfies these restrictions.

The {\em goal of this work} is to identify a class of 
aggregate queries which can be evaluated without the need to compute or materialise any joins and which, nevertheless, cover many practical cases.
The key to this class of queries is the notion of {\em guardedness}. More specifically, we call a query with aggregates on top of 
an acyclic join query {\em guarded} if all attributes involved in a \texttt{GROUP}~\texttt{BY} clause and in any aggregate expression 
are contained in a single relation, referred to as the ``{\em guard}''. Here we allow any aggregate function from the ANSI SQL standard -- including  statistical functions such as \texttt{MEDIAN}, \texttt{VARIANCE}, \texttt{STDDEV}, \texttt{CORR}, etc.

\begin{example}
\label{ex:tpch-query}
We illustrate the basic ideas with the simple query given in Figure~\ref{fig:tpch-query} over the well-known TPC-H schema.
The query asks for the median of the account balance of suppliers from one of the regions 
'Europe' or 'Asia' for parts with above average  price. 
The join-structure of the query is clearly acyclic as is witnessed by the join tree displayed in Figure~\ref{fig:tpch-query-jointree}.
Moreover the query is trivially {\em guarded}, since it has no grouping and the aggregation is over a single attribute.

Note that the subquery is only used to realise a selection (locally) on the 
relation \texttt{part}. After applying this selection on 
the \texttt{part} relation and also the selection on the \texttt{region} relation, 
the query can be evaluated by {\em propagating frequencies} of attribute value combinations 
rather than intermediate join results
up the join tree. The \texttt{MEDIAN}-aggregate can then be evaluated on the resulting relation at the root node.
Indeed, suppose that we have computed all tuples $t_1, \dots, t_n$ of relation \texttt{supplier} together with the
corresponding frequencies $c_1, \dots, c_n$ of these tuples in the full join result of the five relations. Then we can 
order the values $v_1, \dots, v_n$ of these tuples in ascending order and, by taking the frequencies 
$c_1, \dots, c_n$ into account, it is an easy task to read off the median value.
This is in sharp contrast to traditional query evaluation, which would first compute the {\em join of the five relations} and 
evaluate the aggregate on the full join result.
\hfill $\diamond$
\end{example}

\begin{figure}[t]
    \centering
    \begin{minted}{sql}
  SELECT MEDIAN(s_acctbal)
  FROM part, partsupp, supplier,
       nation, region
  WHERE p_partkey   = ps_partkey
    AND s_suppkey   = ps_suppkey
    AND n_nationkey = s_nationkey
    AND r_regionkey = n_regionkey
    AND p_price > 
        (SELECT avg (p_price) FROM part)
    AND r_name IN ('Europe', 'Asia')
    \end{minted}
    \caption{Query over the TPC-H Schema}
    \label{fig:tpch-query}
\end{figure}

\begin{figure}
    \centering
    \scalebox{0.9}{
    \begin{forest}
    for tree={align=center}
        [{\texttt{supplier} 
        }            
            [{\texttt{nation} 
            }
                [{\texttt{region} 
                }]
            ]
            [{\texttt{partsupp} 
            }
                [{\texttt{part} 
                }]
            ]
        ]
    \end{forest}
    }
    \caption{Join tree for the query in Fig. \ref{fig:tpch-query}}
    \label{fig:tpch-query-jointree}
\end{figure}

As we will see in Section \ref{sect:Experiments}, all of the queries in the STATS-CEB and SNAP benchmarks are thus covered and so is
a small number of queries in the other benchmarks studied here. However, for the most commonly used 
aggregate functions 
\texttt{MIN}, \texttt{MAX}, \texttt{COUNT}, \texttt{SUM}, and \texttt{AVG}, the guardedness restriction can be significantly relaxed. 
We thus define {\em \pwg} queries as queries with aggregates on top of 
an acyclic join query, where the  attributes in a \texttt{GROUP}~\texttt{BY} clause 
and the attributes jointly occurring in an aggregate expression {\em each} are  contained in a single relation. That is, 
the \texttt{GROUP}~\texttt{BY} clause and also each aggregate expression has a {\em guard}, but all these guards may be different.
It will turn out that, 
with this relaxed restriction, we can cover 
all JOB queries and a significant number of queries in the TPC-H and TPC-DS benchmarks.

We will show how to realise Yannakakis-style evaluation for guarded and \pwg queries 
by rewriting subtrees in the logical query plan. In this process, joins are either replaced by semi-joins or they are immediately followed by aggregation. Importantly, in the latter case, the number of tuples to be propagated up the join tree is the same as
in case of semi-joins. The only extension needed is to add columns for the total frequencies (corresponding to \texttt{COUNT(*)})
and for the various aggregate expressions contained in the query. 
All of this can, in a very natural way, be implemented as part of the logical optimisation step.
This approach thus applies also to subqueries and automatically works in conjunction with other optimisation techniques such as subquery decorrelation.
An additional benefit of this optimisation is that it requires no cost-based optimisation, 
making it particularly attractive for systems with a limited or no cost model such as Spark SQL.

%

As a further optimisation, we introduce a new physical operator that intuitively implements a semi-join that keeps track of frequencies
and other aggregate values, and which  can be implemented through minimal changes to standard join algorithms (see Section~\ref{sect:physicalOpt}). It thus integrates smoothly in any typical SQL execution engine. 
We have implemented both, the logical optimisation and the new physical operator in Spark SQL, which 
was specifically designed to cope with complex analytical queries. 
The performance gain observed in our experimental evaluation on several standard benchmarks
can reach up to one or two orders of magnitude for 
analytical queries involving aggregates on top of
non-trivial 
join or path queries.
Notably, our method incurs no 
performance degradation even for simple queries where the size of intermediate results never gets too big anyway. 

In summary, our main contributions are as follows: 

\begin{itemize}[topsep=2pt]
\item We introduce the class of {\em guarded} aggregate queries and show that they allow
for an evaluation without materialising any intermediate join result 
for queries involving \texttt{GROUP}~\texttt{BY} and any 
of the aggregate functions contained in the ANSI SQL standard.
We then relax the restrictions imposed by guardedness and introduce the class of
{\em \pwg} aggregate queries. For these queries, 
we show that the favourable property of avoiding the propagation of any
join results applies to the most commonly used aggregate functions, namely 
\texttt{MIN}, \texttt{MAX}, \texttt{COUNT}, \texttt{SUM}, and \texttt{AVG}.

\item We achieve an additional optimisation of 
guarded and \pwg aggregate queries by introducing a 
{\em novel physical operator} that 
allows us to evaluate aggregate expressions and to compute frequencies of attribute combinations 
by an appropriate extension of the semi-join operator.
\item We have implemented our logical and physical optimisations into Spark SQL
and we have carried out an extensive empirical evaluation based on several standard benchmarks. 
It turns out that these benchmarks contain a significant number of queries or subqueries 
from our newly defined query classes.  
Our  experimental results  clearly prove the efficacy of our optimisation techniques.
\end{itemize}

The rest of the paper is organised as follows. 
We start with preliminaries 
in 
Section~\ref{sect:ACQs}. 
In Section 
\ref{sect:RelatedWork}, we recall several paths of related work. 
Our novel query optimisation techniques on logical query plans are presented in 
Section~\ref{sect:OurSystem}, and we discuss the new physical operator in Section~\ref{sect:physicalOpt}. In 
Section~\ref{sect:Experiments}, we  report on our experimental
evaluation, and we conclude with Section~\ref{sect:Conclusion}.
Further details (both on our optimisations and the experimental evaluation)
are provided 
\ifArxiv
in the appendix.
\else
in the full paper~\cite{DBLP:journals/corr/abs-2406-17076}.
\fi
Our implementation and a benchmark environment 
are 
publicly available at~~\url{https://github.com/dbai-tuw/spark-eval}.

\section{Preliminaries}
\label{sect:ACQs}

The basic form of queries considered here are {\em Conjunctive Queries} (CQs), which correspond to 
select-project-join queries in the Relational Algebra. Consider a CQ 
of  the form 
$Q = \pi_U (R_1 \bowtie  \dots \bowtie R_n )$. Here we assume that equi-joins are replaced by 
natural joins via appropriate renaming of attributes. Moreover, we assume that 
selections applying to a single relation have been pushed immediately in front of this relation 
and the $R_i$'s are the result of these selections. The $R_i$'s are not necessarily distinct
and our results are not affected by self-joins.

Such a CQ is called {\em acyclic} (an ACQ, for short), 
if 
it has a {\em join tree}, i.e., 
a rooted, 
labelled tree $\langle T,r,\lambda\rangle$ with root $r$ and node-labelling
function $\lambda$
such that  
(1) for every relation $R_i$ there exists exactly one node $u$ of $T$ with 
$\lambda(u) = R_i$ and
(2) $\lambda$  satisfies the so-called  connectedness condition, i.e., 
if some attribute $A$ occurs in both relations $\lambda(u)$ and $\lambda(v)$ 
for two nodes $u,v$ of $T$, then 
$A$ occurs in the  relation $\lambda(w)$ for every node $w$ on the path between $u$ and $v$.
Checking if a CQ is acyclic and, if so, constructing a join tree, can be done in linear time w.r.t.\ the size of the query by 
the so-called ``GYO reduction'' algorithm~\cite{report/toronto/Gra79}\cite{DBLP:conf/compsac/YuO79}.

Yannakakis \cite{DBLP:conf/vldb/Yannakakis81} has shown that ACQs can be 
efficiently evaluated (that is, essentially, linear w.r.t.\ the input+output data and 
linear w.r.t.\ the size of the query) via 3 traversals of the  join tree: 
(1) a bottom-up traversal of semi-joins, (2) a top-down traversal of semi-joins,
and (3) a bottom-up traversal of joins. 
Formally, let $u$ be a node in $T$ with  
child nodes $u_1, \dots, u_k$  of $u$ and let 
relations $R$, $R_{i_1}, \dots, R_{i_k}$ be associated with 
the nodes  $u$, $u_1, \dots, u_k$ at some stage of the computation. 
Then we set 

\smallskip

(1) $R = (((R  \ltimes R_{i_1}) \ltimes R_{i_2}) \cdots)  \ltimes  R_{i_k}$,

(2) $R_{i_j} = R_{i_j}  \ltimes R$ for every $j \in \{1, \dots, k \}$, and 

(3) $R = (((R  \bowtie R_{i_1}) \bowtie R_{i_2}) \cdots)  \bowtie  R_{i_k}$ 

\smallskip

\noindent
in the 3 traversals (1), (2), and (3).
The final result of the query is the resulting relation associated with 
the root $r$ of $T$. 
Following the SQL-standard, we are assuming bag semantics for the queries. 
Note however that Yannakakis' algorithm can be applied to both, set semantics and bag semantics.

In this work, we are mainly interested in queries that apply aggregates on top of ACQs and that
may contain ``arbitrary'' selections applied to single relations (that is, not only equality conditions, as is usually assumed for CQs \cite{DBLP:books/aw/AbiteboulHV95}).
Moreover, we allow grouping, which can also take care of the projection. 
In other words, we are interested in queries of the form 
\nop{***************
\begin{eqnarray*}
     Q & = &  \gamma[g_1, \dots, g_\ell, \; A_1(a_1), \dots, A_m(a_m)] \Big( 
       \\  &  &  \mbox{} \quad 
     R_1  \bowtie R_2  \bowtie \cdots \bowtie R_n \Big)
\end{eqnarray*}
***************}
\begin{equation}
\label{eq:basicQueryForm}
       Q  =   \gamma[g_1, \dots, g_\ell, \; A_1(a_1), \dots, A_m(a_m)] 
       \big( 
      R_1  \bowtie \cdots \bowtie R_n \big)
\end{equation}
where $\gamma_{g_1, \dots, g_\ell, \; A_1(a_1), \dots, A_m(a_m)}$ denotes the
grouping operation for attributes $g_1, \dots, g_\ell$ and aggregate
expressions $A_1(a_1), \dots, A_m(a_m)$ for some (standard SQL)
aggregate functions $A_1, \dots, A_m$ applied to expressions
$a_1, \dots, a_m$.  The grouping attributes $g_1, \dots, g_\ell$ are
attributes occurring in the relations $R_1, \dots, R_n$ and 
$a_1, \dots, a_m$ are expressions formed over the attributes 
from $R_1, \dots, R_n$.  A simple query of the form shown in
Equation (\ref{eq:basicQueryForm}) is given in Figure~\ref{fig:tpch-query} 
(in SQL-syntax) with a possible join tree of this query in 
Figure~\ref{fig:tpch-query-jointree}.

In the sequel, it will be convenient to use the following notation: 
suppose that we want to assign the result of a query $Q$
of the form according to Equation (\ref{eq:basicQueryForm}) to a relation $S$ with attributes $g_1, \dots, g_\ell, C_1, \dots, C_m)$, 
such that the values of each aggregate expression $A_i(a_i)$ is assigned to the attribute $C_i$, then we will write

\begin{align}
\label{eq:arrowNotation}
       S  & := &   \gamma[g_1, \dots, g_\ell, \; C_1 \leftarrow A_1(a_1), \dots, C_m \leftarrow A_m(a_m)] \nonumber \\
          & & \big(       R_1  \bowtie \cdots \bowtie R_n \big)
\end{align}

Recently  \cite{DBLP:journals/corr/abs-2303-02723},
a particularly favourable class of ACQs with aggregates has been presented: the class 
of 0MA (short for ``zero-materialisation answerable'') queries. 
These are acyclic queries that 
can be evaluated by executing only the first bottom-up traversal of 
Yannakakis' algorithm. That is, we only need to perform the comparatively cheap semi-joins 
and can completely skip the typically significantly more expensive join phase.  
A query of the form given in Equation (\ref{eq:basicQueryForm}) is 0MA if it satisfies the 
following conditions: 

\begin{itemize}
    \item 
{\em Guardedness}, meaning that there exists a relation $R_i$ that 
contains all grouping attributes $g_1, \dots, g_\ell$
and all attributes occurring in the aggregate expressions $A_1(a_1), \dots, A_m(a_m)$. Then $R_i$ is called the {\em guard} of the query.
If several relations satisfy this property, we arbitrarily choose one guard.
\item 
{\em Set-safety}: we call an aggregate function
{\em set-safe}, if its value over any set $S$ of tuples remains unchanged
if duplicates are eliminated from $S$. A query satisfies the 
set-safety condition, if all its aggregate functions $A_1 \dots, A_m$
are set-safe.
\nop{*********************
In particular, 
duplicate elimination applied to the inner expression
$ R_1  \bowtie \cdots \bowtie R_n $
does not alter
the result of the grouping and aggregate expression 
$\gamma_{g_1, \dots, g_\ell, \; A_1(a_1), \dots, A_m(a_m)}$.
*********************}
\end{itemize}

The root of the join tree 
can be arbitrarily chosen. Hence, we may assume that the root is labelled by the
guard and, therefore, all relevant attributes 
are contained in the root node.
Note that the bottom-up semi-join traversal makes sure that 
all value combinations
of the attributes in the root node indeed occur in the answer tuples 
of the inner part $R_1  \bowtie \cdots \bowtie R_n$ of the query.  
Set-safety of an aggregate function $A_i$ means that multiplicities do not
matter for the evaluation of $A_i$. Hence, if all aggregate functions are set-safe, then we can apply the grouping and 
aggregation $\gamma[g_1, \dots, g_\ell, \; A_1(a_1), \dots, A_m(a_m)]$ right after the first bottom-up traversal.
In SQL, the \texttt{MIN} and \texttt{MAX} aggregates are inherently set-safe. Moreover, an aggregate
becomes set-safe when combined with the \texttt{DISTINCT} keyword. For instance,  
\texttt{COUNT}~\texttt{DISTINCT} is clearly a set-safe aggregate function. Note that the 
query in Figure~\ref{fig:tpch-query} is trivially guarded (i.e., there is no grouping and the only aggregate expression is over the single attribute
\texttt{s\_acctbal}) but 
not set-safe, since multiplicities clearly matter for the 
evaluation of the \texttt{MEDIAN} 
function.

\nop{*********************
\begin{figure*}
    \centering
    \includegraphics[]{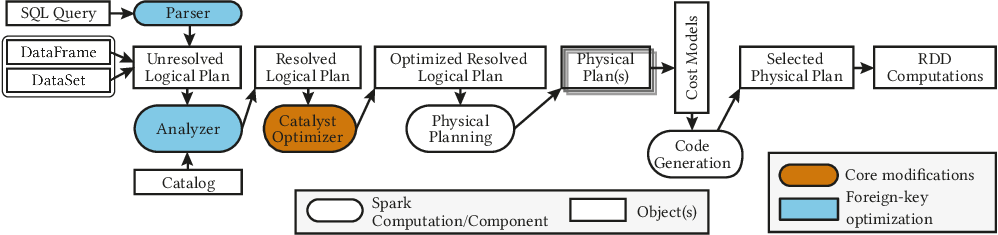}
    \caption{Implementation of the optimisations into the Spark SQL query processing pipeline (adapted from \cite{DBLP:conf/edbt/GrasmannPS23})}
    \label{fig:pipeline_optimisations}
\end{figure*}
*********************}

\section{Related Work}
\label{sect:RelatedWork}

\noindent
{\em Acyclic queries.}
The algorithm for evaluating acyclic queries, presented by Yannakakis over 40 years ago \cite{DBLP:conf/vldb/Yannakakis81}, has long been central to the {\em theory} of query processing.
In recent years, this approach to query evaluation has gained renewed momentum in {\em practice}
as evidenced by several extensions and applications. 
Multiple recent works~\cite{DBLP:conf/sigmod/IdrisUV17,%
DBLP:journals/vldb/IdrisUVVL20,DBLP:journals/corr/abs-2301-04003}, 
 propose extensions of Yannakakis' algorithm for dynamic query evaluation.
 Further research extends and applies Yannakakis' algorithm to comparisons spanning several 
relations~\cite{DBLP:conf/sigmod/0001022},
queries with theta-joins~\cite{DBLP:journals/vldb/IdrisUVVL20},
and privacy preserving query 
processing~\cite{DBLP:conf/sigmod/Wang021}.
An important feature of Yannakakis' algorithm is the 
elimination
of \emph{dangling tuples} (i.e., tuples that 
do not contribute to the 
final result) via semi-joins. 
In a recent paper~\cite{DBLP:journals/corr/TreeTracker}, 
a new join method 
was introduced that 
integrates the detection and elimination of dangling tuples into the join computation.

\smallskip
\noindent
{\em Decompositions.}
An important line of research has extended the applicability of Yannakakis-style query evaluation 
to ``almost acyclic'' queries. Here, ``almost acyclic'' is formalised through 
various notions of decompositions such as (normal, generalised, or fractional) hypertree decompositions 
\cite{DBLP:journals/jcss/GottlobLS02,DBLP:journals/ejc/AdlerGG07,2014grohemarx}. Each of these decompositions
is associated with a notion of ``width'' that measures the distance from acyclicity, 
with acyclic queries having a width of 1. Several works 
\cite{DBLP:journals/tods/AbergerLTNOR17,DBLP:conf/sigmod/Dai0023,%
DBLP:conf/sigmod/PerelmanR15,DBLP:conf/sigmod/TuR15}, 
combine Yannakakis-style query evaluation based on various types of decompositions
with multiway joins and worst-case optimal join techniques. 

\smallskip
\noindent
{\em Aggregate queries.}
Aggregates are commonly used on top of join queries -- 
especially in data analytics. Green et al.~\cite{DBLP:conf/pods/GreenKT07}  gave
a new perspective on aggregate queries by 
by considering $K$-relations, i.e., relations annotated with values from some 
semi-ring  $K$. Join queries over $K$-relations then come down to evaluating
sum-product expressions over the underlying semi-ring. The combination of aggregate queries
with Yannakakis-style query evaluation was studied in the FAQ-framework (Functional Aggregate Queries) \cite{DBLP:conf/pods/KhamisNR16} and, similarly, under the name AJAR (Aggregations and Joins over Annotated Relations)~\cite{DBLP:conf/pods/JoglekarPR16}. 
A crucial problem studied in both papers
is the interplay between the ordering of a sequence of aggregate functions and (generalised or fractional) 
hypertree decompositions. In both papers, the ultimate goal is 
an efficient, Yannankakis-style evaluation algorithm for aggregate queries 
based on finding a good variable order. 
Similar ideas to FAQs and AJAR queries also appear in earlier  works on joins and aggregates over 
factorised databases \cite{DBLP:journals/pvldb/BakibayevKOZ13,DBLP:journals/tods/OlteanuZ15} and on quantified conjunctive queries (QCQs) \cite{DBLP:conf/lics/ChenD12}.
A general framework for hybrid database and linear algebra workloads (as are typical for
machine learning applications) has recently been proposed by 
Shaikhha et al.~\cite{DBLP:journals/pacmpl/ShaikhhaHSO22}. It provides a  performant, unified 
framework for data science pipelines by introducing the purely functional 
language SDQL and combining 
optimisation techniques from  databases (e.g., pushing aggregates past joins) and 
 linear algebra (e.g., matrix chain ordering).


\smallskip
\noindent
{\em Distributed query processing.}
The potential of applying Yannakakis-style query evaluation to distributed processing 
comes from  the fact that the evaluation of ACQs lies in the 
highly parallelisable class \textsf{LogCFL}~\cite{DBLP:journals/jacm/GottlobLS01}.
 This favourable property was later extended to 
``almost acyclic'' queries by establishing the \textsf{LogCFL}-membership also for queries with bounded hypertree width
\cite{DBLP:journals/jcss/GottlobLS02}. 
A realisation of Yannakakis' algorithm 
in MapReduce \cite{DBLP:conf/icdt/AfratiJRSU17}
further emphasised the parallelisability of 
Yannakakis-style query evaluation. 

\smallskip
\noindent
{\em Spark and Spark SQL.}
Spark, 
 a top-level Apache project since 2014, 
is often regarded as a further development 
of the MapReduce processing model. Spark SQL 
\cite{DBLP:conf/sigmod/ArmbrustXLHLBMK15}
provides relational query capability within the Spark framework. 
Query optimisation is a primary focus of Spark SQL, with the powerful Catalyst optimiser being an integral component since its inception \cite{DBLP:conf/sigmod/ArmbrustXLHLBMK15}.
Several later works 
\cite{DBLP:conf/kdd/ShenRLJXP0Z023,DBLP:conf/ipccc/ZhaiSQJW19,DBLP:journals/sp/JiZZW20,%
DBLP:journals/tkde/BaldacciG19,DBLP:conf/ideas/MisegiannisKd22}) have proposed further measures
to speed up query processing in Spark SQL.
\nop{**************
, e.g., by introducing a data cache layer to reduce the cost of random disk I/O or implementing an enhancement of the Spark SQL cost model.
**************}
The recently presented SparkSQL$^{+}$ system~\cite{DBLP:conf/sigmod/Dai0023} combines decompositions and 
worst-case optimal join techniques as well as the
optimisations for CQs with comparisons
 spanning several 
relations
\cite{DBLP:conf/sigmod/0001022}
and 
allows users to 
experiment with different query plans.
Zhang et al.~\cite{DBLP:journals/pvldb/ZhangY0ZC20} recently implemented specific worst-case optimal join algorithms in combination with decomposition-based methods on top of Spark SQL as part of a system focused specifically on subgraph counting. 

\smallskip
\noindent
{\em Reducing the number of join computations.}
Several works have addressed the need to compute a high number of 
joins in different contexts 
and have aimed at reducing this number.  
The work 
by Schleich et al.~\cite{DBLP:conf/sigmod/SchleichOK0N19} 
on LMFAO (Layered Multiple Functional Aggregate Optimization) 
specifically targets 
machine learning applications that require the computation of large batches of 
aggregate queries over the same join query. A dramatic speed-up is achieved 
by decomposing aggregates into views and arranging them at nodes in a join tree
to avoid the re-computation of the same intermediate joins time and again. 
In principle, the need to re-compute similar joins time and again also 
arises in the area of IVM (incremental view maintenance). A revolutionary 
approach to IVM was proposed by Koch et al.~\cite{DBLP:journals/vldb/KochAKNNLS14}
with the 
DBToaster system,
that avoids the re-computation of joins in case of updates to the database
by maintaining ``higher-order'' delta views, i.e., delta queries 
(= first-order deltas), 
delta queries to the deltas (= second-order deltas), etc..
A further performance gain
is achieved with F-IVM (factorised IVM)%
~\cite{DBLP:conf/sigmod/NikolicO18}, 
that groups various aggregates together and thus reduces the number of views to be maintained. Moreover, factorisation is applied, for instance,  to avoid the 
materialisation of Cartesian products in views. Of course, 
independently of IVM, 
factorisation~\cite{DBLP:journals/sigmod/OlteanuS16} is a generally applicable 
method to 
keep the query result in a compressed form and 
avoid its complete materialisation.

\smallskip
\noindent
In summary, a lot of work has been done on optimising aggregate queries -- 
including worst-case optimal join techniques (primarily targeting cyclic queries) and Yannakakis-style query evaluation (for acyclic or almost acyclic queries). 
There also have been very successful 
approaches to {\em reduce} the number of joins that have to be computed when processing 
batches of related aggregate queries. Higher-order incremental view maintenance aims at {\em completely avoiding} 
the need
to (re-) compute joins in case of updates to the database. 
Our goal in this work is also to avoid the need to compute joins -- 
applicable to 
ad hoc query answering rather than IVM. To this end, we identify a class of queries with 
aggregates on top of joins that can be evaluated without actually computing the result 
 of any joins and that covers many relevant cases. 
To the best of our knowledge, 
apart from some severely restricted cases (specifically, Boolean queries and 0MA queries),
this has not been the focus of previous work. 

\section{Rule-Based Optimisations}
\label{sect:OurSystem}

In Section \ref{sect:ACQs}, we have recalled the definition of 0MA (zero-materia\-li\-sation answerable) queries
from Gottlob et al.~\cite{DBLP:journals/corr/abs-2303-02723}. Queries in this class, 
which have to satisfy the set-safety and guardedness conditions, can be evaluated by rooting the 
join tree at the node labelled by the guard and then 
executing the first bottom-up traversal of Yannakakis' algorithm. This means,
that all joins are replaced by semi-joins. The grouping and aggregation can then be evaluated by 
considering only the resulting relation at the root node. 

However, the set-safety condition 
is quite restrictive in that it is only satisfied by a small number of aggregate functions -- primarily 
\texttt{MIN}, \texttt{MAX}, and \texttt{COUNT}~\texttt{DISTINCT}. The vast majority of aggregate functions -- in particular, \texttt{COUNT} (without \texttt{DISTINCT}), \texttt{SUM}, \texttt{AVG}, and the entire collection of
statistical aggregate functions provided by the ANSI SQL standard are thus disallowed. For instance, 
the query in Figure~\ref{fig:tpch-query} involving the \texttt{MEDIAN} aggregate is not 0MA.

In this section, we significantly extend the class of queries with aggregates on top of join queries that can 
be evaluated without actually materialising any joins. To this end, we will first drop the set-safety condition 
in Section \ref{sect:guarded} and then also introduce a relaxation of guardedness 
in Section \ref{sect:piecewise}.
To emphasise the smooth integration of our optimisations into standard SQL execution technology,  
we will describe our optimisations in the form of equivalence-preserving 
transformations of Relational Algebra subexpressions, 
which can be applied anywhere in the logical query plan.

\subsection{Guarded Aggregate Queries}
\label{sect:guarded}

In order to cover {\em all} aggregate functions of the ANSI SQL standard, we now drop the 
safety condition and define the class of guarded aggregate queries as follows:

\begin{definition}
\label{def:guarded}
Let $Q$ be a query of the form given in  Equation (\ref{eq:basicQueryForm}), i.e.,
$Q =  \gamma[g_1, \dots, g_\ell, \; A_1(a_1), \dots, A_m(a_m)] 
       \big( 
      R_1  \bowtie \cdots \bowtie R_n \big)$. 
     We call $Q$ a 
{\em guarded aggregate query} (or simply, ``guarded query''),
if     
$ ( R_1  \bowtie \cdots \bowtie R_n )$ is acyclic and
there exists a relation $R_i$ (= the {\em guard})
that contains 
all attributes that are either part of the grouping or occur 
in one of the aggregate expressions. 
If several relations have this property, 
we arbitrarily choose one as the guard. 
\end{definition}

\noindent
Note that we consider an aggregate expression \texttt{COUNT(*)}
as trivially guarded,  since it contains no attributes at all.
We will now show that, for any aggregate functions of the ANSI SQL standard,  
guarded  queries can be evaluated without propagating any join results up the join tree. 
To this end, we revisit an extension of Yannakakis' algorithm by Pichler and Skritek~\cite{DBLP:journals/jcss/PichlerS13}
to acyclic queries with a \texttt{COUNT(*)} aggregate on top.
We adapt this 
approach to integrate it into the logical query plan of relational query processing and we further extend 
it to all other aggregate functions.

The crucial idea for evaluating a query $Q$ of the form 
given in Equation (\ref{eq:basicQueryForm}) is to propagate frequencies
up the join tree rather than duplicating tuples.
It is convenient to introduce the following notation: 
let $u$ denote a node in the join tree $T$ and let $T_u$ denote the set of all nodes in the subtree rooted at $u$.
Moreover, for any node $u$ in $T$, we write $R(u)$ to denote the relation labelling node $u$ and 
we write $\Attr(u)$ to denote the list of attributes of $R(u)$.  
The goal of the bottom-up 
propagation
of frequencies is to compute, for every node $u$ in $T$, the result of the following query:

\begin{equation}
    \label{eq:frequency}
\gamma[\Attr(u), \texttt{COUNT(*)}] \left( 
\underset{v \in T_u}{\scaleobj{2}{\bowtie}}R(v)\right)    
\end{equation}

This propagation is realised by recursively constructing extended Relational Algebra expressions $\Count(u)$
for every node $u$ of the join tree, such that $\Count(u)$ gives the same result as the query in
Equation~(\ref{eq:frequency}).
Hence, $\Count(u)$ has as attributes all attributes of $R(u)$ plus one additional attribute 
(which we will denote as $c_u$), 
where we store frequency information for each tuple of $R(u)$. 
If $u$ is a leaf node of the join tree, then we initialise the attribute $c_u$ to~1. Formally,
we thus have $\Count(u) = R(u) \times \{(1)\}$. 

Now consider an internal node $u$ of the join tree with child nodes $u_1, \dots, u_k$.
The extended Relational Algebra expression $\Count(u)$ is constructed iteratively by defining 
subexpressions $\Count_i(u)$ with $i  \in \{0, \dots, k\}$. To avoid confusion, we refer to the 
frequency attribute of such a subexpression $\Count_i(u)$ as $c^{i}_{u}$. That is, 
each relation $\Count_i(u)$ consists of the same attributes $\Attr(u)$  as $R(u)$
plus the additional 
frequency attribute $c^{i}_{u}$. 
Then we define 
$\Count_i(u)$ for every $i  \in \{0, \dots, k\}$ and, ultimately, 
$\Count(u)$ as follows:
\begin{tabbing}
$\Count_0(u)$ \= := \= $R(u) \times \{(1)\}$\\
$\Count_i(u)$ \> := \> $\gamma[\Attr(u), c_u^i \leftarrow \texttt{SUM}( c^{i-1}_{u} 
\cdot c_{u_i})] (\Count_{i-1}(u) \bowtie \Count(u_{i}) )$ \\
$\Count(u)$ \>  := \> $\rho_{c_u \leftarrow c_u^k}(\Count_{k}(u))$
\end{tabbing}
\noindent
Intuitively, after initialising 
$c_u^0$ to 1 in $\Count_0(u)$, 
the frequency values $c_u^1, \dots, c_u^k$ are obtained by grouping over the attributes $\Attr(u)$ of $R(u)$
and computing the number of possible extensions of each tuple 
$t \in R(u)$ to the relations labelling the nodes 
in the subtrees rooted at $u_1, \dots, u_k$. 
By the connectedness condition of join trees, 
these extensions are independent of each other, i.e., they share no attributes 
outside $\Attr(u)$. Moreover, the frequency attributes $c_{u}^1, \dots, c_{u}^k$ are functionally dependent on the 
attributes $\Attr(u)$. Hence, by distributivity, the value of  $c_{u}^k$ obtained by iterated summation and multiplication 
for given tuple $t$ of $R(u)$ is equal to computing, for every $i \in \{1, \dots, k\}$ 
the sum $s_i$ of the frequencies of all join partners of 
$t$  in $\Count(u_i)$ and then computing their product, i.e., 
$c_{u} = c_{u}^k = \Pi_{i=1}^k s_i$. 

In the logical query plan of query $Q$, we replace the subexpression corresponding to
the join query $ R_1  \bowtie \cdots \bowtie R_n$ by $\Count(r)$, where $r$ is the root node 
of the
join tree. This root node  was chosen in such a way that 
$R(r)$ contains all grouping attributes $g_1, \dots, g_\ell$.
Hence, the grouping can be applied to $\Count(r)$ in the same way as to the 
original join query. Also the set-safe aggregates (such as \texttt{MIN}, \texttt{MAX},
\texttt{COUNT}~\texttt{DISTINCT}) can be applied to $\Count(r)$ ``as usual''
by simply ignoring the additional attribute $c_r$. 
However, all other (i.e., not set-safe) 
aggregate functions have to be replaced by 
variants that take the  special 
frequency attribute $c_r$ into account.
We thus modify the aggregate functions in expressions
like 
$\texttt{COUNT}(*)$,
$\texttt{COUNT}(B)$,  
$\texttt{SUM}(B)$, and
$\texttt{AVG}(B)$
so that they 
directly operate on tuples with frequencies. For instance, 
let $B$ be an  attribute of the guard $R(r)$
(and, hence, also of $\Count(r)$).
Then, in SQL-notation, we can rewrite 
common aggregate expressions as follows: 

\begin{itemize}
    \item $\texttt{COUNT}(*) \rightarrow \texttt{SUM}(c_r)$
    \item $\texttt{COUNT}(B) \rightarrow \texttt{SUM}(\texttt{IF}(\texttt{ISNULL}(B),0,c_r))$
    \item $\texttt{SUM}(B) \rightarrow \texttt{SUM}(B \cdot c_r)$
    \item $\texttt{AVG}(B) \rightarrow \texttt{SUM}(B \cdot c_r) / \texttt{COUNT}(B)$
\end{itemize}

Recall from Example \ref{ex:tpch-query} that 
the \texttt{MEDIAN} aggregate (like any other statistical function) can be
evaluated by considering $\Count(r)$ as a compressed form of the list of 
all values of attribute $B$ in each group, where the value of the additional attribute $c_r$ indicates
the number of copies of the corresponding value of attribute $B$
in the result of the join query $ R_1  \bowtie \cdots \bowtie R_n$. 
Evaluating an aggregate expression 
\texttt{MEDIAN}(B) or any other statistical function such as \texttt{STDDEV}(B)
can be easily realised for this compressed form of value list. 
Similarly, the evaluation of  aggregate functions on 2 attributes 
such as \texttt{CORR}(B$_1$, B$_2$) or aggregate expressions involving functions on several attributes
such as \texttt{SUM}(f(B$_1$, \dots, B$_k$)) is straightforward by considering 
$\Count(r)$ as a compressed form of the list of all values of the attribute combinations 
$B_1, \dots, B_k$. Again, this crucially depends on the guardedness property, which guarantees that 
all attributes used in aggregate expressions are contained in $R(r)$.

Actually, in Spark SQL, the \texttt{MEDIAN} aggregate has 
a convenient rewriting via  the \texttt{PERCENTILE} function.
The latter is not part of the ANSI SQL standard, but can be found in Spark SQL. 
This function allows one to provide a frequency attribute, which Spark uses to build a map of values and frequencies, sort them, and 
finally find the desired percentile value by an efficient search on the
sorted map. The rewriting of the \texttt{MEDIAN} aggregate looks as
follows: 

\begin{itemize}
    \item $\texttt{MEDIAN}(A) \rightarrow \texttt{PERCENTILE}(0.5, A, c_r)$
\end{itemize}

\begin{example}
\label{exp:median} 
Consider again the query of Figure~\ref{fig:tpch-query}.
The logical query plan generated by Spark SQL 
is shown in  Figure~\ref{fig:unoptimised-query}. There, we write $\sigma_\psi$ and $\sigma_\phi$ to denote 
the selections applied to the relations \texttt{region} and \texttt{part}, respectively. 
That is, $\psi$ checks the condition \texttt{r\_name IN ('Europe', 'Asia')} and
$\phi$ checks the condition \texttt{p\_price $>$ (SELECT avg (p\_price) FROM part)}. The
plan produced by Spark SQL including our optimisation
is shown in Figure~\ref{fig:count-optimised-query}.

We observe that, in the unoptimised query plan, the entire join of all relations is computed before 
the \texttt{MEDIAN} aggregate is applied. In contrast, in the optimised plan, only the additional 
frequency attribute has
to be propagated upwards in the plan. This propagation of frequencies for each join is realised by 2 nodes in the plan directly
above the node realising the join: first, as part of the projection to the attributes which are used 
further up in the plan, the frequency attributes of the two join operands are multiplied with each other. Here we use the 
notation $c_{xy}$ when frequency attributes $c_x$ and $c_y$ are combined. In the second step, these frequency values $c_{xy}$ 
are summed up or, in case of the final result, their median is computed, which 
can be further optimised
by making use of the \texttt{PERCENTILE} function. 
\hfill $\diamond$
\end{example}

\begin{figure*}[t]
    \centering
    \begin{subfigure}[b]{0.35\textwidth}
        \hspace{-2.3em}
        \scalebox{0.85}{ 
                \begin{forest}
        [$\gamma_{\texttt{MEDIAN}(\text{s\_acctbal})}$
        [$\pi_{\text{s\_acctbal}}$
        [$\bowtie_{\text{n\_regionkey = r\_regionkey}}$
         [$\pi_{\text{s\_acctbal, n\_regionkey}}$
          [$\bowtie_{\text{s\_nationkey = n\_nationkey}}$
           [$\pi_{\text{s\_nationkey, s\_acctbal}}$
            [$\bowtie_{\text{s\_suppkey = ps\_suppkey}}$
             [$\pi_{\text{ps\_suppkey}}$
              [$\bowtie_{\text{p\_partkey = ps\_partkey}}$
               [$\pi_{\text{p\_partkey}}$
                [$\sigma_\phi(\texttt{part})$]]
               [$\pi_{\text{ps\_partkey, ps\_suppkey}}$
                [$\texttt{partsupp}$]]
               ]
              ]
              [$\pi_{\text{s\_suppkey, s\_nationkey, s\_acctbal}}$
              [$\texttt{supplier}$]
             ]
             ]
           ]
            [$\pi_{\text{n\_nationkey, n\_regionkey}}$
            [$\texttt{nation}$]
           ]
          ]
         ]
         [$\pi_{\text{r\_regionkey}}$
          [$\sigma_\psi(\texttt{region})$]
         ]
        ]
        ]
        ]
        \end{forest}
        }
        \vspace{3.03em}
        \caption{Query plan generated by Spark SQL}
        \label{fig:unoptimised-query}
    \end{subfigure}
\hfill
    \begin{subfigure}[b]{0.63\textwidth}
        \centering
        \scalebox{0.85}{
        \begin{forest}
        [$\gamma_{\texttt{PERCENTILE}(\text{s\_acctbal}, c_{final}, 0.5)}$
         [$\pi_{\text{s\_acctbal}, c_{final} \leftarrow c_{psp} \cdot c_{snr}}$
          [$\bowtie_{\text{s\_suppkey = ps\_suppkey}}$
          [$\gamma_{\text{s\_suppkey, s\_acctbal}, \texttt{SUM}(c_{snr})}$
            [$\pi_{\text{s\_suppkey, s\_acctbal}, c_{snr} \leftarrow c_{s} \cdot c_{nr}}$
             [$\bowtie_{\text{s\_nationkey = n\_nationkey}}$
              [$\pi_{\text{s\_suppkey, s\_nationkey, s\_acctbal}, c_{s} \leftarrow 1}$, l*=1.5
               [$\texttt{supplier}$]
              ]
              [$\gamma_{\text{n\_nationkey}, \texttt{SUM}(c_{nr})}$, l*=1.8
               [$\pi_{\text{n\_nationkey}, c_{nr} \leftarrow c_n \cdot c_r}$
                [$\bowtie_{\text{n\_regionkey = r\_regionkey}}$
                   [$\pi_{\text{n\_nationkey, n\_regionkey}, c_{n} \leftarrow 1}$
                    [$\texttt{nation}$]
                   ]
                   [$\pi_{\text{r\_regionkey}, c_{r} \leftarrow 1}$, l*=2
                    [$\sigma_\psi(\texttt{region})$]
                   ]
                ]
               ]
              ]
             ]
            ]
           ]
           [$\gamma_{\text{ps\_suppkey}, \texttt{SUM}(c_{psp}) }$, l*=3
            [$\pi_{\text{ps\_suppkey}, c_{psp} \leftarrow c_{ps} \cdot c_{p}}$
             [$\bowtie_{\text{ps\_partkey = p\_partkey}}$
              [$\pi_{\text{ps\_partkey, ps\_suppkey}, c_{ps} \leftarrow 1}$
               [$\texttt{partsupp}$]]
              [$\pi_{\text{p\_partkey}, c_p \leftarrow 1}$
               [$\sigma_\phi(\texttt{part})$]]
              ]
             ]
            ]
          ]
         ]
        ]
        \end{forest}
        }
        \caption{Query plan generated by Spark SQL with rewritten aggregation}
        \label{fig:count-optimised-query}
    \end{subfigure}
    \caption{Query plans for Example~\ref{exp:median}}
\end{figure*}

We conclude this section with an example where we 
display the information that has to be propagated in the optimised evaluation of the query 
from Figure~\ref{fig:tpch-query}.
Actually, it is illustrative to first observe how the tree structure of the join tree is transformed 
into the tree structure of the optimised plan. 
Of course, in the latter, the relations must be at the leaf nodes, whereas, in the former, they also occur at inner nodes. 
Nevertheless, 
the bushy optimised plan clearly reflects the join order from the join tree. That is, 
first, 
\texttt{region} and \texttt{nation} 
are joined to get intermediate result-1, and
\texttt{part} and \texttt{part\_supp} 
are joined to get intermediate result-2. The join of these two intermediate results with 
the relation \texttt{supplier} is then split into two 2-way joins, i.e.: first joining \texttt{supplier} with result-1, which is 
then joined with result-2. 
Hence, for the sake of simplicity, we will discuss the evaluation of this query by looking at the 
relations at each node of the join tree. It is then clear, what the intermediate results at the nodes of the 
logical plan in Figure~\ref{fig:count-optimised-query} look like.

\makeatletter
\newcommand{\oset}[3][0ex]{%
  \mathrel{\mathop{#3}\limits^{
    \vbox to#1{\kern-2\ex@
    \hbox{$\scriptstyle#2$}\vss}}}}
\makeatother

\newcommand*\cjoin{\oset{\tiny \gamma,+,*}{\ltimes}}

\definecolor{relname}{RGB}{200, 200, 240}
\definecolor{relheader}{RGB}{210, 210, 255}

\begin{figure}[h]



\tikzstyle{every picture}+=[remember picture]
\tikzstyle{na} = [shape=rectangle,inner sep=0pt]

\newcommand{\ptStrike}[2]{    
    \tikz[baseline=(#1.base)]\node[na](#1){#2};
}

\newcommand{\strike}[2]{  
    \begin{tikzpicture}[overlay]
        \draw (#1.west) -- (#2.east);
    \end{tikzpicture}
}

\begin{tikzpicture}[scale=0.8, transform shape]
            \node [](supplier) at (3.0,0.0) {
            \begin{tabular}{|c|c|c|c|c|}
            \hline
             \rowcolor{relname} \multicolumn{5}{|c|}{\texttt{supplier}} \\
            \hline
                \rowcolor{relheader} $\mathbf N$ & $\mathbf S$ & $\mathbf A$ & $\cdots$ & $\mathbf c$ \\
                \hline
                $n_1$ & $s_1$ & $20$ & $\cdots$ & $30$ \\
                 \hline
                 $n_1$ & $s_2$ & $40$ & $\cdots$ & $20$ \\
                 \hline
                 \ptStrike{A} {$n_1$} & $s_4$ & $30$ & $\cdots$ & \ptStrike{B}{$1$}   \\
                 \hline
                 $n_2$ & $s_1$ & $10$ & $\cdots$ & $36$  \\
                 \hline
                 $n_2$ & $s_2$ & $30$ & $\cdots$ & $24$  \\
                 \hline
                 \ptStrike{C} {$n_4$} & $s_2$ & $20$ & $\cdots$  & \ptStrike{D}{$1$}  \\
                 \hline
            \end{tabular}

             \strike{A}{B}
              \strike{C}{D}
        };

        \node [](nation) at (1.0,-3.9) {
            \begin{tabular}{|c|c|c|c|}
            \hline
             \rowcolor{relname} \multicolumn{4}{|c|}{\texttt{nation}} \\
            \hline
                \rowcolor{relheader} $\mathbf N$ & $\mathbf R$ & $\cdots$ & $\mathbf c$ \\
                \hline
                $n_1$ & $r_1$ & $\cdots$ & $3$ \\
                 \hline
                 $n_1$ & $r_2$ & $\cdots$ & $2$ \\
                 \hline
                 \ptStrike{A} {$n_1$} & $r_4$ & $\cdots$ & \ptStrike{B}{$1$}   \\
                 \hline
                 $n_2$ & $r_1$ & $\cdots$ & $3$  \\
                 \hline
                 $n_2$ & $r_2$ & $\cdots$ & $2$  \\
                 \hline
                 {$n_2$} & $r_3$ & $\cdots$  & $1$  \\
                 \hline
            \end{tabular}

             \strike{A}{B}
        };

        \node [](partsupplier) at (5.0,-3.9) {
            \begin{tabular}{|c|c|c|c|}
            \hline
             \rowcolor{relname} \multicolumn{4}{|c|}{\texttt{partsupplier}} \\
            \hline
                \rowcolor{relheader} $\mathbf S$ & $\mathbf P$ & $\cdots$ & $\mathbf c$ \\
                \hline
                $s_1$ & $p_1$ & $\cdots$ & $3$ \\
                 \hline
                 $s_1$ & $p_2$ & $\cdots$ & $2$ \\
                 \hline
                {$s_1$} & $p_3$ & $\cdots$ & {$1$}   \\
                 \hline
                 $s_2$ & $p_1$ & $\cdots$ & $3$  \\
                 \hline
                 $s_2$ & $p_3$ & $\cdots$ & $1$  \\
                 \hline
                 {$s_3$} & $p_1$ & $\cdots$  & $3$  \\
                 \hline
            \end{tabular}
        };

                \node [](region) at (1.0,-7.8) {
            \begin{tabular}{|c|c|c|c|}
            \hline
             \rowcolor{relname} \multicolumn{3}{|c|}{\texttt{region}} \\
            \hline
                \rowcolor{relheader} $\mathbf R$ & $\cdots$ & $\mathbf c$ \\
                \hline
                $r_1$ & $\cdots$ & $1$ \\
                 \hline
                 $r_1$ & $\cdots$ & $1$ \\
                 \hline
                 {$r_1$} & $\cdots$ & {$1$}   \\
                 \hline
                 $r_2$ &  $\cdots$ & $1$  \\
                 \hline
                 $r_2$ & $\cdots$ & $1$  \\
                 \hline
                 {$r_3$} & $\cdots$  & $1$  \\
                 \hline
            \end{tabular}

        };

        \node [](part) at (5.0,-7.8) {
             \begin{tabular}{|c|c|c|c|}
            \hline
             \rowcolor{relname} \multicolumn{3}{|c|}{\texttt{part}} \\
            \hline
                \rowcolor{relheader} $\mathbf P$ & $\cdots$ & $\mathbf c$ \\
                \hline
                $p_1$ & $\cdots$ & $1$ \\
                 \hline
                 $p_1$ & $\cdots$ & $1$ \\
                 \hline
                 {$p_1$} & $\cdots$ & {$1$}   \\
                 \hline
                 $p_2$ &  $\cdots$ & $1$  \\
                 \hline
                 $p_2$ & $\cdots$ & $1$  \\
                 \hline
                 {$p_3$} & $\cdots$  & $1$  \\
                 \hline
            \end{tabular}
        };

    \draw [very thick] (nation) -- (supplier);
    \draw [very thick] (partsupplier) -- (supplier);
    \draw [very thick] (nation) -- (region);
    \draw [very thick] (partsupplier) -- (part);

    \end{tikzpicture}

    \caption{Evaluation of the query from Figure~\ref{fig:tpch-query}}
    \label{fig:eval-guarded}
\end{figure}    

\begin{example}
\label{ex:guarded-evaluation}    
Consider again the query from Figure~\ref{fig:tpch-query}. 
Note that all joins in this query are along foreign-key/primary-key relationship. 
This allows us additional optimisations, which we discuss 
\ifArxiv
In Appendix~\ref{app:physical},
\else
in the full paper~\cite{DBLP:journals/corr/abs-2406-17076}.
\fi
For the sake of illustration, let us ignore the primary keys for a while 
and allow multiple occurrences of values in
these attributes. In 
Figure~\ref{fig:eval-guarded}, we illustrate the evaluation of the query on a small sample database.
The tables (with attribute names of the join attributes abbreviated to single characters) 
are arranged in the form of the join tree. Attributes not relevant to our discussion are captured by ``\dots''. 
The original contents of the tables is shown to the left of the ``\dots'' column. In the right-most column, we display
the frequency attribute $c$ for each tuple at the end of the entire bottom-up traversal. For instance, the \texttt{Region} table has
3 tuples with attribute value $R = r_1$. Hence, all tuples in \texttt{Nation} with $R = r_1$ have $c=3$, i.e., the number of possible 
extensions to the subtree below. The tuple with value $R = r_4$ is deleted since it has no join partner below.
In the root node, the sums of the frequency attributes of all join partners to the left and to the right are multiplied. For instance, 
the tuple with attribute values $N = n_1$ and $S = s_1$ has 5 possible extensions to the subtree on the left and 6 on the right. 
Hence, for this tuple, we get $c = 30$. 

For the evaluation of $\texttt{MEDIAN(A)}$ we see at the root node, that the first tuple has attribute value 
$A = 20$ and its frequency in the overall join result is $c = 30$. Likewise, the values 
$A = 40$, 10, 30 occur 20, 36, and 24 times, respectively in the join result. We thus get 
$\texttt{MEDIAN(A)} =20$.
\hfill $\diamond$
\end{example}

\subsection{\Pwg Aggregate Queries}
\label{sect:piecewise}

As we will see in our experimental evaluation in Section \ref{sect:Experiments}, the 
class of guarded queries covers significantly more cases from common benchmarks than 0MA. 
However, the requirement of a single guard for {\em all} attributes occurring in the \texttt{GROUP}~\texttt{BY}
clause or in any of the aggregate expressions is still quite restrictive. In this section,
we show that for the most commonly used aggregate functions \texttt{MIN}, 
 \texttt{MAX},  \texttt{SUM},  \texttt{COUNT}, and \texttt{AVG}, we can further extend
 the class of queries that can be evaluated without materialising any joins. 
We thus introduce the class of \pwg aggregate queries:

\begin{definition}
\label{def:pwguarded}
Let $Q$ be a query of the form given in  Equation~(\ref{eq:basicQueryForm}), i.e.,
$Q =  \gamma[g_1, \dots, g_\ell, \; A_1(a_1), \dots, A_m(a_m)] 
       \big( 
      R_1  \bowtie \cdots \bowtie R_n \big)$. 
     We call $Q$ a 
{\em \pwg aggregate query} (or simply, ``\pwg query''),
if     
$ ( R_1  \bowtie \cdots \bowtie R_n )$ is acyclic and
there exists a relation $R_{i_0}$ that contains all grouping attributes and, for 
every $j \in \{1, \dots, m\}$, the following conditions hold: 
\begin{itemize}
    \item If $A_j \in \{$\texttt{MIN}, 
 \texttt{MAX}, \texttt{SUM}, \texttt{COUNT}, \texttt{AVG}$\}$, then there exists 
 a relation $R_{i_j}$ that contains all attributes occurring in $A_j(a_j)$.  
 \item Otherwise, i.e., $A_j \not\in \{$\texttt{MIN}, 
 \texttt{MAX}, \texttt{SUM}, \texttt{COUNT}, \texttt{AVG}$\}$,
 then 
 $R_{i_0}$  contains all attributes occurring in  $A_j(a_j)$.  
\end{itemize}
Each of these relations $R_{i_0}$ and $R_{i_j}$ is called the 
``guard'' of the corresponding set of attributes. We refer to $R_{i_0}$ as the \emph{root} guard. By slight abuse of notation, 
we also refer to the nodes labelled by $R_{i_0}$ and $R_{i_j}$
as guards.
If several relations 
could be chosen as guard for a group of attributes, we arbitrarily choose one.
\end{definition}

For the evaluation of \pwg queries, we choose the node of the join tree $T$ corresponding to the root guard as the root node of $T$.
The bottom-up propagation of the frequency attribute works exactly as for guarded queries. Hence, also
the evaluation of all aggregate expressions that are guarded by the  root guard
is realised exactly as in case of guarded queries. 
In the rest of this section, we concentrate on the evaluation of aggregate expressions that are 
{\em not 
guarded by the root node} of the join tree and whose
aggregate function is one of \texttt{MIN}, 
 \texttt{MAX}, \texttt{SUM}, \texttt{COUNT}, and \texttt{AVG}. 
Clearly, the evaluation of \texttt{AVG} is based on \texttt{SUM} and \texttt{COUNT}.
Hence, it suffices to describe the evaluation of the remaining four aggregate functions.

Similarly to Koch et al.~\cite{DBLP:journals/vldb/KochAKNNLS14}, we extend the relations by
additional attributes to carry information on aggregate expressions.
Below, we describe which information has to be propagated up the join tree in order to evaluate a single aggregate expression $A_j(a_j)$. 
For the evaluation of $Q$, we add the frequency attribute plus {\em all} these additional attributes 
to the corresponding nodes in the join~tree.

Suppose that $A_j(a_j)$ is of the form  $A_j(f_j(\bar B_j)$ with 
$A_j \in \{$\texttt{MIN}, 
\texttt{MAX}, \texttt{SUM}, \texttt{COUNT}$\}$ and $f_j$ is an arbitrary function on attributes $\bar B_j$
jointly occurring in one of the relations $R_1, \dots, R_n$. We choose as guard of the 
aggregate expression  $A_j(a_j)$  the  
node $w$ that contains all attributes $\bar B_j$ and that is highest up in the join tree $T$ with this property. 
Since we are assuming that $A_j(a_j)$ is not guarded by the 
root node $r$ of $T$, this means that $w$ is different from $r$.
Then we add to all relations along the path
from $w$ to  $r$ an additional attribute $\Aggj$. Analogously to 
Equation~(\ref{eq:frequency}),  
the intended meaning of $\Aggj$ for every node $u$ 
on the path between $w$ and $r$ is as follows:

\begin{equation}
    \label{eq:pwfrequency}
\gamma[\Attr(u), \Aggj \leftarrow A_j(f_j(\bar B_j)] \big( {\Huge \bowtie}_{v \in T_u} (R(v)\big),   
\end{equation}

\nop{Kuerzen*****************
\noindent
where we again write 
$R(v)$ and $\Attr(v)$ to denote the relation labelling node $v$ and its attributes,
and $T_u$ to denote the set of all nodes in the subtree rooted at $u$.
*****************}

For the {\em initialisation} of $\Aggj$, 
suppose that the frequency attribute of 
relation $R(w)$ at node $w$  has already been computed 
as described in Section~\ref{sect:guarded}.
Hence, in particular, $R(w)$ is restricted to the tuples $t$ with 
positive frequency.
For an arbitrary tuple $t$ in $R(w)$, we write $t.c$, $t.\Aggj$, and $t.\bar B_j$ 
to denote the values of $t$ at the frequency attribute $c$, at the aggregate attribute $\Aggj$, and
at the attributes $\bar B_j$, respectively. 
Then we define $t.\Aggj$ as follows: 

\begin{itemize}
    \item If $A_j \in \{$\texttt{MIN}, 
\texttt{MAX}$\}$, then we set $t.\Aggj$ := $f_j(t.\bar B_j)$.
\item If $A_j$ = \texttt{COUNT}, then we distinguish two cases:
If $f_j(t.\bar B_j)$ = \texttt{NULL}, 
then   we set $t.\Aggj$ := 0;
otherwise  $t.\Aggj$ := $t.c$.
\item If $A_j$ = \texttt{SUM}, then we set $t.\Aggj$ := $f_j(t.\bar B_j) * t.c$.
\end{itemize}

\noindent
To verify that $\Aggj$ is equal to the additional attribute according 
to Equation~(\ref{eq:pwfrequency}), we note that
all tuples in a group
defined by a value combination of the original attributes $\Attr(w)$ of $R(w)$ 
(thus, corresponding to a single tuple $t \in R(w)$)
coincide on the attributes $\bar B_j$.
Hence, the \texttt{MAX} and \texttt{MIN} of
$f_j(\bar B_j)$ over the tuples in such a group is 
simply the value of $f_j(t.\bar B_j)$. 
For $A_j \in \{$\texttt{COUNT}, \texttt{SUM}$\}$, we have to take the 
number of 
tuples in each such group into account,
which corresponds to the 
frequency value $t.c$.
For the \texttt{COUNT} aggregate, we also have to consider the 
special case that $f_j(\bar B_j)$ = NULL, which means that \texttt{COUNT}$(f_j(\bar B_j))$
for the entire group is 0.

For the {\em propagation} of the additional attribute $\Aggj$
along the path from $w$ to the root $r$, 
consider an ancestor node $u$ of $w$ and let $u_1, \dots, u_k$ denote the child nodes of 
$u$. W.l.o.g., we assume that the child node $u_1$ is on the path from $w$ to $r$. 
Suppose that the frequency attribute at 
every child node $u_i$ of $u$ and the attribute $\Aggj$ at node $u_1$ have already been computed. 
We are assuming that $w$ is the highest node in the join tree that contains all attributes $\bar B_j$. 
Hence, $u$ does not contain all attributes $\bar B_j$, and, by the connectedness conditions,
neither does any of the nodes $u_2, \dots, u_k$.
For an arbitrary tuple $t \in R(u)$,
let $\{t_1, \dots, t_\alpha\}$ denote the set of all tuples 
in 
$R(u_1)$ that join with $t$.
We compute the value $t.\Aggj$ as follows: 

\begin{itemize}
\item First suppose that $A_j \in \{$\texttt{MIN}, \texttt{MAX}$\}$. 
Then we set $t.\Aggj$ := $A_j(\{t_1.\Aggj, \dots$, $t_\alpha.\Aggj\})$. 
\item 
Now let $A_j \in \{$\texttt{SUM}, \texttt{COUNT}$\}$. 
For every $i \in \{2, \dots, k\}$, let 
$s_i$ denote the sum of the frequencies of all join partners of $t$ in $R(u_i)$.
Then we set $t.\Aggj$ := $\big( \sum_{\lambda = 1}^\alpha t_\lambda.\Aggj[u_1] \big) * \Pi_{i=2}^k s_i$.
\end{itemize}

For the correctness of this propagation of  attribute $\Aggj$, 
recall that we are assuming that the attributes $\bar B_j$ are not fully contained in 
the relation $R(u)$ and, hence, by the connectedness condition, they cannot 
be fully contained in any of the child nodes
$\{u_2, \dots, u_k\}$ either.  
Hence,  the value combinations of $\bar B_j$
in $\big( {\Huge \bowtie}_{v \in T_u} (R(v)\big)$
must already occur in  $\big( {\Huge \bowtie}_{v \in T_{u_1}} (R(v)\big)$.
The  \texttt{MIN} or  \texttt{MAX} of $f_j(\bar B_j)$ of a tuple $t$ 
when grouping over the attributes of $R(u)$ is, therefore, simply 
obtained by grouping over the attributes of $R(u_1)$ and aggregating over the 
join partners of $t$ in $R(u_1)$. Similar considerations apply to 
the computation of $\Aggj$ in case of  
\texttt{COUNT} and  \texttt{SUM}. 
The aggregation 
of $\Aggj$ over the join partners of $t$ in $u_1$ 
yields the value $\big( \sum_{\lambda = 1}^\alpha t_\lambda.\Aggj \big)$.
In contrast to \texttt{MIN} and  \texttt{MAX}, 
we now also  have to take the possible extensions of $t$ to the relations in the subtrees of $T$ rooted at 
the nodes $u_2, \dots, u_k$ into account.  
The number of possible extensions of $t$ to $\big( {\Huge \bowtie}_{v \in T_{u_i}} (R(v)\big)$ 
corresponds to the sum $s_i$ of the frequencies of all 
join partners of $t$ in $R(u_i)$. Hence, 
by the connectedness conditions, the number of extensions of $t$ to 
the relations at {\em all} subtrees of $T_{u_2}, \dots, T_{u_k}$
is  obtained as  $\Pi_{i=2}^k s_i$.

We conclude this section by an important observation concerning the 
{\em size of the 
relations} that we propagate up the join tree $T$: for every node $u$ of $T$, 
the relation $\Count(u)$ contains precisely the tuples of $R(u)$ that one would get by the first 
bottom-up traversal of Yannakakis' algorithm via semi-joins, extended by the frequency attribute $c_u$. 
That is, we never add tuples, we only add one attribute to each relation. 
Similarly, for every aggregate expression $A_j(f_j(\bar B_j))$ that is not guarded by the root $r$ of $T$, 
we add an attribute $\Aggj$ to all nodes along the path between the guard of 
$A_j(f_j(\bar B_j))$  and the root. In other words, the data structures that we have to materialise
and propagate
in the course of our evaluation of \pwg queries is {\em linearly bounded} in the size of the data.

We note that this property would no longer be guaranteed for one of the following two extensions of the \pwg fragment: 
either allowing aggregates other than
\texttt{MIN}, 
\texttt{MAX}, \texttt{SUM}, \texttt{COUNT}, 
\texttt{AVG} to be guarded by a relation different from the root guard
or allowing aggregate expressions  
$A_j(f_j(\bar B_j))$ whose attributes are not guarded by a single relation. 
In both cases, one would have to propagate (all possible) 
individual values of attributes rather than 
aggregated values up the join tree, which would destroy this linear bound. 
A more detailed discussion is given 
\ifArxiv
in Appendix~\ref{app:limits}
\else
in the full paper~\cite{DBLP:journals/corr/abs-2406-17076}.
\fi

\section{Optimised Physical Operators}
\label{sect:physicalOpt}

The optimisations presented in 
Section~\ref{sect:OurSystem}
avoid a 
good deal of materialisation of intermediate results.
But there are still joins needed, namely, between the relations at a parent node and its child nodes.
Only after these joins, we apply the grouping and aggregation and thus 
bring the intermediate relations back to linear size (data complexity).
Similarly to Schleich et al.~\cite{DBLP:conf/sigmod/SchleichOK0N19}, 
we now combine the join computation with aggregation into
a single operation. We thus introduce a new physical operator 
(referred to as AggJoin) that
computes and propagates the 
frequency attribute $c$ from Section~\ref{sect:guarded} 
and the additional aggregate attributes $\Aggj$ from Section~\ref{sect:piecewise}
in a semi-join-like style. 
Below, we describe a possible \emph{join-less} realisation of this operator. 
As a preprocessing step (yet before the AggJoin is called), 
the following operations are carried out: 

Every relation is extended by a {\em frequency attribute} $c$,
and, for every tuple $t$ in every relation $R$, we initialise $t.c$ as $t.c$ := $1$.
Moreover, for every aggregate expression $\Aj$ that is not guarded by the root guard, 
we determine the node $w$ highest up in the join tree that contains all attributes in $\fBj$. 
Then we add an attribute $\Aggj$ to every relation along the path from $w$ to $r$.
For every tuple $t \in R(w)$, we initialise this attribute as follows: 
    \begin{itemize}
        \item If  $A_j \in \{ \texttt{MIN}, \texttt{MAX}, \texttt{SUM} \}$, then 
        we set $t.\Aggj$ := $f_j(t.\bar B_j)$, i.e., we apply $f_j$ to the values of 
        the attributes 
        $\bar B_j$ in tuple $t$.
        \item If  $A_j \in \{ \texttt{COUNT}\}$, then we 
        set $t.\Aggj$ := 1 if $f_j(t.\bar B_j)   \neq \texttt{NULL}$, and 
         $t.\Aggj$ := 0 if $f_j(t.\bar B_j) = \texttt{NULL}$.
    \end{itemize}
For $A_j \in \{ \texttt{SUM}, \texttt{COUNT}\}$, 
we thus deviate from the initialisation of $t.\Aggj$ at node $w$
described in Section~\ref{sect:piecewise} by leaving out 
the multiplication of $t.\Aggj$ with 
the frequency value $t.c$.  This multiplication by $t.c$ has to be 
integrated into the AggJoin operator, which will take care of this multiplication
when it determines $t.c$.

Finally, if  $A_j \in \{ \texttt{SUM}, \texttt{COUNT}\}$, then we set $t.\Aggj$ := 1
for every tuple $t$ in a relation labelling an ancestor node $u$ of $w$. The reason for this is
that it will allow us to uniformly propagate the $\Aggj$ value from one child of $u$ and 
the frequency values from the other children of $u$ in a uniform way via multiplication.

From now on, let $R$ and $S$ denote relations labelling nodes $u_R$ and $u_S$ in the join tree, such that $u_S$ is a child node of $u_R$. 
We first describe the propagation of the frequency attribute by the AggJoin for a tuple $r \in R$. 
\begin{itemize}
    \item check that $r \in R \ltimes S$ holds;
    \item define $S'$ := $S \ltimes \{r\}$, i.e., 
   the tuples in $S$ that join with $r$;
    \item define $sc$ := $\sum_{s \in S'} s.c$, i.e., the sum 
    of the frequencies of all tuples in $S$ that join with $r$;
    \item Finally, we set $r.c$ := $r.c \cdot sc$, i.e., the frequency of $r$ is multiplied by the sum of the frequencies 
    of the join partners in $S$. Here, it makes no difference, if $r.c$ still had its initial value 1 or if $R$ had already gone through 
    calls of AggJoin with relations at other child nodes of $u_R$.
\end{itemize}
It is easy to verify that this new AggJoin operator does 
precisely the work needed to get from $\Count_{i-1}(u_R)$ to  $\Count_{i}(u_R)$
according to 
Section~\ref{sect:guarded}. 
After the initialisation, we have $r.c =1$ for all tuples in $R$. This 
corresponds to $\Count_{0}(u_R)$.
Then we successively execute the AggJoin operator, where $R = \Count_{i-1}(u_R)$
and $S$ is the relation at the $i$-th child node of $u_R$. 
Hence, in each such call, we either delete $r$ (if it has no join partner in $S$) or 
we multiply the current value of $r.c$ by the sum of 
the frequencies of its join partners in $S$.

Let us now consider the aggregate expressions $\Aj$ that are not guarded by the root guard.
For the {\em initialisation} of the attribute $\Aggj$ in case of 
$A_j \in \{ \texttt{SUM},\texttt{COUNT} \}$, 
we proceed as with
the frequency attribute: 
Suppose that relation $R$ is the one, where $\Aggj$ has to be initialised. 
Now, for every tuple $r \in R$, the value initially assigned to $r.\Aggj$ ultimately 
has to be multiplied by $r.c$ to arrive at the initialisation according to Section~\ref{sect:piecewise}. 
Hence, for every relation $S$ at a child node of $u_R$, 
we multiply $r.\Aggj$ with $sc$ =  
the sum of the frequencies of all tuples in $S$ that join with $r$.

For  the {\em propagation}  of the attribute $\Aggj$ in case of 
$A_j \in \{\texttt{MIN}$, \texttt{MAX}, $\texttt{SUM}, \texttt{COUNT}\}$,
we distinguish two cases: 
First suppose that $S$ does not contain the attribute $\Aggj$.
Then, for every tuple $r \in R$  that has at least one join partner in $S$,
we proceed as follows:  
For  $A_j \in \{\texttt{MIN}, \texttt{MAX}\}$, we simply leave the value of $r.\Aggj$ unchanged,
i.e., the $\Aggj$ attribute is propagated to $R$ from the relation at a different child node.
For $A_j \in \{\texttt{SUM}, \texttt{COUNT}\}$, we again proceed 
analogously to the frequency propagation, i.e., 
we multiply $r.\Aggj$ with 
the sum of the frequencies of all tuples in $S$ that join with $r$.
Now suppose that $S$ contains the attribute $\Aggj$.
Then, for every tuple $r \in R$  that has at least one join partner in $S$,
we proceed as follows:  
For  $A_j \in \{\texttt{MIN}, \texttt{MAX}\}$, we assign to $r.\Aggj$ 
the minimum resp.\ maximum value of $s.\Aggj$ over all tuples $s \in S$ that join with $r$.
For $A_j \in \{\texttt{SUM}, \texttt{COUNT}\}$, we determine the sum of the values 
$s.\Aggj$ over all tuples $s \in S$ that join with $r$ and multiply 
the current value of $r.\Aggj$ with this sum.

\begin{algorithm}[t]
\SetKwProg{Fn}{Function}{}{}
\SetKwRepeat{Do}{do}{while}
\KwIn{Two lists $R,S$ of tuples with the same values of the join attributes; \newline
List $I_S = \{s_1, \dots, s_m\}$ of indices of aggregate attributes $\Agg_{s_i}$
present in both $S$ and $R$; \newline
List $I_R = \{r_1, \dots, r_n\}$ of indices of aggregate attributes $\Agg_{r_i}$
present only in $R$;
}

\Fn{AggHashJoin$(R,S, I_S, I_R)$}{
    $sc \gets 0$\;
    \ForEach{$s \in I_S$}{
    \lIf{$A_s \in \{\MIN, \MAX\}$}{$val_s \gets init[s]$}
    \lIf{$A_s \in \{\SUM, \COUNT\}$}{$val_s \gets 0$}
    }
    \ForEach{$t \in S$}{
       $sc \gets sc + t.c$\;
       \ForEach{$s \in I_S$}{
          \lIf{$A_s \in \{\MIN, \MAX\}$}{$val_s \gets A_s (val_s,t.\Agg_s)$}
          \lIf{$A_s \in \{\SUM, \COUNT\}$}{$val_s \gets val_s + t.\Agg_s$}
        }
    }
    \ForEach{$t \in R_v$}{
        $t.c  \gets  t.c \cdot sc$\;
       \ForEach{$s \in I_S$}{
          \lIf{$A_s \in \{\MIN, \MAX\}$}{$t.\Agg_s \gets A_s (val_s,t.\Agg_s)$}
          \lIf{$A_s \in \{\SUM, \COUNT\}$}{$t.\Agg_s \gets t.\Agg_s \cdot val_s$}
        }
       \ForEach{$r \in I_R$}{
          \lIf{$A_s \in \{\SUM, \COUNT\}$}{$t.\Agg_r \gets t.\Agg_r \cdot sc$}
        }
        \textbf{emit} $t$\;
    }
}
\caption{Hash Join with aggregate propagation}
\label{alg:HashJoin}
\end{algorithm}

\nop{Kuerzen***********************
The work done by the AggJoin operator essentially corresponds to
a sum-product over $K$-relations~\cite{DBLP:conf/pods/GreenKT07} for appropriately chosen semi-ring $K$.
While $K$-relations are widely studied, they are not typically connected to standard SQL query evaluation. The rewriting from 
Section~\ref{sect:OurSystem} enables us to naturally connect the two concepts
**********************}
The rewriting from 
Section~\ref{sect:OurSystem} allows for a smooth 
integration of the AggJoin operator into the physical query plan.
For instance, we have extended Spark SQL by 
three different implementations of the AggJoin operator, 
corresponding to the existing three join implementations 
\textit{shuffled-hash join}, {\em sort-merge join}, 
and \textit{broadcast-hash join}. 
In Algorithm~\ref{alg:HashJoin},
we sketch the realisation of 
the AggJoin operator based on the shuffled-hash join.
We use pseudo-code notation to leave out the technical details
so as not to obscure the simplicity of the extension from join computation to 
semi-join-like aggregation. 
As in the explanations above, we write $R$ and $S$ to denote 
pairs of relations whose nodes in the join tree are in parent-child relationship. 
Moreover, 
the AggJoin operator is only called after all the initialisations of additional attributes $t.c$ and 
$t.\Aggj$ have been carried out as described above.
Clearly, the hash-phase
(including the partitioning by Spark SQL)
is left unchanged. 
Only the join-phase is affected, which we briefly discuss next:

The AggHashJoin takes as input a set of tuples from $R$ and of tuples from $S$ that join. Additionally, the indices of the aggregate attributes $I_S$ (which have to be
propagated from $S$ to $R$) and  $I_R$ (which are only
contained in $R$) are taken as input.
In the first step, we initialise $sc$ (that is used to sum up the frequency values
over the tuples in $S$) 
and $val_s$ for every $s \in I_s$ (that is used for aggregating the attribute $\Agg_s$). 
An aggregate attribute $\Agg_s$ is used to propagate values for the 
aggregate expression $\As$. 
For $A_s \in \{ \MIN, \MAX \}$ 
we assume that the (system-dependent) 
\emph{maximal} element for this data type in case of \MIN\ and the 
\emph{minimal} one in case of \MAX, respectively,
is stored in the variable $init[s]$. The foreach-loop over the tuples of $S$ aggregates
the frequency attribute and all the other additional attributes. 
The foreach-loop over the tuples of $R$ uses these aggregated values from the tuples of $S$ 
to update the corresponding attributes of the tuples in $R$. 
The latter foreach-loop also has to 
multiply the initial value of aggregate attributes in case of \SUM\ and \COUNT\
by the sum $sc$ of the frequency attributes.

\ifArxiv
In Appendix~\ref{app:physical},
\else
In the full paper \cite{DBLP:journals/corr/abs-2406-17076},
\fi
we also discuss the realisation of the Agg\-Join in case of
the  {\em sort-merge join}, 
and \textit{broadcast-hash join}. It is then straightforward
to extend these ideas to join types not supported by Spark SQL such as 
the block-nested loops join.

Clearly,  replacing a physical join operation by the respective AggJoin variant does not
introduce any overhead (apart from the computationally cheap
management of the additional frequency and aggregate attributes). 
Moreover, if none of the additional attributes is needed (e.g., if the query is 0MA), then 
our AggJoin operator actually degenerates to a simple semi-join. 

\section{Experimental Evaluation}
\label{sect:Experiments}

\paragraph{Experimental Setup}
In order to evaluate our approach for mate\-rialisation-free query evaluation, we have implemented the methods presented in 
Sections \ref{sect:OurSystem} and \ref{sect:physicalOpt}
in Spark SQL. 
We perform experiments using a wide range of standard benchmarks, namely the
\textit{Join Order Benchmark (JOB)}~\cite{DBLP:journals/pvldb/LeisGMBK015}, \textit{STATS-CEB}~\cite{DBLP:journals/pvldb/HanWWZYTZCQPQZL21}, TPC-H~\cite{tpch}, TPC-DS~\cite{tpcds}, and the \textit{Large-Scale Subgraph Query Benchmark (LSQB)} \cite{DBLP:conf/sigmod/MhedhbiLKWS21}.
In addition, we evaluate performance on simple graph queries evaluated on two real-world graphs from the \textit{SNAP (Stanford Network Analysis Project}
    \cite{snapnets}) dataset.
In particular, we experiment on the following two datasets  which are commonly used in analyses of graph queries (e.g., 
by Hu and Wang~\cite{DBLP:journals/pacmmod/0005023}):

\medskip
{\small\hspace{1em}
    \begin{tabular}{c|c|c|c}
\centering
        \textbf{Graph} & \textbf{Nodes} & \textbf{Edges} & \textbf{(un)directed} \\
        \midrule
         web-Google & 875,713 & 5,105,039 & directed \\
         com-DBLP & 317,080 & 1,049,866 & undirected \\
    \end{tabular}
    }
\medskip

We evaluate the performance of basic graph queries, namely path queries requiring 
between 3 and 8 joins (i.e., between 4 and 9 edges) 
and three small tree queries.
These queries can be viewed as counting the number of homomorphisms from certain patterns (i.e., paths and trees in this case); a task that has recently gained popularity in graph learning where the results of the queries are injected into machine learning models (e.g.,~\cite{DBLP:conf/icml/NguyenM20,DBLP:conf/nips/BarceloGRR21,DBLP:journals/corr/abs-2402-08595,bao2024homomorphismcountsstructuralencodings}).


%
The overall performance of our proposed optimisations on the applicable queries is summarised in Table~\ref{tab:mainresults}. 
We refer to the reference performance of Spark SQL without any alterations as \emph{Ref}. Our experiments on the SNAP graphs specifically are summarised in Table~\ref{tab:snap-results}. 
The fastest execution time achieved for each case is printed in boldface.
The results obtained by applications of the logical optimisations described in Section~\ref{sect:guarded} 
for guarded queries
are referred to
as \emph{Guarded Aggregate Optimisation (GuAO)}. We use \emph{GuAO$^+$} to refer to the further extension of our logical optimisations 
to \pwg queries described in Section~\ref{sect:piecewise} plus the enhancement of the physical query plan
using the AggJoin operator described in Section~\ref{sect:physicalOpt}.
The speed-up achieved by \emph{GuAO$^+$} over \emph{Ref} is explicitly stated in Table~\ref{tab:mainresults} in the column \emph{GuAO$^+$ Speedup}.
For most benchmarks, we report end-to-end (e2e) times for subsequently executing all queries of a given benchmark where our optimisations are applicable (to the full query, or at least one subquery).

In all experiments, we execute each query 6 times, with the first run being a warm-up run to ensure that our measurements are not affected by initial reads of tables into memory. We report statistics gathered from the last 5 runs and report mean query execution time 
as well as the standard deviation over these runs. 
Finally, note that we execute the full query, even if our optimisation applies only to a subquery. In such a case, the plan for the subquery is optimised according to Section~\ref{sect:OurSystem}, and the rest of the query plan remains unchanged.

Full experimental details are provided in the full paper~\cite{DBLP:journals/corr/abs-2406-17076}. Our implementation, with a fully automated setup for reproducibility is available at \url{https://github.com/dbai-tuw/spark-eval}.

\nop{
\begin{table*}
\caption{Overview of the benchmark queries}
        \begin{tabular}{|c|c|c|c|c|c|c|}
    \hline
        \textbf{Benchmark} & \textbf{queries} & \textbf{equi-join agg queries} & \textbf{acyclic} & \textbf{pw-guarded} & \textbf{guarded} & \textbf{0MA} \\
        \hline
        JOB & 113 & 113 & 113 & 113  & 19 & 19 \\
        \hline
        STATS-CEB & 146 & 146 & 146 & 146 & 146 & 0 \\
        \hline
        TPC-H & 22 & 15 & 14 & 7 & 3 & 1 \\
        \hline
        LSQB & 9 & 4 & 2 & 2 & 2 & 0 \\
        \hline
        SNAP & 18 & 18 & 18 & 18 & 18 & 0 \\
        \hline
        TPC-DS & 99 & & & & & \\
        \hline
    \end{tabular}
\end{table*}
}

\begin{table*}
\begin{minipage}{.37\linewidth}
\small
\centering
\caption{Summary of the applicability of our method on benchmarks. We report the number of queries in benchmark (\#), equi-join aggregate queries (\emph{$\bowtie$-agg}), acyclic queries (\emph{acyc}), \pwg queries (\emph{pwg}), guarded queries (\emph{g}), and 0MA queries. Fragments proposed in this work are highlighted in blue.}
\label{tbl:applic}
{\renewcommand{\arraystretch}{1}%
\setlength{\tabcolsep}{3.5pt}
        \begin{tabular}{|c|c|c|c|c|c|c|}
    \hline
        \textbf{Benchmark} & \textbf{\#} & \textbf{$\bowtie$-agg} & \textbf{acyc} & {\color{RoyalBlue}\textbf{pwg}} & {\color{RoyalBlue}\textbf{g}} & \textbf{0MA} \\
        \hline
        JOB & 113 & 113 & 113 & 113  & 19 & 19 \\
        STATS-CEB & 146 & 146 & 146 & 146 & 146 & 0 \\
        TPC-H & 22 & 15 & 14 & 7 & 3 & 1 \\
        LSQB & 9 & 4 & 2 & 2 & 2 & 0 \\
        SNAP & 18 & 18 & 18 & 18 & 18 & 0 \\
        TPC-DS & 99 & 64 & 63 & 30 & 15 & 0 \\
        \hline
    \end{tabular}
}
\end{minipage}
%
%
%
%
%
%
%
\hfill
\begin{minipage}{0.59\linewidth}
%
%
\centering
\caption{Performance on SNAP graphs (t.o. marks timeouts, o.o.m marks out of memory errors).}
\label{tab:snap-results}
{\renewcommand{\arraystretch}{1}%
\setlength{\tabcolsep}{4.2pt}
\begin{tabular}{|c||c|c|c||c|c|c|} 
\hline
\multicolumn{1}{|c||}{} & \multicolumn{3}{c||}{\bf web-Google}                                 & \multicolumn{3}{c|}{\bf com-DBLP}                                 \\
\textbf{Query}         & \bf Spark    &  \bf GuAO       & \bf GuAO$^+$        & \bf Spark     & \bf GuAO      & \bf GuAO$^+$        \\ 
\hhline{|=::==::====|}
path-03                & 27.97\sd{1.5}      & 6.90\sd{0.6}  & 6.08\sd{0.65}  & 6.32\sd{1.1}    & 2.35\sd{0.5} &  1.59\sd{0.12}  \\
path-04                & 449.14\sd{26.9}   & 7.58\sd{0.6}  & \bf 6.89\sd{0.30}  & 50.97\sd{9.8}    & 2.24\sd{0.4} & \bf 1.76\sd{0.16}  \\
path-05                & o.o.m.             & 8.95\sd{1.0}  & \bf 7.53\sd{0.48}  & 400.87\sd{15.2}   & 2.74\sd{0.2} & \bf 2.03\sd{0.25}  \\
path-06                & o.o.m.           & 9.37\sd{1.0}  & \bf 8.80\sd{0.25}  & o.o.m.            & 2.98\sd{0.2} & \bf 2.18\sd{0.14}  \\
path-07                & o.o.m.            & 11.32\sd{0.9} & \bf 9.76\sd{1.21}  & o.o.m.          & 3.64\sd{0.2} & \bf 2.38\sd{0.26}  \\
path-08                & o.o.m.              & 11.30\sd{2.1} & \bf 10.05\sd{1.49} & o.o.m.            & 3.75\sd{0.4} & \bf 2.53\sd{0.30}  \\
tree-01                & 539.11\sd{22.4}   & 7.73\sd{1.0}  & \bf 6.53\sd{1.11}  & 25.96\sd{4.5}    & 1.95\sd{0.1} & \bf 1.47\sd{0.28}  \\
tree-02                & o.o.m.             & 12.43\sd{3.2} & \bf 7.29\sd{0.73}  & 328.88\sd{11.5}  & 3.02\sd{0.7} & \bf 1.69\sd{0.16}  \\
tree-03                & o.o.m.            & 12.21\sd{5.6} & \bf 8.16\sd{0.66}  & o.o.m.         & 3.17\sd{0.2} & \bf 1.99\sd{0.16}  \\
\hline
\end{tabular}

}
\end{minipage}
\end{table*}

{\renewcommand{\arraystretch}{1}%
\begin{table*}[t]
    \caption{Summary of the impact of 
  aggregate optimisation on 
   execution times (seconds). 
    Reported numbers are mean times over 5 runs of the same query with standard deviations given after $\pm$. 
    ``--'' indicates that the query is \pwg and, therefore,  the optimisation from Section \ref{sect:guarded} for guarded queries 
    is not applicable. 
    }
    \label{tab:mainresults}
    \vspace{-1em}
\setlength\doublerulesep{8pt}
    \centering
    \begin{tabular}{|c|c||r|r|r|r|} 
\hhline{--::----}
\textbf{Query}      & \textbf{\# joins (mean)}           & \bf Ref                                                                                                           & \bf GuAO        & \bf GuAO$^+$        & \bf GuAO$^+$ Speedup   \\ 
\hhline{--||----}
STATS-CEB e2e     &      3.33       & 1558\sd{7.3}                                                                                                       & 97.9\sd{6.1}   & \bf 64.8\sd{7.9}       & 24.04 x          \\ 
\hhline{--||----}
JOB e2e            &    7.65      & 3217.84\sd{106}                                                                                                     & -  & \bf 2189.46\sd{76} & 1.47 x       \\
\hhline{--||----}
TPC-H e2e \scriptsize{SF200} & 1.57 & 3757.2                                                                                                      &  - & \bf 3491.06  & 1.08 x           \\
TPC-H Ex.1 \scriptsize{SF200} & 4 & 168.4                                                                                                      &  107.5 & \bf 105.11  & 1.60 x           \\

\hhline{--||----}
LSQB Q1 \scriptsize{SF300} & 9 & 3096\sd{232} 
& \bf 677\sd{23} & 688\sd{23}         & 4.57 x          \\
LSQB Q4 \scriptsize{SF300} & 3 & 602\sd{37}                                                                                                         & 593\sd{15}     & \bf 592\sd{9}          & 1.02x         \\ 
\hhline{--||----}
TPC-DS e2e \scriptsize{SF100} & 2.52 & 5154.5                                                                                                      &  - &  5047.5 & 1.02 x           \\

\hhline{--||----}
\end{tabular}
\end{table*}

}
\label{sect:ExperimentalResults}

\subsubsection*{Applicability}
\phantom{.} \smallskip

\noindent
To enable Yannakakis-style query evaluation in the context of standard query execution engines, we have focused on specific queries, namely 
guarded and piecewise-guarded acyclic aggregate queries (cf., Section~\ref{sect:OurSystem}).
As a first step, we therefore provide a more detailed analysis as to how many of the queries in the studied benchmarks fit into this class, and what factors limit further applicability. The analysis is summarised in Table~\ref{tbl:applic}.

Despite the variety of considered benchmarks, we find that our optimisations for piecewise-guarded queries are widely applicable through all of them. In JOB and STATS-CEB, all queries fall into the schema of piecewise guarded-aggregation, and our method thus applies to these benchmarks as a whole. 
In contrast, the 0MA fragment covers only 19 of the 259 queries in total in these two benchmarks.
Our methods also apply to all the tested basic graph queries (path or tree queries) that were tested on the SNAP dataset.

In LSQB, our approach applies to 2 of the 9 queries. 
But as reported in Table~\ref{tbl:applic}, only 4 of the queries are equi-join queries with aggregation, the others contain joins on inequalities, which requires entirely different techniques (e.g., \cite{DBLP:journals/pvldb/KhayyatLSOPQ0K15}). Of the 4  equi-join queries, 2 are not acyclic.

Our method applies to half of the equi-join aggregate queries in the TPC-DS benchmark. The queries in the benchmark are typically highly complex, often combining multiple subqueries and employing more elaborate SQL features. We observe that in some instances, \emph{GuAO$^+$} even applies to multiple subqueries in the same query.
In TPC-H, our optimisations apply to 7 out of the 15 acyclic equi-join aggregate queries in the benchmark. Notably, TPC-H Q2 contains a 0MA subquery (with \texttt{MIN} aggregation) and TPC-H Q11 contains a guarded sum aggregate subquery. 

TPC-H Q2 is particularly illustrative as the subquery is correlated: the attribute \texttt{p\_partkey} from the outer query is used in the aggregation subquery as follows:
\smallskip
    \begin{minted}{sql}
    SELECT MIN(ps_supplycost) 
    ... WHERE p_partkey = ps_partkey ...
    \end{minted}
\smallskip
The Spark SQL query planner decorrelates this subquery via typical magic decorrelation (see~\cite{DBLP:conf/icde/SeshadriPL96}) -- resulting in the following select statement for the decorrelated subquery. This query is still guarded and thus 0MA. Our rewriting rules then apply naturally after decorrelation, with no need for any special handling of these cases.
\smallskip
\begin{minted}{sql}
    SELECT ps_partkey, MIN(ps_supplycost)
    ... GROUP BY ps_partkey
\end{minted}
\smallskip

We recall that our method is fully integrated into the query optimisation phase. Hence, when our optimisations are not applicable to a query, its execution is not affected. Recognising whether the rewriting rules are applicable is trivial and requires, in our observations, negligible additional time in the query planning phase to perform our rewriting (about 2ms in all of our experiments). Going forward, it is additionally possible to pre-process queries to make them fit into the fragments where our methods are applicable. 
A brief discussion of such approaches is given in the outlook to future work in Section~\ref{sect:Conclusion}.

\subsubsection*{Performance impact of GuAO/GuAO$^+$}
\phantom{.} \smallskip

\noindent
Our  main results over standard benchmarks are summarised in Table~\ref{tab:mainresults}.
We additionally report the mean number of joins of the (sub)queries affected by our optimisation.
We make two key observations with respect to the performance of our methods. 

When queries are challenging -- e.g., they have many joins or the joins are not along PK/FK pairs -- then our method provides enormous potential for speed-up. In JOB, a benchmark where suboptimal join orderings in large queries cause intermediate blow-up, we achieve almost 50\% speed-up. STATS-CEB purposefully introduces joins along FK/FK relationships to challenge query evaluation systems with the resulting large number of intermediate tuples. Our method automatically avoids all of these difficulties and we see an immense 24-fold speed-up. Similarly, for the more difficult of the two LSQB queries (Q1), we see that the large number of joins creates significant intermediate blow-up with standard relational query evaluation. Again our method achieves a very large improvement of about 450\%. Even in the simple query from Example~\ref{ex:tpch-query} we observe 60\% speed-up over unoptimised Spark SQL.

On the other hand, we observe that especially the two TPC benchmarks contain primarily queries where the join evaluation itself is very straightforward. In TPC-H,  4 of the 7 tested queries contain only a single join. In TPC-DS we observe similar patterns. As a result, there is little to no unnecessary materialisation in many of these queries. Our key insight here is that the experiments confirm that our method (and in particular \emph{GuAO$^+$}) does not introduce any overhead in these cases. By not causing performance decrease in those cases where there is no unnecessary materialisation, combined with some gains in the few harder queries, we still see modest overall speed-ups for these benchmarks.

With respect to basic graph queries, we see in Table~\ref{tab:snap-results} that even with significant resources, counting short paths and small trees is effectively impossible on large graphs with current methods. This 
holds for both, Spark SQL and for specialised graph
database systems. In stark contrast, \emph{GuAO} and \emph{GuAO$^+$} effectively trivialise these types of queries even with significantly less resources than are available on our test system  (the highest observed  memory usage for \emph{GuAO$^+$} in our SNAP experiments was roughly 5GB). Since these experiments focus on graph data,  
we additionally compared 
with specialised graph database systems (Neo4j~\cite{neoj}, K{\`{U}}ZU system~\cite{kuzu}, and GraphDB~\cite{graphdb}). The results of these experiments 
are equally
sobering as with standard Spark SQL, in that almost all queries failed with timeout (set to 30 min). 
This is in sharp contrast to 
GuAO and GuAO$^+$, which answer these queries in a matter of a few seconds. We refer to 
\ifArxiv
Appendix~\ref{app:benchmark}
\else
the full paper~\cite{DBLP:journals/corr/abs-2406-17076}
\fi
for detailed results.

In summary, our experiments paint a clear picture. In more challenging queries, our approach offers very significant improvements. At the same time, in cases where little unnecessary materialisation is performed, \emph{GuAO$^+$}  introduces no additional overhead and thus exhibits no performance degradation on simpler queries.

\begin{figure*}[t]
\vspace{-1em} 
    \centering
    \includegraphics[width=1.0\linewidth]{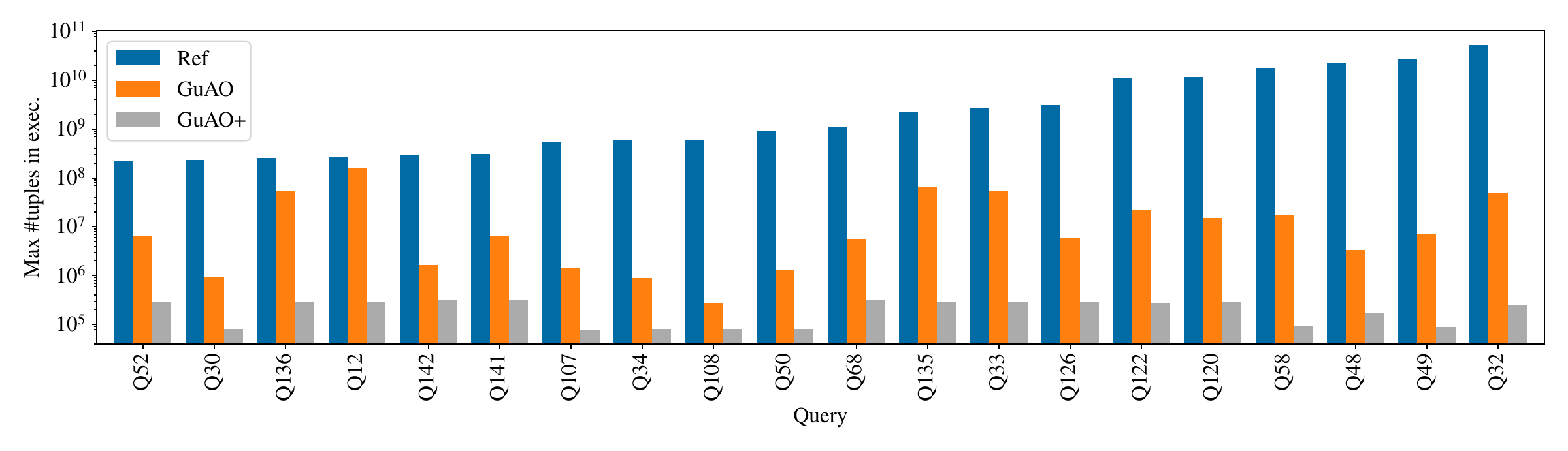}
    \vspace{-2em}
    \caption{Comparison of the maximal number of materialised tuples in a table during query execution for 20 queries of STATS-CEB. Y-axis in logarithmic scale (base 10).}
    \label{fig:maxrows}
\end{figure*}
\subsubsection*{How much materialisation can be avoided?}
\label{sect:material}
\phantom{.}
\smallskip

\noindent
Throughout the paper, 
we have been motivated by the premise that easy to implement logical optimisation rules for query plans can avoid a significant amount of intermediate materialisation in aggregate queries. Moreover, with the addition of natural physical operators, we can avoid any such materialisation altogether. However, this raises the question of how much unnecessary materialisation actually occurs when using standard query planning methods.

To study this question, we compare the maximum number of tuples that occur in an intermediate table during query execution for the STATS-CEB queries. Again, we  report the mean over 5 runs (we omit error bars as the variation between runs is mostly 0 and negligible in other cases).  We note that these intermediate table sizes are naturally closely correlated to overall memory consumption, as well as communication cost in a distributed setting.
Figure~\ref{fig:maxrows} reports the peak number of materialised tuples during query execution for the 20 queries where standard Spark SQL materialises the most intermediate tuples. 
The data clearly shows that an improvement in the order of magnitudes of materialised tuples is often possible.
In particular, we see the well documented effect of classical relational query processing techniques leading to substantial intermediate blow-up. The largest relations in the dataset have in the order of $3 \cdot 10^5$ tuples, an enormous difference to the observed sizes of up to $10^{10}$ intermediate tuples for \emph{Ref}. The data shows that already by rewriting the logical query plans according to 
Section~\ref{sect:OurSystem}, we regularly see a reduction in peak intermediate table size of over 2 orders of magnitude. 

However, 
the optimised logical query plan still requires some mild materialisation 
between aggregation steps, which we manage to eliminate with 
the physical operators described in Section~\ref{sect:physicalOpt}. The resulting \emph{GuAO$^+$} system consistently reduces the number of materialised tuples by \textbf{at least 3 orders of magnitude} on the reported queries in Figure~\ref{fig:maxrows}. In fact, the reported numbers for \emph{GuAO$^+$} are always precisely the cardinality of the largest relation in the query, as the execution using our method 
never introduces any new  tuples (cf.~Section~\ref{sect:physicalOpt}). That is, this number can also inherently not be improved upon. Over the whole benchmark, we observe that the peak number of materialised tuples by \emph{GuAO$^+$} is \emph{at least 10 times less} than that of standard Spark SQL query execution in 118 out of the 146 queries. In all other cases, the peak number of materialised tuples by \emph{Ref} and \emph{GuAO$^+$} is exactly the same, i.e., \emph{Ref} is never better.

\section{Conclusion}
\label{sect:Conclusion}

In this work, we have introduced several optimisations for guarded aggregate queries, enabling significant reductions 
of the need to materialise intermediate results when evaluating analytical queries with aggregates over join or path queries. 
Our approach emphasises seamless integration into standard database systems, requiring only localised modifications to logical query plans. Additionally, we propose the use of new physical operators that extend semi-joins to manage frequencies and other aggregate information
with the aim to completely eliminate  the computation of intermediate joins
and to facilitate straightforward integration into physical query plans.


We have implemented our optimisations into Spark SQL, which has been specifically designed as a 
powerful tool to deal with complex analytical queries. Our experimental evaluation confirms that the proposed techniques can provide significant performance improvements by avoiding costly materialisation, especially in larger queries. 
Furthermore, 
the integration of our optimisations into Spark SQL serves not only as a proof of concept but 
is already a practical enhancement of an important tool in data analytics -- which is even more 
valuable given the limited applicability of advanced query optimisation techniques in Spark SQL due to its limited cost-model.

So far, we have applied our optimisation techniques to acyclic queries with \pwg aggregation.
In principle, both acyclicity and piecewise-guardedness can be enforced by applying appropriate joins upfront. More specifically, if a query uses unguarded aggregation, we can create a guard by joining relations that cover all attributes involved in grouping and/or  aggregation.
Similarly, a cyclic query can be transformed into an acyclic one via (generalised or fractional) hypertree 
decompositions~\cite{DBLP:conf/pods/GottlobLS99,DBLP:journals/ejc/AdlerGG07,2014grohemarx}. Again,
this requires the computation of some joins upfront.
Of course, carrying out such joins upfront to ensure guardedness and/or acyclicity comes at a cost -- 
both in terms of space and time. 
We leave it as an interesting open question for future research to analyse 
when extending our approach to cyclic and/or  unguarded queries actually pays off.

\clearpage

\bibliographystyle{ACM-Reference-Format}
\bibliography{main}


\begin{thebibliography}{61}


\ifx \showCODEN    \undefined \def \showCODEN     #1{\unskip}     \fi
\ifx \showDOI      \undefined \def \showDOI       #1{#1}\fi
\ifx \showISBNx    \undefined \def \showISBNx     #1{\unskip}     \fi
\ifx \showISBNxiii \undefined \def \showISBNxiii  #1{\unskip}     \fi
\ifx \showISSN     \undefined \def \showISSN      #1{\unskip}     \fi
\ifx \showLCCN     \undefined \def \showLCCN      #1{\unskip}     \fi
\ifx \shownote     \undefined \def \shownote      #1{#1}          \fi
\ifx \showarticletitle \undefined \def \showarticletitle #1{#1}   \fi
\ifx \showURL      \undefined \def \showURL       {\relax}        \fi
\providecommand\bibfield[2]{#2}
\providecommand\bibinfo[2]{#2}
\providecommand\natexlab[1]{#1}
\providecommand\showeprint[2][]{arXiv:#2}

\bibitem[\protect\citeauthoryear{Aberger, Lamb, Tu, N{\"{o}}tzli, Olukotun, and R{\'{e}}}{Aberger et~al\mbox{.}}{2017}]%
        {DBLP:journals/tods/AbergerLTNOR17}
\bibfield{author}{\bibinfo{person}{Christopher~R. Aberger}, \bibinfo{person}{Andrew Lamb}, \bibinfo{person}{Susan Tu}, \bibinfo{person}{Andres N{\"{o}}tzli}, \bibinfo{person}{Kunle Olukotun}, {and} \bibinfo{person}{Christopher R{\'{e}}}.} \bibinfo{year}{2017}\natexlab{}.
\newblock \showarticletitle{EmptyHeaded: {A} Relational Engine for Graph Processing}.
\newblock \bibinfo{journal}{\emph{{ACM} Trans. Database Syst.}} \bibinfo{volume}{42}, \bibinfo{number}{4} (\bibinfo{year}{2017}), \bibinfo{pages}{20:1--20:44}.
\newblock
\urldef\tempurl%
\url{https://doi.org/10.1145/3129246}
\showDOI{\tempurl}


\bibitem[\protect\citeauthoryear{Abiteboul, Hull, and Vianu}{Abiteboul et~al\mbox{.}}{1995}]%
        {DBLP:books/aw/AbiteboulHV95}
\bibfield{author}{\bibinfo{person}{Serge Abiteboul}, \bibinfo{person}{Richard Hull}, {and} \bibinfo{person}{Victor Vianu}.} \bibinfo{year}{1995}\natexlab{}.
\newblock \bibinfo{booktitle}{\emph{Foundations of Databases}}.
\newblock \bibinfo{publisher}{Addison-Wesley}.
\newblock
\showISBNx{0-201-53771-0}
\urldef\tempurl%
\url{http://webdam.inria.fr/Alice/}
\showURL{%
\tempurl}


\bibitem[\protect\citeauthoryear{Adler, Gottlob, and Grohe}{Adler et~al\mbox{.}}{2007}]%
        {DBLP:journals/ejc/AdlerGG07}
\bibfield{author}{\bibinfo{person}{Isolde Adler}, \bibinfo{person}{Georg Gottlob}, {and} \bibinfo{person}{Martin Grohe}.} \bibinfo{year}{2007}\natexlab{}.
\newblock \showarticletitle{Hypertree width and related hypergraph invariants}.
\newblock \bibinfo{journal}{\emph{Eur. J. Comb.}} \bibinfo{volume}{28}, \bibinfo{number}{8} (\bibinfo{year}{2007}), \bibinfo{pages}{2167--2181}.
\newblock
\urldef\tempurl%
\url{https://doi.org/10.1016/j.ejc.2007.04.013}
\showDOI{\tempurl}


\bibitem[\protect\citeauthoryear{Afrati, Joglekar, R{\'{e}}, Salihoglu, and Ullman}{Afrati et~al\mbox{.}}{2017}]%
        {DBLP:conf/icdt/AfratiJRSU17}
\bibfield{author}{\bibinfo{person}{Foto~N. Afrati}, \bibinfo{person}{Manas~R. Joglekar}, \bibinfo{person}{Christopher R{\'{e}}}, \bibinfo{person}{Semih Salihoglu}, {and} \bibinfo{person}{Jeffrey~D. Ullman}.} \bibinfo{year}{2017}\natexlab{}.
\newblock \showarticletitle{{GYM:} {A} Multiround Distributed Join Algorithm}. In \bibinfo{booktitle}{\emph{Proceedings {ICDT}}} \emph{(\bibinfo{series}{LIPIcs})}, Vol.~\bibinfo{volume}{68}. \bibinfo{publisher}{Schloss Dagstuhl - Leibniz-Zentrum f{\"{u}}r Informatik}, \bibinfo{pages}{4:1--4:18}.
\newblock
\urldef\tempurl%
\url{https://doi.org/10.4230/LIPIcs.ICDT.2017.4}
\showDOI{\tempurl}


\bibitem[\protect\citeauthoryear{Armbrust, Xin, Lian, Huai, Liu, Bradley, Meng, Kaftan, Franklin, Ghodsi, and Zaharia}{Armbrust et~al\mbox{.}}{2015}]%
        {DBLP:conf/sigmod/ArmbrustXLHLBMK15}
\bibfield{author}{\bibinfo{person}{Michael Armbrust}, \bibinfo{person}{Reynold~S. Xin}, \bibinfo{person}{Cheng Lian}, \bibinfo{person}{Yin Huai}, \bibinfo{person}{Davies Liu}, \bibinfo{person}{Joseph~K. Bradley}, \bibinfo{person}{Xiangrui Meng}, \bibinfo{person}{Tomer Kaftan}, \bibinfo{person}{Michael~J. Franklin}, \bibinfo{person}{Ali Ghodsi}, {and} \bibinfo{person}{Matei Zaharia}.} \bibinfo{year}{2015}\natexlab{}.
\newblock \showarticletitle{Spark {SQL:} Relational Data Processing in Spark}. In \bibinfo{booktitle}{\emph{Proceedings {SIGMOD}}}. \bibinfo{publisher}{{ACM}}, \bibinfo{pages}{1383--1394}.
\newblock
\urldef\tempurl%
\url{https://doi.org/10.1145/2723372.2742797}
\showDOI{\tempurl}


\bibitem[\protect\citeauthoryear{Atserias, Grohe, and Marx}{Atserias et~al\mbox{.}}{2013}]%
        {DBLP:journals/siamcomp/AtseriasGM13}
\bibfield{author}{\bibinfo{person}{Albert Atserias}, \bibinfo{person}{Martin Grohe}, {and} \bibinfo{person}{D{\'{a}}niel Marx}.} \bibinfo{year}{2013}\natexlab{}.
\newblock \showarticletitle{Size Bounds and Query Plans for Relational Joins}.
\newblock \bibinfo{journal}{\emph{{SIAM} J. Comput.}} \bibinfo{volume}{42}, \bibinfo{number}{4} (\bibinfo{year}{2013}), \bibinfo{pages}{1737--1767}.
\newblock
\urldef\tempurl%
\url{https://doi.org/10.1137/110859440}
\showDOI{\tempurl}


\bibitem[\protect\citeauthoryear{Bakibayev, Kocisk{\'{y}}, Olteanu, and Zavodny}{Bakibayev et~al\mbox{.}}{2013}]%
        {DBLP:journals/pvldb/BakibayevKOZ13}
\bibfield{author}{\bibinfo{person}{Nurzhan Bakibayev}, \bibinfo{person}{Tom{\'{a}}s Kocisk{\'{y}}}, \bibinfo{person}{Dan Olteanu}, {and} \bibinfo{person}{Jakub Zavodny}.} \bibinfo{year}{2013}\natexlab{}.
\newblock \showarticletitle{Aggregation and Ordering in Factorised Databases}.
\newblock \bibinfo{journal}{\emph{Proc. {VLDB} Endow.}} \bibinfo{volume}{6}, \bibinfo{number}{14} (\bibinfo{year}{2013}), \bibinfo{pages}{1990--2001}.
\newblock
\urldef\tempurl%
\url{https://doi.org/10.14778/2556549.2556579}
\showDOI{\tempurl}


\bibitem[\protect\citeauthoryear{Baldacci and Golfarelli}{Baldacci and Golfarelli}{2019}]%
        {DBLP:journals/tkde/BaldacciG19}
\bibfield{author}{\bibinfo{person}{Lorenzo Baldacci} {and} \bibinfo{person}{Matteo Golfarelli}.} \bibinfo{year}{2019}\natexlab{}.
\newblock \showarticletitle{A Cost Model for {SPARK} {SQL}}.
\newblock \bibinfo{journal}{\emph{{IEEE} Trans. Knowl. Data Eng.}} \bibinfo{volume}{31}, \bibinfo{number}{5} (\bibinfo{year}{2019}), \bibinfo{pages}{819--832}.
\newblock
\urldef\tempurl%
\url{https://doi.org/10.1109/TKDE.2018.2850339}
\showDOI{\tempurl}


\bibitem[\protect\citeauthoryear{Bao, Jin, Bronstein, İsmail~İlkan Ceylan, and Lanzinger}{Bao et~al\mbox{.}}{2024}]%
        {bao2024homomorphismcountsstructuralencodings}
\bibfield{author}{\bibinfo{person}{Linus Bao}, \bibinfo{person}{Emily Jin}, \bibinfo{person}{Michael Bronstein}, \bibinfo{person}{İsmail~İlkan Ceylan}, {and} \bibinfo{person}{Matthias Lanzinger}.} \bibinfo{year}{2024}\natexlab{}.
\newblock \bibinfo{title}{Homomorphism Counts as Structural Encodings for Graph Learning}.
\newblock
\newblock
\showeprint[arxiv]{2410.18676}~[cs.LG]
\urldef\tempurl%
\url{https://arxiv.org/abs/2410.18676}
\showURL{%
\tempurl}


\bibitem[\protect\citeauthoryear{Barcel{\'{o}}, Geerts, Reutter, and Ryschkov}{Barcel{\'{o}} et~al\mbox{.}}{2021}]%
        {DBLP:conf/nips/BarceloGRR21}
\bibfield{author}{\bibinfo{person}{Pablo Barcel{\'{o}}}, \bibinfo{person}{Floris Geerts}, \bibinfo{person}{Juan~L. Reutter}, {and} \bibinfo{person}{Maksimilian Ryschkov}.} \bibinfo{year}{2021}\natexlab{}.
\newblock \showarticletitle{Graph Neural Networks with Local Graph Parameters}. In \bibinfo{booktitle}{\emph{Proceedings {NeurIPS}}}. \bibinfo{pages}{25280--25293}.
\newblock
\urldef\tempurl%
\url{https://proceedings.neurips.cc/paper/2021/hash/d4d8d1ac7e00e9105775a6b660dd3cbb-Abstract.html}
\showURL{%
\tempurl}


\bibitem[\protect\citeauthoryear{Chen and Dalmau}{Chen and Dalmau}{2012}]%
        {DBLP:conf/lics/ChenD12}
\bibfield{author}{\bibinfo{person}{Hubie Chen} {and} \bibinfo{person}{V{\'{\i}}ctor Dalmau}.} \bibinfo{year}{2012}\natexlab{}.
\newblock \showarticletitle{Decomposing Quantified Conjunctive (or Disjunctive) Formulas}. In \bibinfo{booktitle}{\emph{Proceedings {LICS}}}. \bibinfo{publisher}{{IEEE} Computer Society}, \bibinfo{pages}{205--214}.
\newblock
\urldef\tempurl%
\url{https://doi.org/10.1109/LICS.2012.31}
\showDOI{\tempurl}


\bibitem[\protect\citeauthoryear{Dai, Wang, and Yi}{Dai et~al\mbox{.}}{2023}]%
        {DBLP:conf/sigmod/Dai0023}
\bibfield{author}{\bibinfo{person}{Binyang Dai}, \bibinfo{person}{Qichen Wang}, {and} \bibinfo{person}{Ke Yi}.} \bibinfo{year}{2023}\natexlab{}.
\newblock \showarticletitle{SparkSQL+: Next-generation Query Planning over Spark}. In \bibinfo{booktitle}{\emph{Companion of the 2023 International Conference on Management of Data, {SIGMOD/PODS} 2023}}. \bibinfo{publisher}{{ACM}}, \bibinfo{pages}{115--118}.
\newblock
\urldef\tempurl%
\url{https://doi.org/10.1145/3555041.3589715}
\showDOI{\tempurl}


\bibitem[\protect\citeauthoryear{Durand and Mengel}{Durand and Mengel}{2015}]%
        {DBLP:journals/mst/0001M15}
\bibfield{author}{\bibinfo{person}{Arnaud Durand} {and} \bibinfo{person}{Stefan Mengel}.} \bibinfo{year}{2015}\natexlab{}.
\newblock \showarticletitle{Structural Tractability of Counting of Solutions to Conjunctive Queries}.
\newblock \bibinfo{journal}{\emph{Theory Comput. Syst.}} \bibinfo{volume}{57}, \bibinfo{number}{4} (\bibinfo{year}{2015}), \bibinfo{pages}{1202--1249}.
\newblock
\urldef\tempurl%
\url{https://doi.org/10.1007/S00224-014-9543-Y}
\showDOI{\tempurl}


\bibitem[\protect\citeauthoryear{Gottlob, Lanzinger, Longo, Okulmus, Pichler, and Selzer}{Gottlob et~al\mbox{.}}{2023a}]%
        {DBLP:conf/amw/GottlobLLOPS23}
\bibfield{author}{\bibinfo{person}{Georg Gottlob}, \bibinfo{person}{Matthias Lanzinger}, \bibinfo{person}{Davide~Mario Longo}, \bibinfo{person}{Cem Okulmus}, \bibinfo{person}{Reinhard Pichler}, {and} \bibinfo{person}{Alexander Selzer}.} \bibinfo{year}{2023}\natexlab{a}.
\newblock \showarticletitle{Reaching Back to Move Forward: Using Old Ideas to Achieve a New Level of Query Optimization (short paper)}. In \bibinfo{booktitle}{\emph{Proceedings {AMW}}} \emph{(\bibinfo{series}{{CEUR} Workshop Proceedings})}, Vol.~\bibinfo{volume}{3409}. \bibinfo{publisher}{CEUR-WS.org}.
\newblock
\urldef\tempurl%
\url{https://ceur-ws.org/Vol-3409/paper6.pdf}
\showURL{%
\tempurl}


\bibitem[\protect\citeauthoryear{Gottlob, Lanzinger, Longo, Okulmus, Pichler, and Selzer}{Gottlob et~al\mbox{.}}{2023b}]%
        {DBLP:journals/corr/abs-2303-02723}
\bibfield{author}{\bibinfo{person}{Georg Gottlob}, \bibinfo{person}{Matthias Lanzinger}, \bibinfo{person}{Davide~Mario Longo}, \bibinfo{person}{Cem Okulmus}, \bibinfo{person}{Reinhard Pichler}, {and} \bibinfo{person}{Alexander Selzer}.} \bibinfo{year}{2023}\natexlab{b}.
\newblock \showarticletitle{Structure-Guided Query Evaluation: Towards Bridging the Gap from Theory to Practice}.
\newblock \bibinfo{journal}{\emph{CoRR}}  \bibinfo{volume}{abs/2303.02723} (\bibinfo{year}{2023}).
\newblock
\urldef\tempurl%
\url{https://doi.org/10.48550/arXiv.2303.02723}
\showDOI{\tempurl}
\showeprint[arXiv]{2303.02723}


\bibitem[\protect\citeauthoryear{Gottlob, Leone, and Scarcello}{Gottlob et~al\mbox{.}}{2001}]%
        {DBLP:journals/jacm/GottlobLS01}
\bibfield{author}{\bibinfo{person}{Georg Gottlob}, \bibinfo{person}{Nicola Leone}, {and} \bibinfo{person}{Francesco Scarcello}.} \bibinfo{year}{2001}\natexlab{}.
\newblock \bibinfo{journal}{\emph{J. {ACM}}} \bibinfo{volume}{48}, \bibinfo{number}{3} (\bibinfo{year}{2001}), \bibinfo{pages}{431--498}.
\newblock
\urldef\tempurl%
\url{https://doi.org/10.1145/382780.382783}
\showDOI{\tempurl}


\bibitem[\protect\citeauthoryear{Gottlob, Leone, and Scarcello}{Gottlob et~al\mbox{.}}{2002a}]%
        {DBLP:journals/jcss/GottlobLS02}
\bibfield{author}{\bibinfo{person}{Georg Gottlob}, \bibinfo{person}{Nicola Leone}, {and} \bibinfo{person}{Francesco Scarcello}.} \bibinfo{year}{2002}\natexlab{a}.
\newblock \showarticletitle{Hypertree Decompositions and Tractable Queries}.
\newblock \bibinfo{journal}{\emph{J. Comput. Syst. Sci.}} \bibinfo{volume}{64}, \bibinfo{number}{3} (\bibinfo{year}{2002}), \bibinfo{pages}{579--627}.
\newblock
\urldef\tempurl%
\url{https://doi.org/10.1006/jcss.2001.1809}
\showDOI{\tempurl}


\bibitem[\protect\citeauthoryear{Gottlob, Leone, and Scarcello}{Gottlob et~al\mbox{.}}{2002b}]%
        {DBLP:conf/pods/GottlobLS99}
\bibfield{author}{\bibinfo{person}{Georg Gottlob}, \bibinfo{person}{Nicola Leone}, {and} \bibinfo{person}{Francesco Scarcello}.} \bibinfo{year}{2002}\natexlab{b}.
\newblock \showarticletitle{Hypertree Decompositions and Tractable Queries}.
\newblock \bibinfo{journal}{\emph{J. Comput. Syst. Sci.}} \bibinfo{volume}{64}, \bibinfo{number}{3} (\bibinfo{year}{2002}), \bibinfo{pages}{579--627}.
\newblock
\urldef\tempurl%
\url{https://doi.org/10.1006/JCSS.2001.1809}
\showDOI{\tempurl}


\bibitem[\protect\citeauthoryear{Graham}{Graham}{1979}]%
        {report/toronto/Gra79}
\bibfield{author}{\bibinfo{person}{Marc~H. Graham}.} \bibinfo{year}{1979}\natexlab{}.
\newblock \bibinfo{booktitle}{\emph{{On The Universal Relation}}}.
\newblock \bibinfo{type}{{T}echnical {R}eport}. \bibinfo{institution}{University of Toronto}.
\newblock


\bibitem[\protect\citeauthoryear{GraphDB}{GraphDB}{2024}]%
        {graphdb}
\bibfield{author}{\bibinfo{person}{GraphDB}.} \bibinfo{year}{(accessed July 2024)}\natexlab{}.
\newblock \bibinfo{howpublished}{https://graphdb.ontotext.com/}.
\newblock


\bibitem[\protect\citeauthoryear{Green, Karvounarakis, and Tannen}{Green et~al\mbox{.}}{2007}]%
        {DBLP:conf/pods/GreenKT07}
\bibfield{author}{\bibinfo{person}{Todd~J. Green}, \bibinfo{person}{Gregory Karvounarakis}, {and} \bibinfo{person}{Val Tannen}.} \bibinfo{year}{2007}\natexlab{}.
\newblock \showarticletitle{Provenance semirings}. In \bibinfo{booktitle}{\emph{Proceedings {PODS}}}. \bibinfo{publisher}{{ACM}}, \bibinfo{pages}{31--40}.
\newblock
\urldef\tempurl%
\url{https://doi.org/10.1145/1265530.1265535}
\showDOI{\tempurl}


\bibitem[\protect\citeauthoryear{Grohe and Marx}{Grohe and Marx}{2014}]%
        {2014grohemarx}
\bibfield{author}{\bibinfo{person}{Martin Grohe} {and} \bibinfo{person}{D{\'{a}}niel Marx}.} \bibinfo{year}{2014}\natexlab{}.
\newblock \showarticletitle{Constraint Solving via Fractional Edge Covers}.
\newblock \bibinfo{journal}{\emph{{ACM} Trans. Algorithms}} \bibinfo{volume}{11}, \bibinfo{number}{1} (\bibinfo{year}{2014}), \bibinfo{pages}{4:1--4:20}.
\newblock


\bibitem[\protect\citeauthoryear{Han, Wu, Wu, Zhu, Yang, Tan, Zeng, Cong, Qin, Pfadler, Qian, Zhou, Li, and Cui}{Han et~al\mbox{.}}{2021}]%
        {DBLP:journals/pvldb/HanWWZYTZCQPQZL21}
\bibfield{author}{\bibinfo{person}{Yuxing Han}, \bibinfo{person}{Ziniu Wu}, \bibinfo{person}{Peizhi Wu}, \bibinfo{person}{Rong Zhu}, \bibinfo{person}{Jingyi Yang}, \bibinfo{person}{Liang~Wei Tan}, \bibinfo{person}{Kai Zeng}, \bibinfo{person}{Gao Cong}, \bibinfo{person}{Yanzhao Qin}, \bibinfo{person}{Andreas Pfadler}, \bibinfo{person}{Zhengping Qian}, \bibinfo{person}{Jingren Zhou}, \bibinfo{person}{Jiangneng Li}, {and} \bibinfo{person}{Bin Cui}.} \bibinfo{year}{2021}\natexlab{}.
\newblock \showarticletitle{Cardinality Estimation in {DBMS:} {A} Comprehensive Benchmark Evaluation}.
\newblock \bibinfo{journal}{\emph{Proc. {VLDB} Endow.}} \bibinfo{volume}{15}, \bibinfo{number}{4} (\bibinfo{year}{2021}), \bibinfo{pages}{752--765}.
\newblock
\urldef\tempurl%
\url{https://doi.org/10.14778/3503585.3503586}
\showDOI{\tempurl}


\bibitem[\protect\citeauthoryear{Hu and Wang}{Hu and Wang}{2023}]%
        {DBLP:journals/pacmmod/0005023}
\bibfield{author}{\bibinfo{person}{Xiao Hu} {and} \bibinfo{person}{Qichen Wang}.} \bibinfo{year}{2023}\natexlab{}.
\newblock \showarticletitle{Computing the Difference of Conjunctive Queries Efficiently}.
\newblock \bibinfo{journal}{\emph{Proc. {ACM} Manag. Data}} \bibinfo{volume}{1}, \bibinfo{number}{2} (\bibinfo{year}{2023}), \bibinfo{pages}{153:1--153:26}.
\newblock
\urldef\tempurl%
\url{https://doi.org/10.1145/3589298}
\showDOI{\tempurl}


\bibitem[\protect\citeauthoryear{Hu and Miranker}{Hu and Miranker}{2024}]%
        {DBLP:journals/corr/TreeTracker}
\bibfield{author}{\bibinfo{person}{Zeyuan Hu} {and} \bibinfo{person}{Daniel~P. Miranker}.} \bibinfo{year}{2024}\natexlab{}.
\newblock \showarticletitle{TreeTracker Join: Turning the Tide When a Tuple Fails to Join}.
\newblock \bibinfo{journal}{\emph{CoRR}}  \bibinfo{volume}{abs/2403.01631} (\bibinfo{year}{2024}).
\newblock
\showeprint[arXiv]{2403.01631}
\urldef\tempurl%
\url{http://arxiv.org/abs/2403.01631}
\showURL{%
\tempurl}


\bibitem[\protect\citeauthoryear{Idris, Ugarte, and Vansummeren}{Idris et~al\mbox{.}}{2017}]%
        {DBLP:conf/sigmod/IdrisUV17}
\bibfield{author}{\bibinfo{person}{Muhammad Idris}, \bibinfo{person}{Mart{\'{\i}}n Ugarte}, {and} \bibinfo{person}{Stijn Vansummeren}.} \bibinfo{year}{2017}\natexlab{}.
\newblock \showarticletitle{The Dynamic Yannakakis Algorithm: Compact and Efficient Query Processing Under Updates}. In \bibinfo{booktitle}{\emph{Proceedings {SIGMOD}}}. \bibinfo{publisher}{{ACM}}, \bibinfo{pages}{1259--1274}.
\newblock
\urldef\tempurl%
\url{https://doi.org/10.1145/3035918.3064027}
\showDOI{\tempurl}


\bibitem[\protect\citeauthoryear{Idris, Ugarte, Vansummeren, Voigt, and Lehner}{Idris et~al\mbox{.}}{2020}]%
        {DBLP:journals/vldb/IdrisUVVL20}
\bibfield{author}{\bibinfo{person}{Muhammad Idris}, \bibinfo{person}{Mart{\'{\i}}n Ugarte}, \bibinfo{person}{Stijn Vansummeren}, \bibinfo{person}{Hannes Voigt}, {and} \bibinfo{person}{Wolfgang Lehner}.} \bibinfo{year}{2020}\natexlab{}.
\newblock \showarticletitle{General dynamic Yannakakis: conjunctive queries with theta joins under updates}.
\newblock \bibinfo{journal}{\emph{{VLDB} J.}} \bibinfo{volume}{29}, \bibinfo{number}{2-3} (\bibinfo{year}{2020}), \bibinfo{pages}{619--653}.
\newblock
\urldef\tempurl%
\url{https://doi.org/10.1007/s00778-019-00590-9}
\showDOI{\tempurl}


\bibitem[\protect\citeauthoryear{Ji, Zhao, Zhai, and Wu}{Ji et~al\mbox{.}}{2020}]%
        {DBLP:journals/sp/JiZZW20}
\bibfield{author}{\bibinfo{person}{Xuechun Ji}, \bibinfo{person}{Mao{-}Xian Zhao}, \bibinfo{person}{Mingyu Zhai}, {and} \bibinfo{person}{Qingxi Wu}.} \bibinfo{year}{2020}\natexlab{}.
\newblock \showarticletitle{Query Execution Optimization in Spark {SQL}}.
\newblock \bibinfo{journal}{\emph{Sci. Program.}}  \bibinfo{volume}{2020} (\bibinfo{year}{2020}), \bibinfo{pages}{6364752:1--6364752:12}.
\newblock
\urldef\tempurl%
\url{https://doi.org/10.1155/2020/6364752}
\showDOI{\tempurl}


\bibitem[\protect\citeauthoryear{Jin, Bronstein, Ceylan, and Lanzinger}{Jin et~al\mbox{.}}{2024}]%
        {DBLP:journals/corr/abs-2402-08595}
\bibfield{author}{\bibinfo{person}{Emily Jin}, \bibinfo{person}{Michael Bronstein}, \bibinfo{person}{{\.I}smail~{\.I}lkan Ceylan}, {and} \bibinfo{person}{Matthias Lanzinger}.} \bibinfo{year}{2024}\natexlab{}.
\newblock \showarticletitle{Homomorphism Counts for Graph Neural Networks: All About That Basis}.
\newblock \bibinfo{journal}{\emph{CoRR}}  \bibinfo{volume}{abs/2402.08595} (\bibinfo{year}{2024}).
\newblock
\urldef\tempurl%
\url{https://doi.org/10.48550/ARXIV.2402.08595}
\showDOI{\tempurl}
\showeprint[arXiv]{2402.08595}


\bibitem[\protect\citeauthoryear{Jin, Feng, Chen, Liu, and Salihoglu}{Jin et~al\mbox{.}}{2023}]%
        {kuzu}
\bibfield{author}{\bibinfo{person}{Guodong Jin}, \bibinfo{person}{Xiyang Feng}, \bibinfo{person}{Ziyi Chen}, \bibinfo{person}{Chang Liu}, {and} \bibinfo{person}{Semih Salihoglu}.} \bibinfo{year}{2023}\natexlab{}.
\newblock \showarticletitle{K{\`{U}}ZU Graph Database Management System}. In \bibinfo{booktitle}{\emph{13th Conference on Innovative Data Systems Research, {CIDR} 2023, Amsterdam, The Netherlands, January 8-11, 2023}}. \bibinfo{publisher}{www.cidrdb.org}.
\newblock
\urldef\tempurl%
\url{https://www.cidrdb.org/cidr2023/papers/p48-jin.pdf}
\showURL{%
\tempurl}


\bibitem[\protect\citeauthoryear{Joglekar, Puttagunta, and R{\'{e}}}{Joglekar et~al\mbox{.}}{2016}]%
        {DBLP:conf/pods/JoglekarPR16}
\bibfield{author}{\bibinfo{person}{Manas~R. Joglekar}, \bibinfo{person}{Rohan Puttagunta}, {and} \bibinfo{person}{Christopher R{\'{e}}}.} \bibinfo{year}{2016}\natexlab{}.
\newblock \showarticletitle{{AJAR:} Aggregations and Joins over Annotated Relations}. In \bibinfo{booktitle}{\emph{Proceedings {PODS}}}. \bibinfo{publisher}{{ACM}}, \bibinfo{pages}{91--106}.
\newblock
\urldef\tempurl%
\url{https://doi.org/10.1145/2902251.2902293}
\showDOI{\tempurl}


\bibitem[\protect\citeauthoryear{Khamis, Ngo, and Rudra}{Khamis et~al\mbox{.}}{2016}]%
        {DBLP:conf/pods/KhamisNR16}
\bibfield{author}{\bibinfo{person}{Mahmoud~Abo Khamis}, \bibinfo{person}{Hung~Q. Ngo}, {and} \bibinfo{person}{Atri Rudra}.} \bibinfo{year}{2016}\natexlab{}.
\newblock \showarticletitle{{FAQ:} Questions Asked Frequently}. In \bibinfo{booktitle}{\emph{Proceedings {PODS}}}. \bibinfo{publisher}{{ACM}}, \bibinfo{pages}{13--28}.
\newblock
\urldef\tempurl%
\url{https://doi.org/10.1145/2902251.2902280}
\showDOI{\tempurl}


\bibitem[\protect\citeauthoryear{Khayyat, Lucia, Singh, Ouzzani, Papotti, Quian{\'{e}}{-}Ruiz, Tang, and Kalnis}{Khayyat et~al\mbox{.}}{2015}]%
        {DBLP:journals/pvldb/KhayyatLSOPQ0K15}
\bibfield{author}{\bibinfo{person}{Zuhair Khayyat}, \bibinfo{person}{William Lucia}, \bibinfo{person}{Meghna Singh}, \bibinfo{person}{Mourad Ouzzani}, \bibinfo{person}{Paolo Papotti}, \bibinfo{person}{Jorge{-}Arnulfo Quian{\'{e}}{-}Ruiz}, \bibinfo{person}{Nan Tang}, {and} \bibinfo{person}{Panos Kalnis}.} \bibinfo{year}{2015}\natexlab{}.
\newblock \showarticletitle{Lightning Fast and Space Efficient Inequality Joins}.
\newblock \bibinfo{journal}{\emph{Proc. {VLDB} Endow.}} \bibinfo{volume}{8}, \bibinfo{number}{13} (\bibinfo{year}{2015}), \bibinfo{pages}{2074--2085}.
\newblock
\urldef\tempurl%
\url{https://doi.org/10.14778/2831360.2831362}
\showDOI{\tempurl}


\bibitem[\protect\citeauthoryear{Koch, Ahmad, Kennedy, Nikolic, N{\"{o}}tzli, Lupei, and Shaikhha}{Koch et~al\mbox{.}}{2014}]%
        {DBLP:journals/vldb/KochAKNNLS14}
\bibfield{author}{\bibinfo{person}{Christoph Koch}, \bibinfo{person}{Yanif Ahmad}, \bibinfo{person}{Oliver Kennedy}, \bibinfo{person}{Milos Nikolic}, \bibinfo{person}{Andres N{\"{o}}tzli}, \bibinfo{person}{Daniel Lupei}, {and} \bibinfo{person}{Amir Shaikhha}.} \bibinfo{year}{2014}\natexlab{}.
\newblock \showarticletitle{DBToaster: higher-order delta processing for dynamic, frequently fresh views}.
\newblock \bibinfo{journal}{\emph{{VLDB} J.}} \bibinfo{volume}{23}, \bibinfo{number}{2} (\bibinfo{year}{2014}), \bibinfo{pages}{253--278}.
\newblock
\urldef\tempurl%
\url{https://doi.org/10.1007/S00778-013-0348-4}
\showDOI{\tempurl}


\bibitem[\protect\citeauthoryear{Lanzinger, Pichler, and Selzer}{Lanzinger et~al\mbox{.}}{2024}]%
        {DBLP:journals/corr/abs-2406-17076}
\bibfield{author}{\bibinfo{person}{Matthias Lanzinger}, \bibinfo{person}{Reinhard Pichler}, {and} \bibinfo{person}{Alexander Selzer}.} \bibinfo{year}{2024}\natexlab{}.
\newblock \showarticletitle{Avoiding Materialisation for Guarded Aggregate Queries}.
\newblock \bibinfo{journal}{\emph{CoRR}}  \bibinfo{volume}{abs/2406.17076} (\bibinfo{year}{2024}).
\newblock
\urldef\tempurl%
\url{https://doi.org/10.48550/ARXIV.2406.17076}
\showDOI{\tempurl}
\showeprint[arXiv]{2406.17076}


\bibitem[\protect\citeauthoryear{Leis, Gubichev, Mirchev, Boncz, Kemper, and Neumann}{Leis et~al\mbox{.}}{2015}]%
        {DBLP:journals/pvldb/LeisGMBK015}
\bibfield{author}{\bibinfo{person}{Viktor Leis}, \bibinfo{person}{Andrey Gubichev}, \bibinfo{person}{Atanas Mirchev}, \bibinfo{person}{Peter~A. Boncz}, \bibinfo{person}{Alfons Kemper}, {and} \bibinfo{person}{Thomas Neumann}.} \bibinfo{year}{2015}\natexlab{}.
\newblock \showarticletitle{How Good Are Query Optimizers, Really?}
\newblock \bibinfo{journal}{\emph{Proc. {VLDB} Endow.}} \bibinfo{volume}{9}, \bibinfo{number}{3} (\bibinfo{year}{2015}), \bibinfo{pages}{204--215}.
\newblock
\urldef\tempurl%
\url{https://doi.org/10.14778/2850583.2850594}
\showDOI{\tempurl}


\bibitem[\protect\citeauthoryear{Leskovec and Krevl}{Leskovec and Krevl}{2014}]%
        {snapnets}
\bibfield{author}{\bibinfo{person}{Jure Leskovec} {and} \bibinfo{person}{Andrej Krevl}.} \bibinfo{year}{2014}\natexlab{}.
\newblock \bibinfo{title}{{SNAP Datasets}: {Stanford} Large Network Dataset Collection}.
\newblock \bibinfo{howpublished}{\url{http://snap.stanford.edu/data}}.
\newblock


\bibitem[\protect\citeauthoryear{Mancini, Karthik, Chandra, Mageirakos, and Ailamaki}{Mancini et~al\mbox{.}}{2022}]%
        {DBLP:conf/sigmod/ManciniKCMA22}
\bibfield{author}{\bibinfo{person}{Riccardo Mancini}, \bibinfo{person}{Srinivas Karthik}, \bibinfo{person}{Bikash Chandra}, \bibinfo{person}{Vasilis Mageirakos}, {and} \bibinfo{person}{Anastasia Ailamaki}.} \bibinfo{year}{2022}\natexlab{}.
\newblock \showarticletitle{Efficient {Massively} {Parallel} {Join} {Optimization} for {Large} {Queries}}. In \bibinfo{booktitle}{\emph{Proceedings {SIGMOD}}}. \bibinfo{publisher}{{ACM}}, \bibinfo{pages}{122--135}.
\newblock
\urldef\tempurl%
\url{https://doi.org/10.1145/3514221.3517871}
\showDOI{\tempurl}


\bibitem[\protect\citeauthoryear{Mhedhbi, Lissandrini, Kuiper, Waudby, and Sz{\'{a}}rnyas}{Mhedhbi et~al\mbox{.}}{2021}]%
        {DBLP:conf/sigmod/MhedhbiLKWS21}
\bibfield{author}{\bibinfo{person}{Amine Mhedhbi}, \bibinfo{person}{Matteo Lissandrini}, \bibinfo{person}{Laurens Kuiper}, \bibinfo{person}{Jack Waudby}, {and} \bibinfo{person}{G{\'{a}}bor Sz{\'{a}}rnyas}.} \bibinfo{year}{2021}\natexlab{}.
\newblock \showarticletitle{{LSQB:} a large-scale subgraph query benchmark}. In \bibinfo{booktitle}{\emph{Proceedings {GRADES}}}. \bibinfo{publisher}{{ACM}}, \bibinfo{pages}{8:1--8:11}.
\newblock
\urldef\tempurl%
\url{https://doi.org/10.1145/3461837.3464516}
\showDOI{\tempurl}


\bibitem[\protect\citeauthoryear{Misegiannis, Kantere, and d'Orazio}{Misegiannis et~al\mbox{.}}{2022}]%
        {DBLP:conf/ideas/MisegiannisKd22}
\bibfield{author}{\bibinfo{person}{Michail~Georgoulakis Misegiannis}, \bibinfo{person}{Vasiliki Kantere}, {and} \bibinfo{person}{Laurent d'Orazio}.} \bibinfo{year}{2022}\natexlab{}.
\newblock \showarticletitle{Multi-objective query optimization in Spark {SQL}}. In \bibinfo{booktitle}{\emph{Proceedings {IDEAS}}}. \bibinfo{publisher}{{ACM}}, \bibinfo{pages}{70--74}.
\newblock
\urldef\tempurl%
\url{https://doi.org/10.1145/3548785.3548800}
\showDOI{\tempurl}


\bibitem[\protect\citeauthoryear{Neo4J}{Neo4J}{2024}]%
        {neoj}
\bibfield{author}{\bibinfo{person}{Neo4J}.} \bibinfo{year}{(accessed July 2024)}\natexlab{}.
\newblock \bibinfo{howpublished}{https://neo4j.com/}.
\newblock


\bibitem[\protect\citeauthoryear{Nguyen and Maehara}{Nguyen and Maehara}{2020}]%
        {DBLP:conf/icml/NguyenM20}
\bibfield{author}{\bibinfo{person}{Hoang Nguyen} {and} \bibinfo{person}{Takanori Maehara}.} \bibinfo{year}{2020}\natexlab{}.
\newblock \showarticletitle{Graph Homomorphism Convolution}. In \bibinfo{booktitle}{\emph{Proceedings {ICML}}} \emph{(\bibinfo{series}{Proceedings of Machine Learning Research})}, Vol.~\bibinfo{volume}{119}. \bibinfo{publisher}{{PMLR}}, \bibinfo{pages}{7306--7316}.
\newblock
\urldef\tempurl%
\url{http://proceedings.mlr.press/v119/nguyen20c.html}
\showURL{%
\tempurl}


\bibitem[\protect\citeauthoryear{Nikolic and Olteanu}{Nikolic and Olteanu}{2018}]%
        {DBLP:conf/sigmod/NikolicO18}
\bibfield{author}{\bibinfo{person}{Milos Nikolic} {and} \bibinfo{person}{Dan Olteanu}.} \bibinfo{year}{2018}\natexlab{}.
\newblock \showarticletitle{Incremental View Maintenance with Triple Lock Factorization Benefits}. In \bibinfo{booktitle}{\emph{Proceedings of the 2018 International Conference on Management of Data, {SIGMOD} Conference 2018, Houston, TX, USA, June 10-15, 2018}}, \bibfield{editor}{\bibinfo{person}{Gautam Das}, \bibinfo{person}{Christopher~M. Jermaine}, {and} \bibinfo{person}{Philip~A. Bernstein}} (Eds.). \bibinfo{publisher}{{ACM}}, \bibinfo{pages}{365--380}.
\newblock
\urldef\tempurl%
\url{https://doi.org/10.1145/3183713.3183758}
\showDOI{\tempurl}


\bibitem[\protect\citeauthoryear{Olteanu and Schleich}{Olteanu and Schleich}{2016}]%
        {DBLP:journals/sigmod/OlteanuS16}
\bibfield{author}{\bibinfo{person}{Dan Olteanu} {and} \bibinfo{person}{Maximilian Schleich}.} \bibinfo{year}{2016}\natexlab{}.
\newblock \showarticletitle{Factorized Databases}.
\newblock \bibinfo{journal}{\emph{{SIGMOD} Rec.}} \bibinfo{volume}{45}, \bibinfo{number}{2} (\bibinfo{year}{2016}), \bibinfo{pages}{5--16}.
\newblock
\urldef\tempurl%
\url{https://doi.org/10.1145/3003665.3003667}
\showDOI{\tempurl}


\bibitem[\protect\citeauthoryear{Olteanu and Z{\'{a}}vodn{\'{y}}}{Olteanu and Z{\'{a}}vodn{\'{y}}}{2015}]%
        {DBLP:journals/tods/OlteanuZ15}
\bibfield{author}{\bibinfo{person}{Dan Olteanu} {and} \bibinfo{person}{Jakub Z{\'{a}}vodn{\'{y}}}.} \bibinfo{year}{2015}\natexlab{}.
\newblock \showarticletitle{Size Bounds for Factorised Representations of Query Results}.
\newblock \bibinfo{journal}{\emph{{ACM} Trans. Database Syst.}} \bibinfo{volume}{40}, \bibinfo{number}{1} (\bibinfo{year}{2015}), \bibinfo{pages}{2:1--2:44}.
\newblock
\urldef\tempurl%
\url{https://doi.org/10.1145/2656335}
\showDOI{\tempurl}


\bibitem[\protect\citeauthoryear{Perelman and R{\'{e}}}{Perelman and R{\'{e}}}{2015}]%
        {DBLP:conf/sigmod/PerelmanR15}
\bibfield{author}{\bibinfo{person}{Adam Perelman} {and} \bibinfo{person}{Christopher R{\'{e}}}.} \bibinfo{year}{2015}\natexlab{}.
\newblock \showarticletitle{DunceCap: Compiling Worst-Case Optimal Query Plans}. In \bibinfo{booktitle}{\emph{Proceedings {SIGMOD}}}. \bibinfo{publisher}{{ACM}}, \bibinfo{pages}{2075--2076}.
\newblock
\urldef\tempurl%
\url{https://doi.org/10.1145/2723372.2764945}
\showDOI{\tempurl}


\bibitem[\protect\citeauthoryear{Pichler and Skritek}{Pichler and Skritek}{2013}]%
        {DBLP:journals/jcss/PichlerS13}
\bibfield{author}{\bibinfo{person}{Reinhard Pichler} {and} \bibinfo{person}{Sebastian Skritek}.} \bibinfo{year}{2013}\natexlab{}.
\newblock \showarticletitle{Tractable counting of the answers to conjunctive queries}.
\newblock \bibinfo{journal}{\emph{J. Comput. Syst. Sci.}} \bibinfo{volume}{79}, \bibinfo{number}{6} (\bibinfo{year}{2013}), \bibinfo{pages}{984--1001}.
\newblock
\urldef\tempurl%
\url{https://doi.org/10.1016/j.jcss.2013.01.012}
\showDOI{\tempurl}


\bibitem[\protect\citeauthoryear{Schleich, Olteanu, Khamis, Ngo, and Nguyen}{Schleich et~al\mbox{.}}{2019}]%
        {DBLP:conf/sigmod/SchleichOK0N19}
\bibfield{author}{\bibinfo{person}{Maximilian Schleich}, \bibinfo{person}{Dan Olteanu}, \bibinfo{person}{Mahmoud~Abo Khamis}, \bibinfo{person}{Hung~Q. Ngo}, {and} \bibinfo{person}{XuanLong Nguyen}.} \bibinfo{year}{2019}\natexlab{}.
\newblock \showarticletitle{A Layered Aggregate Engine for Analytics Workloads}. In \bibinfo{booktitle}{\emph{Proceedings of the 2019 International Conference on Management of Data, {SIGMOD} Conference 2019, Amsterdam, The Netherlands, June 30 - July 5, 2019}}, \bibfield{editor}{\bibinfo{person}{Peter~A. Boncz}, \bibinfo{person}{Stefan Manegold}, \bibinfo{person}{Anastasia Ailamaki}, \bibinfo{person}{Amol Deshpande}, {and} \bibinfo{person}{Tim Kraska}} (Eds.). \bibinfo{publisher}{{ACM}}, \bibinfo{pages}{1642--1659}.
\newblock
\urldef\tempurl%
\url{https://doi.org/10.1145/3299869.3324961}
\showDOI{\tempurl}


\bibitem[\protect\citeauthoryear{Seshadri, Pirahesh, and Leung}{Seshadri et~al\mbox{.}}{1996}]%
        {DBLP:conf/icde/SeshadriPL96}
\bibfield{author}{\bibinfo{person}{Praveen Seshadri}, \bibinfo{person}{Hamid Pirahesh}, {and} \bibinfo{person}{T.~Y.~Cliff Leung}.} \bibinfo{year}{1996}\natexlab{}.
\newblock \showarticletitle{Complex Query Decorrelation}. In \bibinfo{booktitle}{\emph{Proceedings {ICDE}}}. \bibinfo{publisher}{{IEEE} Computer Society}, \bibinfo{pages}{450--458}.
\newblock
\urldef\tempurl%
\url{https://doi.org/10.1109/ICDE.1996.492194}
\showDOI{\tempurl}


\bibitem[\protect\citeauthoryear{Shaikhha, Huot, Smith, and Olteanu}{Shaikhha et~al\mbox{.}}{2022}]%
        {DBLP:journals/pacmpl/ShaikhhaHSO22}
\bibfield{author}{\bibinfo{person}{Amir Shaikhha}, \bibinfo{person}{Mathieu Huot}, \bibinfo{person}{Jaclyn Smith}, {and} \bibinfo{person}{Dan Olteanu}.} \bibinfo{year}{2022}\natexlab{}.
\newblock \showarticletitle{Functional collection programming with semi-ring dictionaries}.
\newblock \bibinfo{journal}{\emph{Proc. {ACM} Program. Lang.}} \bibinfo{volume}{6}, \bibinfo{number}{{OOPSLA1}} (\bibinfo{year}{2022}), \bibinfo{pages}{1--33}.
\newblock
\urldef\tempurl%
\url{https://doi.org/10.1145/3527333}
\showDOI{\tempurl}


\bibitem[\protect\citeauthoryear{Shen, Ren, Lu, Jiang, Xu, Peng, Li, Zhang, and Cui}{Shen et~al\mbox{.}}{2023}]%
        {DBLP:conf/kdd/ShenRLJXP0Z023}
\bibfield{author}{\bibinfo{person}{Yu Shen}, \bibinfo{person}{Xinyuyang Ren}, \bibinfo{person}{Yupeng Lu}, \bibinfo{person}{Huaijun Jiang}, \bibinfo{person}{Huanyong Xu}, \bibinfo{person}{Di Peng}, \bibinfo{person}{Yang Li}, \bibinfo{person}{Wentao Zhang}, {and} \bibinfo{person}{Bin Cui}.} \bibinfo{year}{2023}\natexlab{}.
\newblock \showarticletitle{Rover: An Online Spark {SQL} Tuning Service via Generalized Transfer Learning}. In \bibinfo{booktitle}{\emph{Proceedings {KDD}}}. \bibinfo{publisher}{{ACM}}, \bibinfo{pages}{4800--4812}.
\newblock
\urldef\tempurl%
\url{https://doi.org/10.1145/3580305.3599953}
\showDOI{\tempurl}


\bibitem[\protect\citeauthoryear{TPC-DS}{TPC-DS}{[n.d.]}]%
        {tpcds}
\bibfield{author}{\bibinfo{person}{TPC-DS}.} \bibinfo{year}{[n.d.]}\natexlab{}.
\newblock \bibinfo{title}{TPC-DS Benchmark}.
\newblock \bibinfo{howpublished}{\url{https://www.tpc.org/tpcds/}}.
\newblock


\bibitem[\protect\citeauthoryear{TPC-H}{TPC-H}{[n.d.]}]%
        {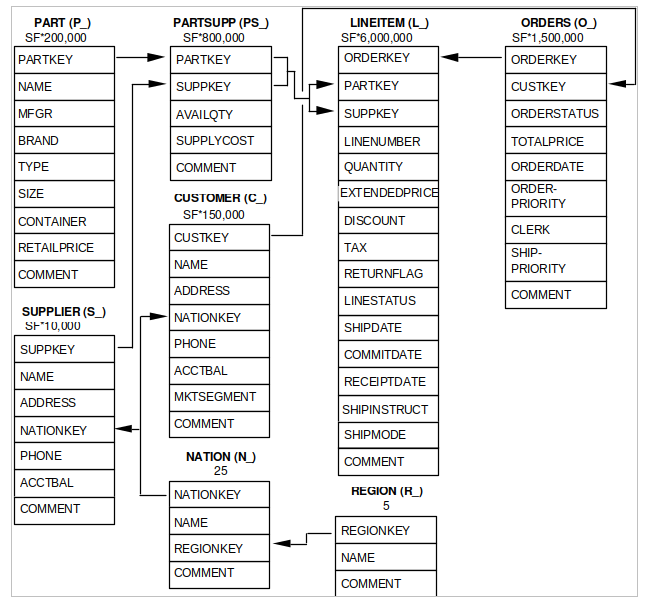}
\bibfield{author}{\bibinfo{person}{TPC-H}.} \bibinfo{year}{[n.d.]}\natexlab{}.
\newblock \bibinfo{title}{TPC-H Benchmark}.
\newblock \bibinfo{howpublished}{\url{https://www.tpc.org/tpch/}}.
\newblock


\bibitem[\protect\citeauthoryear{Tu and R{\'{e}}}{Tu and R{\'{e}}}{2015}]%
        {DBLP:conf/sigmod/TuR15}
\bibfield{author}{\bibinfo{person}{Susan Tu} {and} \bibinfo{person}{Christopher R{\'{e}}}.} \bibinfo{year}{2015}\natexlab{}.
\newblock \showarticletitle{DunceCap: Query Plans Using Generalized Hypertree Decompositions}. In \bibinfo{booktitle}{\emph{Proceedings {SIGMOD}}}. \bibinfo{publisher}{{ACM}}, \bibinfo{pages}{2077--2078}.
\newblock
\urldef\tempurl%
\url{https://doi.org/10.1145/2723372.2764946}
\showDOI{\tempurl}


\bibitem[\protect\citeauthoryear{Wang, Hu, Dai, and Yi}{Wang et~al\mbox{.}}{2023}]%
        {DBLP:journals/corr/abs-2301-04003}
\bibfield{author}{\bibinfo{person}{Qichen Wang}, \bibinfo{person}{Xiao Hu}, \bibinfo{person}{Binyang Dai}, {and} \bibinfo{person}{Ke Yi}.} \bibinfo{year}{2023}\natexlab{}.
\newblock \showarticletitle{Change Propagation Without Joins}.
\newblock \bibinfo{journal}{\emph{CoRR}}  \bibinfo{volume}{abs/2301.04003} (\bibinfo{year}{2023}).
\newblock
\urldef\tempurl%
\url{https://doi.org/10.48550/arXiv.2301.04003}
\showDOI{\tempurl}
\showeprint[arXiv]{2301.04003}


\bibitem[\protect\citeauthoryear{Wang and Yi}{Wang and Yi}{2022}]%
        {DBLP:conf/sigmod/0001022}
\bibfield{author}{\bibinfo{person}{Qichen Wang} {and} \bibinfo{person}{Ke Yi}.} \bibinfo{year}{2022}\natexlab{}.
\newblock \showarticletitle{Conjunctive Queries with Comparisons}. In \bibinfo{booktitle}{\emph{{SIGMOD} '22: International Conference on Management of Data}}. \bibinfo{publisher}{{ACM}}, \bibinfo{pages}{108--121}.
\newblock
\urldef\tempurl%
\url{https://doi.org/10.1145/3514221.3517830}
\showDOI{\tempurl}


\bibitem[\protect\citeauthoryear{Wang and Yi}{Wang and Yi}{2021}]%
        {DBLP:conf/sigmod/Wang021}
\bibfield{author}{\bibinfo{person}{Yilei Wang} {and} \bibinfo{person}{Ke Yi}.} \bibinfo{year}{2021}\natexlab{}.
\newblock \showarticletitle{Secure Yannakakis: Join-Aggregate Queries over Private Data}. In \bibinfo{booktitle}{\emph{{SIGMOD} '21: International Conference on Management of Data}}. \bibinfo{publisher}{{ACM}}, \bibinfo{pages}{1969--1981}.
\newblock
\urldef\tempurl%
\url{https://doi.org/10.1145/3448016.3452808}
\showDOI{\tempurl}


\bibitem[\protect\citeauthoryear{Yannakakis}{Yannakakis}{1981}]%
        {DBLP:conf/vldb/Yannakakis81}
\bibfield{author}{\bibinfo{person}{Mihalis Yannakakis}.} \bibinfo{year}{1981}\natexlab{}.
\newblock \showarticletitle{Algorithms for Acyclic Database Schemes}. In \bibinfo{booktitle}{\emph{Proceedings {VLDB}}}. \bibinfo{publisher}{VLDB}, \bibinfo{pages}{82--94}.
\newblock


\bibitem[\protect\citeauthoryear{Yu and Özsoyoğlu}{Yu and Özsoyoğlu}{1979}]%
        {DBLP:conf/compsac/YuO79}
\bibfield{author}{\bibinfo{person}{C.~T. Yu} {and} \bibinfo{person}{M.~Z. Özsoyoğlu}.} \bibinfo{year}{1979}\natexlab{}.
\newblock \showarticletitle{An algorithm for tree-query membership of a distributed query}. In \bibinfo{booktitle}{\emph{The {IEEE} Computer Society's Third International Computer Software and Applications Conference, {COMPSAC} 1979}}. \bibinfo{pages}{306--312}.
\newblock


\bibitem[\protect\citeauthoryear{Zhai, Song, Qiu, Ji, and Wu}{Zhai et~al\mbox{.}}{2019}]%
        {DBLP:conf/ipccc/ZhaiSQJW19}
\bibfield{author}{\bibinfo{person}{Mingyu Zhai}, \bibinfo{person}{Aibo Song}, \bibinfo{person}{Jingyi Qiu}, \bibinfo{person}{Xuechun Ji}, {and} \bibinfo{person}{Qingxi Wu}.} \bibinfo{year}{2019}\natexlab{}.
\newblock \showarticletitle{Query optimization Approach with Shuffle Intermediate Cache Layer for Spark {SQL}}. In \bibinfo{booktitle}{\emph{Proceedings {IPCCC}}}. \bibinfo{publisher}{{IEEE}}, \bibinfo{pages}{1--6}.
\newblock
\urldef\tempurl%
\url{https://doi.org/10.1109/IPCCC47392.2019.8958719}
\showDOI{\tempurl}


\bibitem[\protect\citeauthoryear{Zhang, Yu, Zhang, Zhao, and Cheng}{Zhang et~al\mbox{.}}{2020}]%
        {DBLP:journals/pvldb/ZhangY0ZC20}
\bibfield{author}{\bibinfo{person}{Hao Zhang}, \bibinfo{person}{Jeffrey~Xu Yu}, \bibinfo{person}{Yikai Zhang}, \bibinfo{person}{Kangfei Zhao}, {and} \bibinfo{person}{Hong Cheng}.} \bibinfo{year}{2020}\natexlab{}.
\newblock \showarticletitle{Distributed Subgraph Counting: {A} General Approach}.
\newblock \bibinfo{journal}{\emph{Proc. {VLDB} Endow.}} \bibinfo{volume}{13}, \bibinfo{number}{11} (\bibinfo{year}{2020}), \bibinfo{pages}{2493--2507}.
\newblock
\urldef\tempurl%
\url{http://www.vldb.org/pvldb/vol13/p2493-zhang.pdf}
\showURL{%
\tempurl}


\end{thebibliography}

\clearpage

\ifArxiv

\appendix

\section{Further Details on the Logical Optimisation}
\label{app:logical}

In Example~\ref{ex:guarded-evaluation}, we have already shown our logical optimisation at work 
in case of the guarded aggregate query from Figure~\ref{fig:tpch-query}. 
In Section~\ref{app:ExamplePiecewise}, we provide additional illustration of our optimisations by 
looking at the evaluation of a concrete \pwg aggregate query. Moreover, 
in Section~\ref{app:FK/PK}, we present a further optimisation that may arise from the presence of primary keys 
or unique attributes. Finally, in Section~\ref{app:limits}, we discuss the consequences, when the 
restrictions imposed by piecewise-guardedness are relaxed. More precisely, 
we provide more details on why the two extensions mentioned at the end of Section~\ref{sect:piecewise} would 
destroy the linear space property (w.r.t.\ the size of the data) enjoyed by \pwg queries.

\subsection{\PWG Optimisation at Work}
\label{app:ExamplePiecewise}

We now illustrate the evaluation of \pwg aggregate queries by
extending Example~\ref{ex:guarded-evaluation}.

\begin{figure}[h]

\tikzstyle{every picture}+=[remember picture]
\tikzstyle{na} = [shape=rectangle,inner sep=0pt]

\newcommand{\ptStrike}[2]{    
    \tikz[baseline=(#1.base)]\node[na](#1){#2};
}

\newcommand{\strike}[2]{  
    \begin{tikzpicture}[overlay]
        \draw (#1.west) -- (#2.east);
    \end{tikzpicture}
}

\begin{tikzpicture}[scale=0.8, transform shape]
            \node [](supplier) at (3.0,0.0) {
            \begin{tabular}{|c|c|c|c|c|c|}
            \hline
             \rowcolor{relname} \multicolumn{6}{|c|}{\texttt{supplier}} \\
            \hline
                \rowcolor{relheader} $\mathbf N$ & $\mathbf S$ & $\cdots$ & $\mathbf c$ & $\mathbf \mathrm{Agg}_1$ & $\mathbf \mathrm{Agg}_2$ \\
                \hline
                $n_1$ & $s_1$ & $\cdots$ & $30$ & $10$ & $625$ \\
                 \hline
                 $n_1$ & $s_2$ & $\cdots$ & $20$ & $10$ & $375$ \\
                 \hline
                 \ptStrike{A} {$n_1$} & $s_4$ & $\cdots$ & $1$ & - & \ptStrike{B}{-}  \\
                 \hline
                 $n_2$ & $s_1$ & $\cdots$ & $36$ & $5$ & $750$ \\
                 \hline
                 $n_2$ & $s_2$ & $\cdots$ & $24$ & $5$ & $450$ \\
                 \hline
                 \ptStrike{C} {$n_4$} & $s_2$ & $\cdots$ & $1$  & - & \ptStrike{D}{-} \\
                 \hline
            \end{tabular}

             \strike{A}{B}
              \strike{C}{D}
        };

        \node [](nation) at (1.0,-3.9) {
            \begin{tabular}{|c|c|c|c|c|}
            \hline
             \rowcolor{relname} \multicolumn{5}{|c|}{\texttt{nation}} \\
            \hline
                \rowcolor{relheader} $\mathbf N$ & $\mathbf R$ & $\cdots$ & $\mathbf c$ & $\mathbf \mathrm{Agg}_1$ \\
                \hline
                $n_1$ & $r_1$ & $\cdots$ & $3$ & $10$ \\
                 \hline
                 $n_1$ & $r_2$ & $\cdots$ & $2$ & $20$ \\
                 \hline
                 \ptStrike{A} {$n_1$} & $r_4$ & $\cdots$ & $1$ & \ptStrike{B}{-}  \\
                 \hline
                 $n_2$ & $r_1$ & $\cdots$ & $3$ & $10$ \\
                 \hline
                 $n_2$ & $r_2$ & $\cdots$ & $2$ & $20$ \\
                 \hline
                 {$n_2$} & $r_3$ & $\cdots$  & $1$ & $5$ \\
                 \hline
            \end{tabular}

             \strike{A}{B}
        };

        \node [](partsupplier) at (5.0,-3.9) {
            \begin{tabular}{|c|c|c|c|c|}
            \hline
             \rowcolor{relname} \multicolumn{5}{|c|}{\texttt{partsupplier}} \\
            \hline
                \rowcolor{relheader} $\mathbf S$ & $\mathbf P$ & $\cdots$ & $\mathbf c$ & $\mathbf \mathrm{Agg}_2$ \\
                \hline
                $s_1$ & $p_1$ & $\cdots$ & $3$ & $45$ \\
                 \hline
                 $s_1$ & $p_2$ & $\cdots$ & $2$ & $50$ \\
                 \hline
                {$s_1$} & $p_3$ & $\cdots$ & {$1$} & $30$  \\
                 \hline
                 $s_2$ & $p_1$ & $\cdots$ & $3$ & $45$ \\
                 \hline
                 $s_2$ & $p_3$ & $\cdots$ & $1$ & $30$ \\
                 \hline
                 {$s_3$} & $p_1$ & $\cdots$  & $3$ & $45$ \\
                 \hline
            \end{tabular}
        };

                \node [](region) at (1.0,-7.8) {
            \begin{tabular}{|c|c|c|c|c|c|}
            \hline
             \rowcolor{relname} \multicolumn{5}{|c|}{\texttt{region}} \\
            \hline
                \rowcolor{relheader} $\mathbf R$ & $\mathbf X$ & $\cdots$ & $\mathbf c$ & $\mathbf \mathrm{Agg}_1$ \\
                \hline
                $r_1$ & $10$ & $\cdots$ & $1$ & $10$ \\
                 \hline
                 $r_1$ & $20$ & $\cdots$ & $1$ & $20$ \\
                 \hline
                 {$r_1$} & $15$ & $\cdots$ & {$1$} & $15$  \\
                 \hline
                 $r_2$ & $20$ & $\cdots$ & $1$ & $20$ \\
                 \hline
                 $r_2$ & $25$ & $\cdots$ & $1$ & $25$ \\
                 \hline
                 {$r_3$} & $5$ & $\cdots$  & $1$ & $5$ \\
                 \hline
            \end{tabular}

        };

        \node [](part) at (5.0,-7.8) {
             \begin{tabular}{|c|c|c|c|c|c|}
            \hline
             \rowcolor{relname} \multicolumn{5}{|c|}{\texttt{part}} \\
            \hline
                \rowcolor{relheader} $\mathbf P$ & $\mathbf Y$ & $\cdots$ & $\mathbf c$ & $\mathbf \mathrm{Agg}_2$ \\
                \hline
                $p_1$ & $20$ & $\cdots$ & $1$ & $20$ \\
                 \hline
                 $p_1$ & $15$ & $\cdots$ & $1$ & $15 $\\
                 \hline
                 {$p_1$} & $10$ & $\cdots$ & {$1$} & $10$  \\
                 \hline
                 $p_2$ & $30$ & $\cdots$ & $1$ & $30$ \\
                 \hline
                 $p_2$ & $20$ & $\cdots$ & $1$ & $20$ \\
                 \hline
                 {$p_3$} & $30$ & $\cdots$  & $1$ & $30$ \\
                 \hline
            \end{tabular}
        };

    \draw [very thick] (nation) -- (supplier);
    \draw [very thick] (partsupplier) -- (supplier);
    \draw [very thick] (nation) -- (region);
    \draw [very thick] (partsupplier) -- (part);

    \end{tikzpicture}

    \caption{Evaluation of the query from Example~\ref{ex:piecewise-evaluation}}
    \label{fig:eval-piecewise}
\end{figure}    

\begin{example}
\label{ex:piecewise-evaluation}
Consider again the query from Figure~\ref{fig:tpch-query}, whose join 
tree is shown in Figure~\ref{fig:tpch-query-jointree}. 
But now we assume that relation \texttt{Region} has an additional
integer attribute $X$ and  relation \texttt{Part}
has an additional integer attribute $Y$. 
Finally, suppose that we 
want to evaluate the query ``{\sf SELECT MIN(X), SUM(Y) \dots GROUP BY N}''.
The query is trivially \pwg, since we have single attributes in the \texttt{GROUP}~\texttt{BY} 
clause and in each aggregate expression.
On the other hand, it is not guarded, since the attributes $X, Y, N$ do not 
jointly occur in a common relation. 

In 
Figure~\ref{fig:eval-guarded}, we illustrate the evaluation of the query on a small sample database.
The tables (with attribute names of the join attributes abbreviated to single characters) 
are arranged in the form of the join tree. Note that we again choose the \texttt{Supplier} relation as root guard. 
Alternatively, we could have chosen the \texttt{Nation} relation,
since it also contains the grouping attribute $N$.  

As in Figure~\ref{fig:eval-guarded}, attributes not relevant to our discussion are captured by ``\dots''. 
The original contents of the tables is shown to the left of the ``\dots'' column. To the right of the ``\dots'' column, 
we now have 
the frequency attribute $c$ plus the aggregate attributes $\Agg_1$ and $\Agg_2$ for the evaluation of the 
\MIN(X) and \SUM(Y) aggregates as described  in Section~\ref{sect:physicalOpt}. The $\Agg_1$ attribute 
(for the \MIN(X) aggregate expression) is added 
to all relations along the path from the node with relation \texttt{Region} up to the root node, while the 
$\Agg_2$ attribute 
(for the  \SUM(Y) aggregate expression) is added 
to all relations along the path from the node with relation \texttt{Part} up to the root node.

The frequency attribute $c$ is propagated up the join tree exactly as in Example~\ref{ex:guarded-evaluation}.
For the aggregate attributes $\Agg1$ and $\Agg_2$, we proceed as follows:
At the leaf nodes (in this simple example, these are  the highest nodes up in the tree that contain the attributes $X$ and
$Y$, respectively), we initialise,  $\Agg_1$ in every tuple to the value of $X$ and $\Agg_2$ to the value of $Y$.
The value of $\Agg_1$ is propagated from the \texttt{Region} relation to its parent node by taking, for every tuple in \texttt{Nation}, the \MIN 
over all its join partners in \texttt{Region}. Analogously,
the value of $\Agg_2$ is propagated to the parent node by taking, for every tuple in \texttt{Partsupplier}, the \SUM 
over all its join partners in \texttt{Part}.

Now let us look at relation \texttt{Supplier} at the root node. The attribute $\Agg_1$ is propagated from  \texttt{Nation}
to \texttt{Supplier} by taking, for every tuple at the parent node, the \MIN over the $\Agg_1$ values of its join partners at the child node.
For instance, the \texttt{Supplier}-tuple with attribute values $N = n_1$ and $S = s_1$ has two join partners in the  \texttt{Nation} table; the \MIN-value 
of their $\Agg_1$-attributes is 10. The same holds for the tuple with $N = n_1$ and $S = s_2$. On the other hand, the tuples in \texttt{Supplier}
with $N = n_2$ have the \MIN-value 5 in the $\Agg_1$-attribute of their join partners.  As was described in Section $\ref{sect:piecewise}$, 
the other child node of the (node labelled by the) \texttt{Supplier} relation plays no role when propagating an aggregate attribute for 
a \MIN or \MAX aggregate expression. 

This is in sharp contrast to the $\Agg_2$ attribute, where we propagate, for 
every tuple in the root node, 
the \SUM over the $\Agg_2$ values 
of the join partners from the right child node. But then we also have to multiply this value by the sum
over the frequency-values of all join partners in the left child node. For instance, take the tuple with  attribute values $N = n_1$ and $S = s_1$ in \texttt{Supplier}.
On the one hand, the sum over the $\Agg_2$-attributes of its join partners in \texttt{Partsupplier} (i.e., the tuples with 
attribute value $S = s_1$) is 125. But then we have to multiply this value by the 
the sum over the $c$-attributes of its join partners in \texttt{Nation} (i.e., the tuples with 
attribute value $N = n_1$), which is 5. 
Hence, the $\Agg_2$-value of the \texttt{Supplier}-tuple is 125 * 5 = 625. Analogously, 
the tuples in the \texttt{Supplier} relation with attribute values ($N = n_1$ and $S = s_2$), ($N = n_2$ and $S = s_1$),
and ($N = n_2$ and $S = s_2$) are obtained as 75*5 = 375, 125*6 = 750, and 745*6 = 450, respectively. 

For the evaluation of $\texttt{MIN(X)}$ and  $\texttt{SUM(Y)}$ for each $N$-group, we then just need to aggregate the $\Agg_1$ and $\Agg_2$ attributes
for each group. That is, we have two groups $n_1$ and $n_2$. There respective values of  $(\texttt{MIN(X)}$, $\texttt{SUM(Y)})$
are $(10, 1000)$ and $(5,1200)$, respectively.
\hfill $\diamond$
\end{example}

\subsection{Leveraging FK/PK-Relationships}
\label{app:FK/PK}

The logical and physical optimisations presented in Sections~\ref{sect:OurSystem} and  \ref{sect:physicalOpt}
allow us to avoid the computation and materialisation of any joins for queries that are guarded or 
\pwg. Nevertheless, the semi-join-like propagation of frequencies and of other aggregate attributes remains
the main cost-factor of evaluating queries in these classes. 
Some computational overhead may still be avoidable when further information about the database constraints is known.
Joins are frequently performed along foreign-key/primary-key (FK/PK) relationships. Knowledge about these relationships 
may actually allow us to replace joins in the frequency-propagation optimisation by semi-joins when 
the joins go along an FK/PK relationship
such that the relation labelling the parent node in the join tree holds the FK and the relation at a child node holds the PK.
This is due to the fact that, in this case,
we know that every tuple of the relation at the parent node can have at most one join partner in the relation at the child node. 

In particular, suppose that in the relation at the child node, 
all tuples have frequency 1. This is guaranteed if the child 
is a leaf node in the join tree. Then the frequency propagation from the child node to the parent node comes down to a 
semi-join. That is, we have to check for each tuple in the relation at the parent node if it has a join partner in the relation
at the child node
(which essentially means that it has no NULL value in an FK-attribute). 
If so, the parent inherits the frequency 1 from the child node; otherwise, we may simply discard this tuple from 
the relation at the parent node. Of course, then the same consideration may be iterated also for the join with the parent of the parent, if 
this is again along an FK/PK-relationship. We illustrate this additional optimisation 
by revisiting Example~\ref{exp:median}.

\begin{example}
\label{exp:FKPK} 
Consider again the query from Figure~\ref{fig:tpch-query}.
An inspection of the join tree in Figure~\ref{fig:tpch-query-jointree} and of the TPC-H schema 
reveals that all joins are along FK/PK-relationships from the relation at the parent node to the relation at the 
child node. 
We can, therefore,  be sure that all frequency attributes
in our optimisation
can only take the value 1. Hence, all joins can be replaced by semi-joins. 
The logical query plan produced by Spark SQL when implementing
this additional optimisation is shown in Figure~\ref{fig:FKPK-optimisation}. 
Here, for each FK/PK-relationship, the referencing relation is shown as the left child of the $\ltimes$-node 
in the query plan and 
the referenced relation as the right child of the $\ltimes$-node. The semi-join is always
from the referenced relation into the referencing relation, i.e., from right to left.
\hfill$\diamond$
\end{example}

\begin{figure}
    \hspace{-2em}
    \scalebox{0.8}{
    \begin{forest}
    [$\gamma_{\texttt{MEDIAN}(\text{s\_acctbal})}$
     [$\pi_{\text{s\_acctbal}}$
      [$\ltimes_{\text{s\_suppkey = ps\_suppkey}}$
      [$\pi_{\text{s\_suppkey, s\_acctbal}}$
        [$\ltimes_{\text{s\_nationkey = n\_nationkey}}$
         [$\pi_{\text{s\_suppkey, s\_nationkey, s\_acctbal}}$
          [$\texttt{supplier}$]
         ]
         [$\pi_{\text{n\_nationkey}}$
          [$\ltimes_{\text{n\_regionkey = r\_regionkey}}$
             [$\pi_{\text{n\_nationkey, n\_regionkey}}$
              [$\texttt{nation}$]
             ]
             [$\pi_{\text{r\_regionkey}}$
              [$\sigma_\psi(\texttt{region})$]
             ]
          ]
         ]
        ]
       ]
       [$\pi_{\text{ps\_suppkey}}$
        [$\ltimes_{\text{ps\_partkey = p\_partkey}}$
         [$\pi_{\text{ps\_partkey, ps\_suppkey}}$
          [$\texttt{partsupp}$]]
         [$\pi_{\text{p\_partkey}}$
          [$\sigma_\phi(\texttt{part})$]]
         ]
        ]
      ]
     ]
    ]
    \end{forest}
    }
    \caption{Query plan generated by Spark SQL with FK/PK optimisation}
    \label{fig:FKPK-optimisation}
\end{figure}

The information on primary keys or, more generally, on 
unique attributes can be exploited for 
 a further optimisation: 
Recall from Example~\ref{exp:median} and Figure~\ref{fig:count-optimised-query}
that, in the optimised query plan, the multiplication of the frequency $c_x$ at the parent with the 
frequency $c_y$ at the child is followed by a projection (actually, a grouping) and a summation of the $c_{xy}$ values. Now suppose that the 
grouping attributes (i.e., the attributes to which we project) contain an attribute or a set of 
attributes with unique values. Then 
each value combination of the grouping attributes occurs at most once. Hence, the 
summation of the $c_{xy}$ values can be omitted in this case.

Finally, the information on primary keys or unique attributes
can be used for yet another optimisation: as was explained 
in Section~\ref{sect:OurSystem}, 
our optimisation of guarded aggregates
starts with 
adding frequency 1 
as an additional attribute to each tuple in the relation $R$ labelling a node 
$u$ in the join tree, i.e., the computation of the optimised Relational Algebra expression 
$\Count(u)$ starts with the initialisation $\Count_0(u) := R \times \{(1)\}$.
Before joining this relation $\Count_0(u)$ with the relations 
at the child nodes, we can group the relation at each child node 
over the attribute(s) relevant
further up in the query plan and sum up the frequency values for each group.
In cases where 
value combinations of the grouping attributes occur frequently in 
such a relation, 
this significantly reduces the size of one join operand and, hence, 
the cost of the join. But, of course, such an 
additional grouping operation would cause useless effort if the grouping attributes 
contain a PK or unique attribute since, in this case, each group would consist of a single tuple.  
In our optimisation, we therefore apply this grouping only if no attribute with 
unique value is involved in the grouping.

\subsection{Discussion of Further Extensions}
\label{app:limits}

As was already mentioned at the end of Section~\ref{sect:piecewise}, our logical optimisations for 
guarded and \pwg aggregate queries guarantee that only linearly big (data complexity) information has to be 
propagated up the join tree. We now briefly discuss, why two natural extensions of our \pwg fragment would 
destroy this favourable property:

First, one may want to extend \pwg queries in such a way that also aggregate expressions with 
aggregates different from 
\texttt{MIN}, 
\texttt{MAX}, \texttt{SUM}, \texttt{COUNT}, and
\texttt{AVG} are allowed to be guarded by a relation different from the root guard. For instance, 
suppose that we modify the query in Example~\ref{ex:piecewise-evaluation} to
``{\sf SELECT MEDIAN(X) \dots GROUP BY N}''.  
In this case, it is no longer clear which aggregated information to propagate from \texttt{Region} (= the highest
node containing attribute $X$) 
along the path to the root $r$ of the join tree 
in order to evaluate \texttt{MEDIAN}($X$)  for each $N$-group.
Instead, we would have to propagate, for every tuple in a relation  
along the path from $\texttt{Region}$ to the root $r$, all possible extensions to the attribute $X$ plus the frequency of such extensions. 
For instance, in order to evaluate \texttt{MEDIAN}($X$) for the group defined by the attribute value $N = n_1$ at the 
root node, we need the information that the tuples with attribute values $(N = n_1, S= s_1)$ and $(N = n_1, S= s_2)$ can both be extended to
$X= 10$, $X= 15$, and $X= 25$  with frequency 1 and to $X= 20$  with frequency 2. That is, in total, we 
have to consider 10 possible extensions of the tuples in this group to correctly compute the  value \texttt{MEDIAN}($X$) = 20
at the root $r$. Analogous considerations apply to all of the other statistical functions in the ANSI standard, 
such as \texttt{VARIANCE}, \texttt{STDDEV}, etc. 
For all of these aggregate functions, for every tuple $t$ along the path to the root, we have to propagate the concrete values (plus their frequency) 
of the attributes in the aggregate expression to which $t$ can be extended. In the worst case, we thus end up with 
{\em quadratically} big (data complexity) data structures along such a path in the join tree.

Similar considerations apply to the requirement that each aggregate expression 
$A_j(f_j(\bar B_j))$
must have a guard. For instance, 
 suppose that we modify the query in Example~\ref{ex:piecewise-evaluation} to
``{\sf SELECT \texttt{SUM}($X*Y$) \dots GROUP BY N}''.
Clearly, it is not enough to propagate aggregated values such as $\SUM(X)$ and $\SUM(Y)$ up the join tree.
Instead, we again need, for every tuple $t$ in every $N$-group at the root node,  
all possible extensions to the attribute $X$ plus the frequency of such extensions
and also all possible extensions to the attribute $Y$ plus the frequency of such extensions.
For instance, we need the information that the tuple with attribute values $(N = n_1, S= s_1)$ has actually 5 extensions 
on the left branch, namely  to
$X= 10$, $X= 15$, and $X= 25$ with frequency 1 and to  $X= 20$ with frequency 2.
Analogously, this tuple has 6 extensions 
on the right branch, namely  to 
$Y= 20$ and $Y = 30$ with frequency 2 and to  $Y= 15$  and $Y= 10$  with frequency 1.
In other words, this tuple has 30 possible extensions in total. Consequently, it
contributes 30 summands (multiplying each of the 5 possible extensions to $X$ with each of the 6 possible extensions to $Y$) 
to the value of \texttt{SUM}($X*Y$) for the group of $N = n_1$. 
Generally speaking, in case of an aggregate expression with attributes $\bar B$ 
covered by (relations labelling) different nodes, 
we essentially have to propagate for every tuple $t$ along the path to the nearest common ancestor 
all  values (plus their frequency) 
of the attributes $\bar B$ to which $t$ can be extended.
In the worst case, we end up with 
{\em quadratically} big (data complexity) or even bigger data structures along such paths in the join tree.

\section{Further Details on the Physical Optimisation}
\label{app:physical}

In Section~\ref{sect:physicalOpt}, we have already shown how to realise the AggJoin operator 
based on Spark's shuffle-hash-join. 
We now also describe the realisation 
of the AggJoin operator 
based on Spark's sort-merge-join. 
As with the AggHashJoin, we write $R$ and $S$ to denote 
pairs of relations whose nodes in the join tree are in parent-child relationship. 
Moreover, we again assume that the 
AggJoin operator is only called after all the initialisations of the 
additional attributes $t.c$ and 
$t.\Aggj$ have been carried out for every tuple $t \in R$
as described in Section~\ref{sect:physicalOpt}.
Clearly, the sort phase
is left unchanged. 
Only the merge-phase is affected, which we briefly discuss next:

\begin{algorithm}[t]
\SetKwProg{Fn}{Function}{}{}
\SetKwRepeat{Do}{do}{while}
\KwIn{Two lists $R,S$ of tuples with the same values of the join attributes; \newline
List $I_S = \{s_1, \dots, s_m\}$ of indices of aggregate attributes $\Agg_{s_i}$
present in both $S$ and $R$; \newline
List $I_R = \{r_1, \dots, r_n\}$ of indices of aggregate attributes $\Agg_{r_i}$
present only in $R$;
}
\Fn{AggMergeJoin$(R,S,I_S,I_R)$}{
$i,j \gets 1,1$\;
    \While{$i \leq |R|$ and $j \leq |S|$}{
      \Repeat{$R[i].A = S[j].A$}{
        \While{$i < |R|$ {\bf and} $R[i].A < S[j].A$}{
            $i \gets i + 1$\;
        }
        \lIf{$R[i].A < S[j].A$}{ {\bf return}} 
        \While{$j < |S|$ {\bf and} $R[i].A > S[j].A$}{
                $j \gets j + 1$
        }
        \lIf{$R[i].A > S[j].A$}{ {\bf return}} 
      }  
        $sc \gets 0$\;
        \ForEach{$s \in I_S$}{
        \lIf{$A_s \in \{\MIN, \MAX\}$}{$val_s \gets init[s]$}
        \lIf{$A_s \in \{\SUM, \COUNT\}$}{$val_s \gets 0$}
        }
        \While{$j \leq |S|$ {\bf and} $R[i].A = S[j].A$}{
                $sc \gets sc + S[j].c$\;
                \ForEach{$s \in I_S$}{
                  \lIf{$A_s \in \{\MIN, \MAX\}$}{$val_s \gets A_s (val_s,S[j].\Agg_s)$}
                  \lIf{$A_s \in \{\SUM, \COUNT\}$}{$val_s \gets val_s + S[j].\Agg_s$}
                }              
                $j \gets j + 1$
        }
          \Do{$i \leq |R|$ {\bf and} $R[i].A = R[i-1].A$}{
            $R[i].c \gets sc \cdot R[i].c$ \;       
            \ForEach{$s \in I_S$}{
                \lIf{$A_s \in \{\MIN, \MAX\}$}{$R[i].\Agg_s \gets A_s (val_s,R[i].\Agg_s)$}
                \lIf{$A_s \in \{\SUM, \COUNT\}$}{$R[i].\Agg_s \gets R[i].\Agg_s \cdot val_s$}
            }
            \ForEach{$r \in I_R$}{
                \lIf{$A_s \in \{\SUM, \COUNT\}$}{$R[i].\Agg_r \gets R[i].\Agg_r \cdot sc$}
            }               
            \textbf{emit} $R[i]$\;
            $i \gets i + 1$\;            
          }    
    }
}
\caption{Sort-Merge Join with aggregate propagation}
\label{alg:SortMergeJoin}
\end{algorithm}

In Algorithms~\ref{alg:SortMergeJoin},
we sketch the realisation of 
the AggJoin operator in the style of a merge join.
Again, we use pseudo-code notation to leave out the technical details. 
Here, we assume that relations $R$ and $S$ are joined over a common attribute $A$. 
The generalisation to attributes with different names or to a join over a compound attribute is immediate.
Similarly to the AggHashJoin, also the AggMergeJoin has four input parameters: 
the sorted (on the join attribute $A$
in ascending order) relations $R$ and $S$ as well as the set of indices of the aggregate attributes $I_S$ 
(which have to be
propagated from $S$ to $R$) and the set of indices of the aggregate attributes $I_R$ (which are only
contained in $R$).

The outer while-loop makes sure that we process all tuples of $R$ and of $S$. 
Inside this while loop, we first have a repeat--loop to find the first resp.\ next pair of join partners 
(i.e., the minimal resp.\ next indices
$i$ and $j$ with $R[i].A = S[j].A$). 
Next comes the initialisation of $sc$ and $val_s$ for every $s \in I_S$ exactly as for the AggHashJoin.
The while-loop with the condition $R[i].A = S[j].A$ iterates through all join partners of $R[i]$ in $S$
and computes $sc$ and $val_s$ for every $s \in I_S$ 
exactly as in case of the AggHashJoin. 
In the do-while-loop, we use these local variables $sc$ and $val_s$ to update 
the frequency attribute $c$ and all aggregate attributes $\Agg_j$ with $j \in I_S \cup I_R$ for all tuples $R[i']$ of $R$ 
with $i' \geq i$ and $R[i].A = R[i'].A$, i.e., all tuples in $R$ that coincide on $A$ with $R[i]$. Again,
this is done exactly as in case of the AggHashJoin. That is, for every such tuple $R[i']$, we first
update the frequency attribute $R[i'].c$ and the aggregate attributes $R[i'].\Agg_s$ for every $s \in I_S$
by making use of the local variables $sc$ and $val_s$. We then also update all aggregate attributes $\Agg_r$
(with aggregate function \SUM\ or \COUNT) for all attributes $r \in I_R$. 
When the frequency attribute $c$ and all aggregate attributes of a tuple $R[i']$ have thus been updated,
the tuple $R[i']$  is emitted.

The broadcast-hash join in Spark SQL sends one of the relations to every node that contains a
split of the other relation, and determines the matching tuples on each node separately. 
This last step is realised by a hash join.
We have implemented our optimisation by choosing $S$ as the 
relation that is sent to all nodes where a split of $R$ is located. Then the AggJoin operator
is realised by executing the AggHashJoin between the local split of $R$ and 
the entire relation $S$. 

Of course, it is equally straightforward to implement the AggJoin operator for
join types not supported by Spark SQL such as the block-nested-loops join. In this case, when considering $R$ as the outer relation and 
$S$ as the inner relation, the frequency attribute $c$ and the aggregate attributes $\Aggj$ 
of each tuple 
$r$ in the current block 
of $R$ can be updated as follows: 
we provisionally add an $sc$ attribute and $val_s$ attributes for every 
$s \in I_S$ to every tuple $r$
of the current block of $R$. 
Then relation $S$ is traversed and, for every tuple $s \in S$, the $sc$ and $val_s$ attributes
of each tuple in the current block of $R$
that joins with $s$ is updated analogously to the while-loop with the condition $R[i].A = S[j].A$ 
in case of the sort-merge-join.
When all of $S$ has been processed, then 
the frequency attribute and all aggregate attributes $\Aggj$ of  every tuple in the current block of $R$ 
are updated analogously to the do-while-loop in case of the sort-merge-join.
If the $sc$ attribute of a tuple in $r \in R$ is $0$ (i.e., $r$ has no join partner in $S$), 
then we simply delete this tuple of $R$.



\section{Detailed Benchmark Results}
\label{app:benchmark}
\subsection{Extended Setup Details}
We perform the experiments on a machine with two AMD EPYC 75F3 32-Core CPUs and 960 GB RAM.
On top of this, we use a VM with 60 cores running Ubuntu 22.04.2 LTS and Spark SQL (Version 3.5.0).
For our experiments, we implement the optimisations presented in Section~\ref{sect:OurSystem} 
and the physical operator from Section~\ref{sect:physicalOpt} natively in Spark SQL. 
Our experimental setup, including the versions of all dependencies,
is reproducible through a docker-compose environment available at \url{https://github.com/dbai-tuw/spark-eval}. 

In order to import the benchmark databases into Spark SQL, we run a second container with PostgreSQL 16, from where the data is fetched over JDBC.
We configure Spark with 900 GB maximum executor memory and 60 available cores. Off-heap storage (to disk) is disabled in order to avoid performance degradation caused by unexpected use of disk storage for intermediate results. 
In our experiments, we encountered some instances where the excessive memory consumption caused the query to fail. However, these cases exhibited such extreme performance differences that they do not impact the conclusions drawn from our experimental evaluation. Specifically,  queries where Spark SQL failed due to requiring more than 900GB of memory are easily solved by our implementation, using  only a small fraction of the available resources.

{\renewcommand{\arraystretch}{1.1}%
\begin{table*}[h]
\centering
\caption{Performance on SNAP graphs, compared to graph database systems}
\label{tab:snap-results.app}
\begin{tabular}{|c||c|c|c|c|c||c|c|c|c|c|} 
\hline
\multicolumn{1}{|c||}{} & \multicolumn{5}{c||}{\bf web-Google}                                 & \multicolumn{5}{c|}{\bf com-DBLP}                                 \\
\textbf{Query}         & \bf Spark    & \bf K{\`{U}}ZU   & \bf Neo4j & \bf GuAO       & \bf GuAO$^+$        & \bf Spark       & \bf K{\`{U}}ZU & \bf Neo4j & \bf GuAO      & \bf GuAO$^+$        \\ 
\hhline{|=::====::======|}
path-03                & 27.97\sd{1.5}   &  \bf 2.14\sd{0.0}   &   1686.68\sd{114.3}    & 6.90\sd{0.6}  & 6.08\sd{0.65}  & 6.32\sd{1.1}    &  \bf 0.297\sd{0.0}   &   97.80\sd{0.7}    & 2.35\sd{0.5} &  1.59\sd{0.12}  \\
path-04                & 449.14\sd{26.9} &   747.15\sd{54.4}   &   t.o.    & 7.58\sd{0.6}  & \bf 6.89\sd{0.30}  & 50.97\sd{9.8}   &    42.63\sd{2.7}    &  1725.10\sd{81.6}   & 2.24\sd{0.4} & \bf 1.76\sd{0.16}  \\
path-05                & o.o.m.               &   t.o.   &    t.o.   & 8.95\sd{1.0}  & \bf 7.53\sd{0.48}  & 400.87\sd{15.2} &  762.14\sd{8.1}      & t.o.    & 2.74\sd{0.2} & \bf 2.03\sd{0.25}  \\
path-06                & o.o.m.               &    t.o.   &   t.o.   & 9.37\sd{1.0}  & \bf 8.80\sd{0.25}  & o.o.m.               &    t.o.     &  t.o.  & 2.98\sd{0.2} & \bf 2.18\sd{0.14}  \\
path-07                & o.o.m.               &   t.o.   &    t.o.   & 11.32\sd{0.9} & \bf 9.76\sd{1.21}  & o.o.m.               &     t.o.   &  t.o.   & 3.64\sd{0.2} & \bf 2.38\sd{0.26}  \\
path-08                & o.o.m.               &    t.o.  &   t.o.    & 11.30\sd{2.1} & \bf 10.05\sd{1.49} & o.o.m.               &    t.o.    &  t.o.   & 3.75\sd{0.4} & \bf 2.53\sd{0.30}  \\
tree-01                & 539.11\sd{22.4} &   t.o.   &   t.o.    & 7.73\sd{1.0}  & \bf 6.53\sd{1.11}  & 25.96\sd{4.5}   &   99.16\sd{0.9}   &  t.o.    & 1.95\sd{0.1} & \bf 1.47\sd{0.28}  \\
tree-02                & o.o.m.               &   t.o.  &   t.o.    & 12.43\sd{3.2} & \bf 7.29\sd{0.73}  & 328.88\sd{11.5} &    t.o.    &  t.o.   & 3.02\sd{0.7} & \bf 1.69\sd{0.16}  \\
tree-03                & o.o.m.               &   t.o.   &   t.o.   & 12.21\sd{5.6} & \bf 8.16\sd{0.66}  & o.o.m.               &    t.o,    &  t.o.   & 3.17\sd{0.2} & \bf 1.99\sd{0.16}  \\
\hline
\end{tabular}
\end{table*}
}

We evaluate the performance of the optimisation over 6 benchmark datasets 
with different characteristics. As already mentioned before, the problems faced by
graph database systems on analytical queries with aggregates on top of path queries are 
similar to the ones faced by relational database systems in case of 
aggregates on top of joins. Hence, in addition to widely used relational benchmarks 
(JOB, STATS/STATS-CEB, TCP-H, TPC-DS), we have also chosen a popular benchmark for graph queries
(SNAP) and one that combines features from relational and graph databases (LSQB):  

\begin{itemize}
    \item The \textit{Join Order Benchmark (JOB)}~\cite{DBLP:journals/pvldb/LeisGMBK015} is a benchmark based on the IMDB real-world data set and a large number of realistic join-aggregation queries featuring many joins and various filtering conditions. It was introduced to study the join ordering problem and evaluate the performance of query optimisers.
    \item The \textit{STATS / STATS-CEB}~\cite{DBLP:journals/pvldb/HanWWZYTZCQPQZL21} benchmark serves a similar purpose as the JOB benchmark but with the explicit addition of joins that do not follow FK/PK relationships (but rather FK/FK to keep with typical usage patterns). 
    \item \textit{TPC-H}~\cite{tpch} and TPC-DS~\cite{tpcds} are standard benchmarks for relational databases that cover a wide range of complex real-world workloads.  In addition, we report performance of our running example query from Example~\ref{exp:median} under the name "TPC-H Ex.1". 
    We use scale factor 200 in the generation of the TPC-H data and scale factor 100 for TPC-DS.
    \item The \textit{Large-Scale Subgraph Query Benchmark (LSQB)} \cite{DBLP:conf/sigmod/MhedhbiLKWS21} is a benchmark of 9 graph queries, designed to test the query optimiser and executor of graph databases as well as relational databases. Its schema represents a social network scenario with relations for, e.g., persons, posts, and comments. 
    We generate data with scale factor 300 for LSQB.
    \item The  \textit{SNAP (Stanford Network Analysis Project}
    \cite{snapnets}) dataset is commonly used to benchmark queries on graph data (e.g.,~\cite{DBLP:journals/pacmmod/0005023}).
In particular, we experiment on the following two popular graphs of various sizes:

\medskip
{\small\hspace{1em}
    \begin{tabular}{c|c|c|c}
\centering
        \textbf{Graph} & \textbf{Nodes} & \textbf{Edges} & \textbf{(un)directed} \\
        \midrule
         web-Google & 875,713 & 5,105,039 & directed \\
         com-DBLP & 317,080 & 1,049,866 & undirected \\
    \end{tabular}
    }
\medskip

For our experiments, we evaluate the performance of basic graph queries, namely path queries requiring 
between 3 and 8 joins (i.e., between 4 and 9 edges) 
and three small tree queries. For example, the path with 3 joins (path-03) 
over the web-Google graph is expressed in SQL as

\medskip

{ \small
\begin{minted}{sql}
    SELECT COUNT(*) FROM
    edge e1, edge e2, edge e3, edge e4
    WHERE e1.toNode = e2.fromNode
     AND  e2.toNode = e3.fromNode
     AND  e3.toNode = e4.fromNode
\end{minted}
}

\medskip

\noindent
These queries can be viewed as counting the number of homomorphisms from certain patterns (i.e., paths and trees in this case). This task has recently gained popularity in graph machine learning where the results of the queries are injected into machine learning models (e.g.,~\cite{DBLP:conf/icml/NguyenM20,DBLP:conf/nips/BarceloGRR21,DBLP:journals/corr/abs-2402-08595,bao2024homomorphismcountsstructuralencodings}).

\end{itemize}

For cases where FK/PK relationships exist in the data, the columns \emph{GuAO + FK/PK} 
and  \emph{GuAO$^+$ + FK/PK} report the performance where this information is provided to enable  the 
additional optimisations outlined in Section~\ref{app:FK/PK}. 
The open source version of Spark SQL (Version 3.5.0) that our implementation is based on does not support specifying information about keys directly.
We provide the necessary FK/PK-information via Spark SQL hints in these cases.

\subsection{Extended SNAP Experiments}
The extended results for our SNAP graph query experiments, including graph database systems are shown in Table~\ref{tab:snap-results.app}. 

We see that also specialised graph databases can only solve the easiest queries tested. While the specialised evaluation of edge relations in K\'{U}ZU does show significant benefits on \texttt{path-03}, this disappears already with one more join. This perfectly illustrates the underlying problem of pair-wise join evaluation in these types of queries, as it is inherently bound the fail, regardless of any further optimisations. 

\subsection{FK/PK Evaluation}
To evaluate the effect of FK/PK information we extended a subset of our tests from Section~\ref{sect:Experiments}, by manually providing FK/PK information, and applying the further optimisations from Section~\ref{app:FK/PK}. The results for these experiments are summarised in Table~\ref{tab:my_label_app}. 
We make two important observations from our experimental results for \emph{GuAO + FK/PK} and \emph{GuAO$^+$ + FK/PK}. First, in the case of \emph{GuAO$^+$}, there is little to no gain from the simplification to semi-joins. This confirms that in cases where joins follow FK/PK relationships, the physical operators from Section~\ref{sect:physicalOpt} are in practical terms as efficient as semi-joins. In a sense, this means that using \emph{GuAO$^+$} makes it unnecessary to be aware of FK/PK relationships on join attributes, as the execution is implicitly optimised appropriately in this scenario anyway.

Detailed analysis of the data shows that the small performance differences, in both directions, are primarily due to the potential additional initial grouping operations as described in Section~\ref{app:FK/PK}. The only case where we observe noteworthy improvement is for \emph{GuAO+FK/PK} on STATS-CEB, where the additional use of FK/PK information yields a 60\% speed-up over \emph{GuAO}. The potential for improvements through FK/PK information seems highly data- and query-dependent. Overall, we conclude that, in the context of our method,
FK/PK information is less relevant than might be expected. 

\subsection{TPC-H Query-wise Details}
In Table~\ref{tab:tpchdetails} we additionally report the details for all individual TPC-H queries that are summarised in the end-to-end experiment of 
Section~\ref{sect:Experiments}. We again see a clear correlation between number of joins/relations in the query and achieved speed-up.

{\renewcommand{\arraystretch}{1.1}%
\begin{table*}[t]
    \caption{Comparison of GuAO, GuAO$^+$, with and without the FK/PK optimisations, where applicable 
    }
    \label{tab:my_label_app}
    
\setlength\doublerulesep{8pt}
    \centering
    \begin{tabular}{|c||r|r|r|r||r|r|} 
\hhline{-::----::--}
\textbf{Query}                 & \bf Ref                                                                                                           & \bf GuAO        & \bf GuAO$^+$        & \bf GuAO$^+$ Speedup & \bf GuAO + FK/PK & \bf GuAO$^+$ + FK/PK  \\ 
\hhline{-||----||--}
\hhline{-||----||--}
STATS-CEB e2e                  & 1558\sd{7.3}                                                                                                       & 97.9\sd{6.1}   & \bf 64.8\sd{7.9}       & 24.04 x             & \bf 58.9\sd{0.6}    & 64.0\sd{7.7}     \\ 
\hhline{-||----||--}
TPC-H Q11 \scriptsize{SF200} & 361.0\sd{13.3}                                                                                                     & 346.5\sd{9.4}  & \bf 341.6\sd{19.2}    & 1.06 x              & 350.1\sd{11.6}   & 344.7\sd{13.9}       \\
TPC-H V.1 \scriptsize{SF200}   & 168.4\sd{4.4}                                                                                                      & 107.5\sd{8.9}  & \bf 105.11\sd{3.9}     & 1.6 x               & 106.1\sd{1.1}   & \bf 102.04\sd{3.9}   \\ 

\hhline{-||----||--}
LSQB Q1 \scriptsize{SF300}   & 3096\sd{232} 
& \bf 677\sd{23} & 688\sd{23}         & 4.57 x              & 688\sd{41}      & 689\sd{35}           \\
LSQB Q4 \scriptsize{SF300}   & 602\sd{37}                                                                                                         & 593\sd{15}     & \bf 592\sd{9}          & 1.02x               & \bf 587\sd{11}  & 600\sd{21}           \\ 
\hhline{-||----||--}
\end{tabular}
\end{table*}

{\renewcommand{\arraystretch}{1.1}%
\begin{table*}[t]
    \caption{Detailed runtimes of the TPC-H queries, including relation count and query type 
    }
    \label{tab:tpchdetails}
    
\setlength\doublerulesep{8pt}
    \centering
    \begin{tabular}{|c||r|r|r|r||r|r|} 
\hhline{-::----::--}
\textbf{Query}                 & \bf Ref                                                                                                           & \bf GuAO        & \bf GuAO$^+$        & \bf GuAO$^+$ Speedup & \bf \# joins & \bf class  \\ 
\hhline{-||----||--}
TPC-H Q2 \scriptsize{SF200}  & 179.4\sd{6.5}                                                                                                      & 164.2\sd{4.7}  & \bf 160.6\sd{3.7}  & 1.12 x              & 3               & 0MA                    \\
TPC-H Q11 \scriptsize{SF200} & 361.0\sd{13.3}                                                                                                     & 346.5\sd{9.4}  & \bf 341.6\sd{19.2}    & 1.06 x              & 2  & guarded   \\
TPC-H Ex.1 \scriptsize{SF200}   & 168.4\sd{4.4}                                                                                                      & 107.5\sd{8.9}  & \bf 105.11\sd{3.9}     & 1.6 x               & 4  & guarded  \\ 
TPC-H Q3 \scriptsize{SF200}   & 798.2\sd{179}                                                                                                      & -  & \bf 728.4\sd{27}     & 1.1 x               & 2  & pw-guarded  \\ 
TPC-H Q12 \scriptsize{SF200}   & \bf 377.0\sd{13.1}                                                                                                      & -  & 381.0\sd{11.8}     & 0.99 x               & 1   & pw-guarded   \\ 
TPC-H Q13 \scriptsize{SF200}   & 440.5\sd{11.1}                                                                                                      & -  & \bf 440.8\sd{4.8}     & 1 x               & 1   & pw-guarded   \\ 
TPC-H Q16 \scriptsize{SF200}   & 106.1\sd{0.95}                                                                                                     & -  & \bf 103.16\sd{3.6}     & 1.03 x               & 1   & pw-guarded   \\ 
TPC-H Q17 \scriptsize{SF200}   & 1495.0\sd{80.5}                                                                                                      & -  & \bf 1335.5\sd{65.2}     & 1.12 x               & 1   & pw-guarded   \\ 

\hhline{-||----||--}
\end{tabular}
\end{table*}

}

\end{document}